\RequirePackage{lineno} 

\documentclass[twocolumn,preprintnumbers,amsmath,amssymb,nofootinbib]{revtex4}

\usepackage{graphicx}
\usepackage{dcolumn}
\usepackage{bm}
 
\usepackage{epstopdf}

\def\bea{\begin{eqnarray}}
\def\eea{\end{eqnarray}}

\def\pp{\mbox{$p$-$p$}}

\def\pa{\mbox{$p$-A}}

\def\pbpb{\mbox{Pb-Pb}}
\def\ppb{\mbox{$p$-Pb}}

\def\pn{\mbox{$p$-N}}
\def\aa{\mbox{A-A}}
\def\nn{\mbox{N-N}}

\def\pt{$p_t$}
\def\mt{$m_t$}
\def\yt{$y_t$}

\def\nch{$n_{ch}$}
\def\mmpt{$\bar p_t$}

\begin{document} 

\setlength{\pdfpagewidth}{8.5in}
\setlength{\pdfpageheight}{11in}

\setpagewiselinenumbers
\modulolinenumbers[5]

\addtolength{\footnotesep}{-10mm}\

\preprint{version 1.5\textsl{}}

\title{Systematic analysis of identified-hadron $\bf p_t$ spectra from 13 TeV p-p collisions
}

\author{Thomas A.\ Trainor}\affiliation{University of Washington, Seattle, WA 98195}


\date{\today}

\begin{abstract}
Identified-hadron (PID) $p_t$ spectra from 13 TeV $p$-$p$ collisions are compared with a two-component (soft+hard) model (TCM) that accurately distinguishes jet-related hadron production (hard component) from nonjet projectile-nucleon dissociation (soft component). The present $p$-$p$ study is similar to and is guided by recent TCM studies of PID spectra from 5 TeV $p$-Pb collisions. The combined analyses serve to establish a well-understood quantitative description of PID hadron production in small collision systems as a control experiment. The control can then be contrasted with conventional interpretations of collision data from more-central A-A collisions as indicating formation of a quark-gluon plasma (QGP). PID $p_t$ spectra from 13 TeV $p$-$p$ collisions exhibit simple consistency with spectra from 5 TeV $p$-Pb collisions. Hadron species abundances are consistent with statistical-model trends predicted prior to commencement of the large hadron collider program. Differential spectrum structure and various ratio measures are quantitatively explained by the TCM, including its jet contribution, and admit no room for claims of hydrodynamic flows in small collision systems.
\end{abstract}


\maketitle

 \section{Introduction}
 
This article reports a study of the systematic variation of identified-hadron (PID) \pt\ spectra from 13 TeV \pp\ collisions reported by Ref.~\cite{alicepppid}. The analysis method follows from several recent PID spectrum studies based on a two-component (soft+hard) model (TCM) of hadron production in high-energy nuclear collisions~\cite{ppbpid,pidpart1,pidpart2}. A basic goal is to extract all available information from PID spectra down to the level of statistical uncertainties and to report it in formats that are suggestive of proper physical interpretation and quantitatively comparable with previous particle-physics results (e.g.\ jet properties).

A more immediate goal is providing substantive response to claims of ``collectivity'' (i.e.\ flows) based on data features associated with small collision systems (e.g.\ \pp, \pa\ collisions). The apparent presence of collective motion within collisions is then interpreted to support  claims of quark-gluon plasma (QGP) formation in those systems~\cite{nature0}. Such claims are counterintuitive based on conventional understanding of QCD and thus merit careful examination of arguments and evidence in their favor.

Reference~\cite{alicepppid} observes that at low \pt\ ``...collective phenomena are observed in...[\pp, \ppb\ and \aa] collisions...'' with several examples given. It then asserts that ``...in order to describe bulk particle production in \aa\ collisions, one usually relies on hydrodynamic and thermodynamic modeling....'' Such statements typically set the stage for argument by analogy: If certain data features were previously interpreted to signal collectivity and QGP formation in more-central \aa\ collisions, and those same features appear in small collision systems, then it is reasonable to claim collectivity (flows) and QGP in small systems as well, no matter that particle and energy densities there are much smaller. 

That framework then provides context for presentation and interpretation of PID \pt\ spectra in Ref.~\cite{alicepppid}. ``We observe that the measured $p_T$ spectra become harder [i.e.\ slope magnitude at low \pt\ decreases] with increasing [event multiplicity], and the effect is more pronounced for protons.'' Such variation ``...is also observed in \pbpb\ collisions...where it is usually associated with the hydrodynamical evolution of the system.'' It is further stated that in \pbpb\ collisions such trends, interpreted in terms of radial expansion (flow), are ``....studied in the context of the Boltzmann-Gibbs Blast-Wave model.'' Since the \pp\ spectra ``...are highly reminiscent to those in \ppb\ and \pbpb, it is interesting to check whether the Blast-Wave model can be extended to describe \pp\ collisions.''

The present analysis addresses those arguments and claims in the following way. The PID TCM developed in Refs.~\cite{ppbpid,pidpart1,pidpart2} is introduced. As a starting point, TCM parameter values from 5 TeV \ppb\ collisions are presented. Previously established $\sqrt{s}$ trends for TCM parameters are used to extrapolate from 5 TeV to 13 TeV. A correction for proton inefficiency developed for 5 TeV \ppb\ collisions in Ref.~\cite{pidpart1} is updated to accommodate 13 TeV \pp\ spectra. Parameters that describe the fractions of total charge densities belonging to each hadron species are derived directly from spectrum data. The quality of the resulting PID TCM description is then evaluated based on Z-scores. With the exception of pions the TCM describes spectrum data within statistical uncertainties.

Given an accurate TCM data representation, and the now well-understood physical interpretation of TCM components, various data presentations and interpretations from Ref.~\cite{alicepppid} are confronted. As in previous cases spectrum data (and data-ratio) features conventionally interpreted as representing collectivity or flows are quantitatively described in terms of minimum-bias jet production. In particular, PID yield- and spectrum-ratio results are quantitatively examined and interpreted within a jet context. Blast-wave (BW) model fits are considered and rejected via Z-scores based on poor fit quality over limited \pt\ intervals. PID spectrum data from Ref.~\cite{alicepppid} do carry much information but do not appear to support claims of collectivity and QGP formation in small systems. 

This article is arranged as follows:
Section~\ref{alicedata} introduces PID spectrum data from 13 TeV \pp\ collisions reported in Ref.~\cite{alicepppid}.
Section~\ref{spectrumtcm} defines a corresponding PID spectrum TCM.
Section~\ref{pppidtcm} describes the process of estimating TCM model parameters. 
Section~\ref{tcmfinal} reports final PID TCM parameter values for 13 TeV \pp\ collisions.
Section~\ref{quality} evaluates TCM data description quality based on Z-scores.
Section~\ref{pidratios} discusses PID yield and spectrum ratios in a TCM context.
Section~\ref{bwave} evaluates the quality and interpretability of BW fits to PID \pp\ spectra.
Section~\ref{sys} discusses systematic uncertainties. 
Sections~\ref{disc} and~\ref{summ} present discussion and summary. 

\section{13 $\bf TeV$ $\bf p$-$\bf p$ PID Spectrum data} \label{alicedata}

The 13 TeV \pp\ PID spectrum data used for this analysis were reported by Ref.~\cite{alicepppid}. Spectrum data were derived from 143 million \pp\ collision events satisfying an INEL $> 0$ minimum-bias trigger (at least one charged particle within $|\eta| < 1$). Events were sorted into ten multiplicity classes based on charge accumulated within a V0 detector (V0M amplitude). Mean charge densities $dN_{ch} / d\eta \rightarrow \bar \rho_0$  as integrated within $|\eta| < 0.5$ or angular acceptance $\Delta \eta = 1$ are 26.0, 20.0 16.2, 13.8, 12.0 10.0 7.95, 6.3, 4.5 and 2.55 for $n \in [1,10]$ respectively. 

\subsection{$\bf p$-$\bf p$ PID  spectrum data} \label{piddataa}

\begin{figure}[h]
	\includegraphics[width=1.65in]{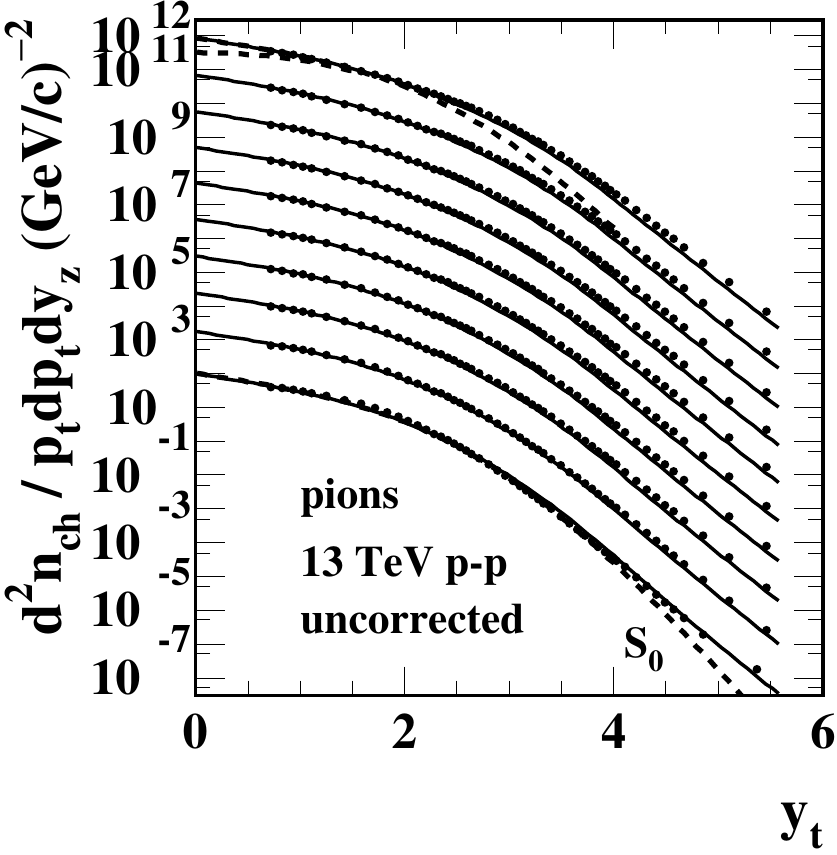}
  	\includegraphics[width=1.65in]{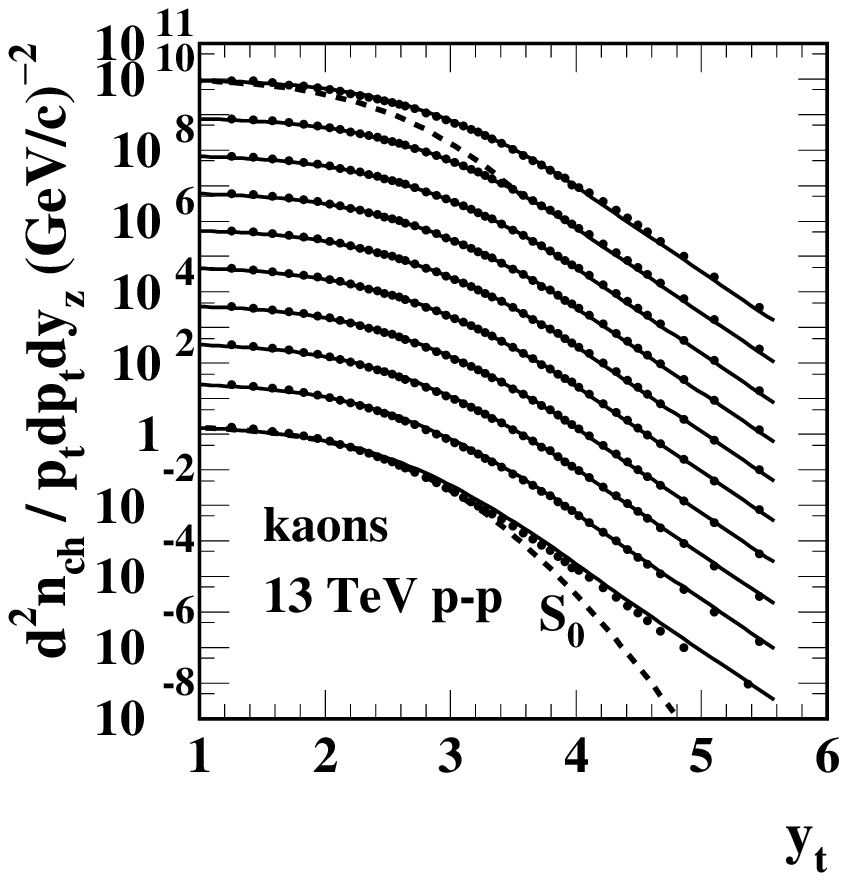}
\put(-145,105) {\bf (a)}
\put(-23,105) {\bf (b)}
\\
  	\includegraphics[width=1.65in]{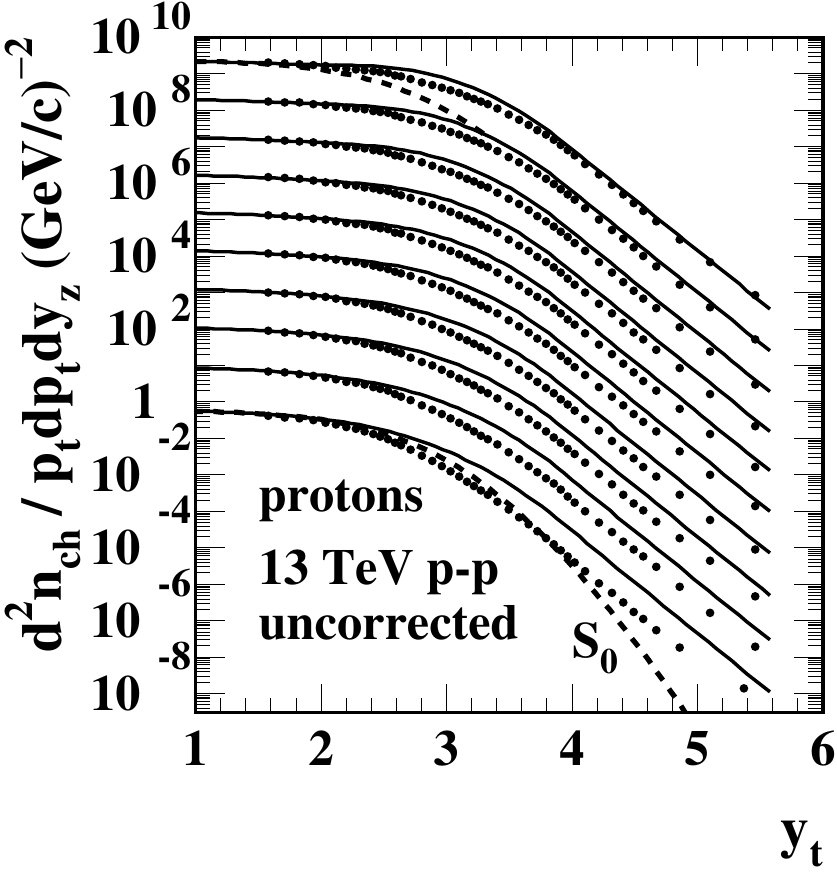}
 \includegraphics[width=1.65in]{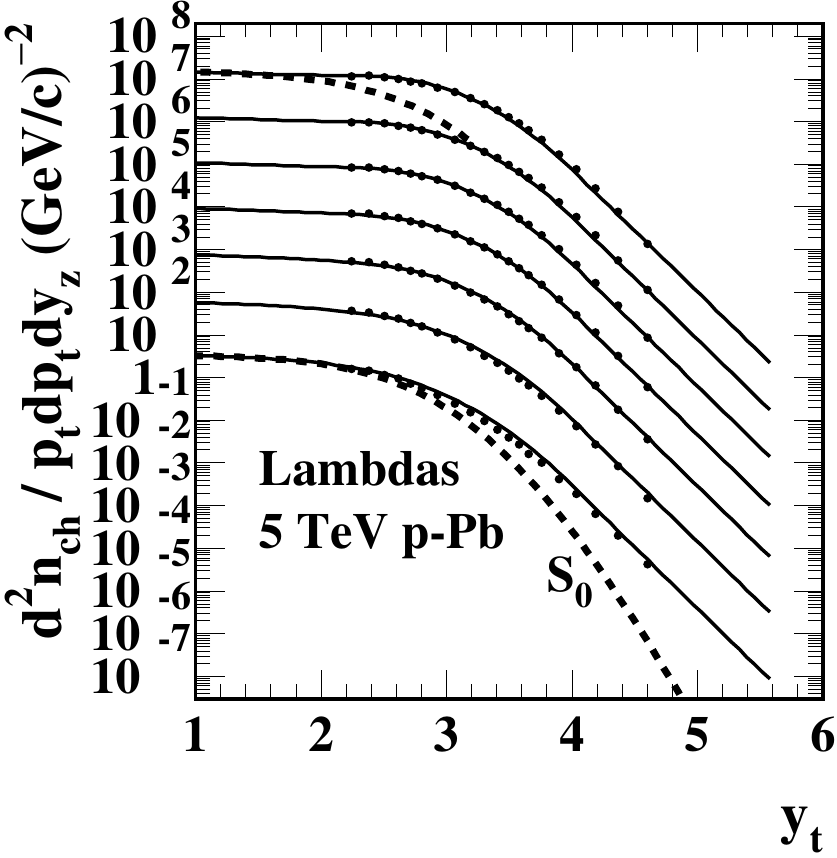}
\put(-145,105) {\bf (c)}
 \put(-23,105) {\bf (d)}
 \\
	\caption{\label{piddata}
\pt\ spectra for identified hadrons:
(a) pions,
(b) charged kaons and
(c) protons from 13 TeV \pp\ collisions~\cite{alicepppid} and
(d) Lambdas from 5 TeV \ppb\ collisions~\cite{aliceppbpid}.
Solid curves represent the PID spectrum TCM from Sec.~\ref{tcmfinal}. Data-model discrepancies in (a) and (c) are discussed in Sec.~\ref{pionproton}. 
} 
\end{figure}

Figure~\ref{piddata} shows PID spectrum data from 13 TeV \pp\ collisions reported in Ref.~\cite{alicepppid} (points) as densities on \pt\ vs transverse rapidity \yt\ with pion mass assumed. Published spectra have been divided by $p_t$ relative to the presentation in Fig.~1 of Ref.~\cite{alicepppid}. The spectra have been scaled by powers of 10 according to $10^{n-1}$ where $n \in [1,10]$ is the centrality class index and $n = 10$ is here most central, following the practice in Ref.~\cite{alicepppid}. Elsewhere in this paper event class $n = 1$ is most central (see $\bar \rho_0$ values listed above). Solid curves are full TCM parametrizations finalized in Sec.~\ref{paramfinal}. Dashed curves are TCM soft components $ z_{si }\bar \rho_s \hat S_0(y_t)$. The differences represent minimum-bias jet contributions $ z_{hi} \bar \rho_h \hat H_0(y_t)$.

The baryon data in panels (c) and (d) (the Lambda data from 5 TeV \ppb\ collisions~\cite{pidpart2} are included as a reference) are expected to correspond closely. However, the proton data in panel (c) fall well below the TCM proton prediction (solid curves). A similar apparent proton inefficiency for 5 TeV \ppb\ collisions was analyzed and corrected in Sec.~III B of Ref.~\cite{pidpart1}. The same  procedure is applied to 13 TeV \pp\ proton data  in Sec.~\ref{correct} below. Pion spectra in panel (a) fall well {\em above} the TCM prediction. Those deviations and a possible relation to the proton inefficiency are discussed in Sec.~\ref{pionproton}.

The plotting format in Fig.~\ref{piddata} is desirable for two reasons: 
(a) Soft component $\hat S_0(y_t)$ that (for example) describes low-\pt\ data within their statistical uncertainties for 5 TeV \ppb\ collisions~\cite{pidpart1,pidpart2} closely approximates a Boltzmann exponential on \mt\ at lower \mt\ and thus follows an $A - m_i \cosh(y_t)/T$ trend on this semilog format on \yt\ (where $m_i$ is the mass for hadron species $i$ and A is some constant). Pions are an exception because of a resonance contribution but that is easily accommodated~\cite{pidpart1}.
(b) Hard component $\hat H_0(y_t)$ that describes high-\pt\ data within their statistical uncertainties for 5 TeV \ppb\ collisions follows an exponential $\exp(-q\, y_t)$ at higher \yt\ equivalent to a power law on \mt\ or \pt\ that manifests in this format as a straight line for $y_t > 4$. 

In contrast, the plotting format chosen for Fig.~1 of Ref.~\cite{alicepppid} shows spectra in the form $d^2 N_{ch} / dp_t dy_z$ that is missing a factor $1/p_t$ appropriate for momentum in the transverse plane and thus does not approximate a Boltzmann exponential for low \mt. The data are presented on linear \pt\ rather than logarithmic \yt, and data binning is also defined on \pt\ requiring dramatic variation of bin widths to control statistical uncertainties. {\em Equal bin widths on \yt} would accomplish the same goal. On linear \pt\ there is no hint of the simple data trends evident on \yt. Ratios to minimum-bias INEL $> 0$ confuse several issues but are dominated by the strongly varying hard/soft $\leftrightarrow$ jet/nonjet ratio in spectra that is of fundamental importance for the understanding of \pp\ collision dynamics.

\subsection{$\bf p$-$\bf p$ PID spectrum data interpretation}

Reference~\cite{alicepppid} interprets spectrum data as follows: ``We observe that the measured $p_T$ spectra become harder with increasing [$\bar \rho_0$], and the effect is more pronounced for protons,'' 
where ``harder'' corresponds to reduced spectrum slope (``flattening'') at lower \pt\ and is said to be similar to observations in \aa\ collisions: ``...the mass dependence of spectral shape modification is also observed in \pbpb\ collisions...{where it is usually associated with the hydrodynamical evolution of the system}.'' 
 
In Ref.~\cite{aliceppbpid} it is suggested that commonalities between \pp\ data and those from \pbpb\ collisions imply the presence of collective flow also in \ppb\ collisions: ``In heavy-ion [\aa] collisions, the flattening of transverse momentum distribution and its mass ordering find their {\em natural explanation in the collective radial expansion of the system} [emphasis added].'' Reference~\cite{alicepppid} presents a similar argument concerning \pp\ collisions: ``In large collision systems such as \pbpb\, multiplicity-dependent modifications of hadron $p_T$ spectra can be interpreted as the hydrodynamical radial expansion of the system and studied in the context of the Boltzmann-Gibbs Blast-Wave model. ... As the  trends...measured in pp collisions are highly reminiscent to those in p-Pb and Pb-Pb, it is interesting to check whether the Blast-Wave model can be extended to describe pp collisions.'' Section~\ref{bwave} below provides a response to that proposal.

Concerning the high-\pt\ region: ``At higher $p_T$ ($\geq 8$ GeV/c), we find that slopes of particle spectra become independent of the multiplicity class considered, as expected from pQCD calculations [Ref.~\cite{kretzer} is cited].'' That characteristic of \pt\ spectra is abundantly clear from the straight-line trends in the format of Fig.~\ref{piddata} above, and the power-law trend clearly begins near 4 GeV/c ($y_t \approx 4$) in all \pp\ and \pa\ collision systems. The same trend has been reported in Refs.~\cite{hardspec,fragevo,mbdijets,ppbpid,pidpart1,pidpart2} for example. Reference~\cite{kretzer} does not speak to that aspect of single-particle spectrum properties since it deals only with fragmentation functions (FFs) characterizing individual reconstructed jets. Jet contributions to high-energy \pp\ \pt\ spectra and angular correlations have been studied in detail (e.g.\ Ref.~\cite{pptheory}). The approximate power-law trend at higher \pt\ for single-particle A-B spectra results from the {\em underlying jet energy spectrum} that is a separate issue~\cite{jetspec2}. The spectrum hard component (what dominates spectra at higher \pt) is {\em quantitatively predicted} by a convolution of measure FFs with a measured jet energy spectrum~\cite{fragevo}. Biases resulting from event-selection methods may cause variation of power-law trends~\cite{tomnewppspec}.

Whatever the current popular interpretation of spectrum trends in \aa\ collisions may be, the interpretation of \pp\ collisions {\em as a fundamental reference system} should be undertaken {\em sui generis} employing the full understanding of elementary nuclear collisions established over forty years by the high-energy (particle-physics) community.

\section{$\bf p$-$\bf p$ PID Spectrum TCM}  \label{spectrumtcm}

The TCM for \pp\ collisions utilized in this study is the product of phenomenological analysis of data from a variety of collision systems and data formats~\cite{ppprd,ppquad,alicetomspec,tommpt}. As such it does not represent imposition of {\em a priori} physical models. Physical interpretation of TCM soft and hard components has been derived {\em a posteriori} by comparing inferred TCM characteristics with other relevant measurements~\cite{hardspec,fragevo}, in particular measured jet characteristics~\cite{eeprd,jetspec2}. Development of the TCM contrasts with data models based on {\em a priori} physical assumptions such as the BW model~\cite{blastwave}.  The TCM does not result from fits to individual spectra (or other data formats), which would require many parameter values. The few TCM parameters have simple $\log(\sqrt{s})$ trends on collision energy and extrapolations from minimum-bias \pp\ trends and are required to be quantitatively consistent across multiple A-B collision systems.

In what follows, a PID \pp\ spectrum TCM is defined, TCM parameters derived for 5 TeV \ppb\ collisions from Ref.~\cite{pidpart1} are described, the energy dependence of TCM parameters for nonPID \pp\ spectra is introduced from Refs.~\cite{tomnewppspec} and \cite{alicetomspec}, and those results are combined to produce {\em predicted} PID TCM spectrum parameters for 13 TeV \pp\ collisions. Those parameter values are then refined based on comparison of TCM and data in Sec.~\ref{tcmfinal}.

\subsection{$\bf p$-$\bf p$ spectrum TCM for unidentified hadrons} \label{unidspec}

The \pt\ or \yt\ spectrum TCM is by definition the sum of soft and hard model components, their details being inferred from data (e.g.\ Ref.~\cite{ppprd}). For \pp\ collisions
\bea  \label{rhotcm}
\bar \rho_{0}(y_t,n_{s}) &\approx& \bar \rho_{s}(n_{s}) \hat S_{0}(y_t) + \bar \rho_{h}(n_{s}) \hat H_{0}(y_t),
\eea
where $n_s$ serves as an event-class index, and factorization of the dependences on \yt\ and \nch\ is a central feature of the spectrum TCM inferred from 200 GeV \pp\ spectrum data in Ref.~\cite{ppprd}. The motivation for transverse rapidity $y_{ti} \equiv \ln[(p_t + m_{ti})/m_i]$ (applied to hadron species $i$) is described in Sec.~\ref{tcmmodel}. The \yt\ integral of Eq.~(\ref{rhotcm}) is $\bar \rho_0 =  \bar \rho_s + \bar \rho_h$, a sum of soft and hard charge densities with $\bar \rho_x = n_x / \Delta \eta$. $\hat S_{0}(y_t)$ and $\hat H_{0}(y_t)$ are unit-normal model functions approximately independent of \nch, and the centrally-important  relation $\bar \rho_{h} \approx \alpha \bar \rho_{s}^2$ with $\alpha \approx O(0.01)$ is inferred from \pp\ spectrum data~\cite{ppprd,ppquad,alicetomspec}. Equation $\bar \rho_0 \approx \bar \rho_s + \alpha \bar \rho_s^2$ is solved to obtain $\bar \rho_s$ from measured $\bar \rho_0$.

To define model functions and other aspects of the \pp\ spectrum TCM, measured hadron spectra are rescaled by charge-density soft component $\bar \rho_s$ to have the form
\bea  \label{norm}
X(y_t,n_s) &\equiv & \frac{\bar \rho_{0}(y_t;n_{ch})}{\bar \rho_{s}}  
=  \hat S_{0}(y_t) +   x(n_s)  \hat H_{0}(y_t),
\eea
where $x(n_s) \equiv \bar \rho_{h}/\bar \rho_{s} \approx \alpha \bar \rho_{s}$. The form of model $\hat S_{0}(y_t)$ is defined by data expressed as $X(y_t)$ in the limit $n_{ch} \rightarrow 0$. The form of $\hat H_{0}(y_t)$ is defined by  spectrum data contributions complementary to soft-component model $\hat S_{0}(y_t)$.

\subsection{$\bf p$-$\bf p$ spectrum TCM for identified hadrons}  \label{pidspec}

Given the \pp\ spectrum TCM for unidentified-hadron spectra in Eq.~(\ref{rhotcm}) a corresponding TCM for identified hadrons can be generated by assuming that each hadron species $i$ comprises certain {\em fractions} of soft and hard TCM components denoted by $z_{si}$ and $z_{hi}$  (both $\leq 1$). The PID spectrum TCM can then be written as
\bea \label{pidspectcm}
\bar \rho_{0i}(y_t,n_s) &=& S_i(y_t,n_s) + H_i(y_t,n_s)
\\ \nonumber
&\approx&  z_{si}(n_s) \bar \rho_{s} \hat S_{0i}(y_t) +   z_{hi}(n_s) \bar \rho_{h} \hat H_{0i}(y_t) 
\eea
with rescaled spectra
\bea \label{xi}
 X_i(y_t,n_s) &\equiv& \frac{\bar \rho_{0i}(y_t,n_s)}{ \bar \rho_{si}(n_s)}
\nonumber \\
&\approx&  \hat S_{0i}(y_t) +  \tilde z_i(n_s) x(n_s) \hat H_{0i}(y_t),
\eea
where $\tilde z_i(n_s) \equiv z_{hi}(n_s)/z_{si}(n_s)$ and unit-integral model functions $\hat S_{0i}(y_t)$ and $\hat H_{0i}(y_t)$ may depend on hadron species $i$.
For identified hadrons of species $i$ the rescale factor $ \bar \rho_{si} = z_{si}(n_s)\bar \rho_{s}$ can be expressed in terms of factor $z_{si}(n_s)$ defined in Eq.~(\ref{zsix}), but for \pp\ collisions $\nu \rightarrow 1$ in that equation. Unit-normal model functions $\hat S_{0i}(y_t)$ and $\hat H_{0i}(y_t)$ must be determined for each hadron species, but close correspondence to unidentified-hadron models is expected. The further differential spectrum quantity
\bea \label{yi}
Y_i(y_t,n_s) &\equiv& \frac{1}{\tilde z_i(n_s) x(n_s)} [X_i(y_t,n_s) - \hat S_{0i}(y_t)]
\\ \nonumber
&\approx & \hat H_{0i}(y_t)
\eea
may be compared with model functions $ \hat H_{0i}(y_t)$. Those expressions have been used in a previous PID TCM spectrum analysis~\cite{ppbpid}. A more precise strategy developed in Refs.~\cite{pidpart1,pidpart2} is utilized in Sec.~\ref{pppidtcm}.

\subsection{PID TCM model functions} \label{tcmmodel}

For spectra structured as in Eq.~(\ref{pidspectcm}) and the trend $x(n_s) \sim n_s  \sim n_{ch}$ the soft-component model function is defined as the limit of $X_i(y_t,n_s)$ as \nch\ (or $n_s$) goes to zero. As noted, hard components of {\em data} spectra are then defined as complementary to {\em model} soft components.

The data soft component for a specific hadron species $i$ (except pions) is typically well described by a L\'evy distribution as a density on $m_{ti}  = \sqrt{p_t^2 + m_i^2}$. The unit-integral soft-component model is 
\bea \label{s00}
\hat S_{0i}(m_{ti}) &=& \frac{A_i}{[1 + (m_{ti} - m_i) / n_i T_i]^{n_i}},
\eea
where $m_{ti}$ is the transverse mass for hadrons $i$ of mass $m_i$, $n_i$ is the L\'evy exponent, $T_i$ is the slope parameter and coefficient $A_i$ is determined by the unit-integral condition. Parameters $(T,n)$ for 5 TeV \ppb\  data in Ref.~\cite{pidpart1} are slightly adjusted for 13 TeV \pp\ data (see Table~\ref{engparamsy}). As defined, the soft-component model is a density on \pt\ or \mt\ (since $m_t dm_t = p_t dp_t$) which may be plotted vs \yt.
 
The unit-integral hard-component model is simply defined on pion $y_{t\pi} \equiv \ln((p_t + m_{t\pi})/m_\pi)$ as a Gaussian, with exponential (on $y_t$) or power-law (on $p_t$) tail at higher \yt\
\bea \label{h00}
\hat H_{0}(y_t) &\approx & B \exp\left\{ - \frac{(y_t - \bar y_t)^2}{2 \sigma^2_{y_t}}\right\}~~~\text{near mode $\bar y_t$}
\\ \nonumber
&\propto &  \exp(- q y_t)~~~\text{for higher $y_t$ -- the tail},
\eea
where the transition from Gaussian to exponential on \yt\ is determined by slope matching~\cite{fragevo}. The $\hat H_0$ tail density varies on \pt\ approximately as power law $1/p_t^{q + 2.2}$. Coefficient $B$ is determined by the unit-integral condition. Initial PID model parameters $(\bar y_t,\sigma_{y_t},q)$ as in Table~\ref{pidparams} are also derived from \ppb\  data in Ref.~\cite{pidpart1} (see Table~\ref{engparamsy}).

Spectra for the present study are presented as densities on \pt\ plotted vs pion rapidity $y_{t\pi}$ with pion mass assumed.  $\hat S_{0i}(m_{ti})$ for species $i$ is defined by Eq.~(\ref{s00}). $\hat H_{0}(y_{t\pi})$ in Eq.~(\ref{h00}) is defined as a density on $y_{t\pi}$ where it has a simple form and is then converted to $\hat H_{0}(p_t)$ via the Jacobian factor $y_{t\pi} / m_{t\pi} p_t $. In general, plotting spectra as densities on \pt\ against logarithmic variable \yt\ permits superior visual access to important low-\pt\ structure where the {\em majority of jet fragments appears}. A further motivation is comparison of spectrum hard components interpreted to arise from a common underlying jet spectrum on \pt\ \cite{hardspec,fragevo}, in which case $y_{t\pi}$ serves simply as a logarithmic measure of hadron \pt\ with well-defined zero. 

\subsection{5 TeV $\bf p$-$\bf Pb$ TCM PID spectrum parameters} \label{pidfracdata}

In this subsection TCM model parameters for PID spectra from 5 TeV \ppb\ collisions reported in Ref.~\cite{pidpart1} are presented. If \ppb\ collisions are linear superpositions of \pn\ collisions, as seems apparent from centrality dependence of \ppb\ spectrum data, then this parametrization is a starting point for a PID TCM for \pp\ collisions.

Table~\ref{pidparams} shows TCM model parameters for hard component $\hat H_0(y_t)$ (first three) and soft component $\hat S_0(y_t)$ (last two). Hard-component parameters vary slowly but significantly with hadron species. Modes $\bar y_t$ shift to higher \yt\ with increasing mass. Widths $\sigma_{y_t}$ are greater for mesons than for baryons. Only $K_s^0$ and $\Lambda$ \ppb\ data extend to sufficiently high \pt\ to determine exponent $q$ which is substantially greater for baryons than for mesons.

Parameter values in Table~\ref{pidparams} for 5 TeV \ppb\ collisions define a {\em fixed} TCM reference independent of centrality that describes the {\em most central} event class (wherein $\bar y_t \approx 3.0$ for baryons)~\cite{pidpart1}. In Ref.~\cite{pidpart2} variation of some hard-component model parameters is determined so as to describe all event classes within statistical uncertainties (see Fig.~4 of Ref.~\cite{pidpart2}). The required variations are linear on hard/soft ratio $x(n_s) \nu(n_s)$: hard-component modes shift to higher \yt\ for baryons while hard-component widths above the mode decrease for mesons.

\begin{table}[h]
	\caption{TCM model parameters for identified hadrons from 5 TeV \ppb\ collisions from Table~VI of Ref~\cite{pidpart1}: hard-component parameters $(\bar y_t,\sigma_{y_t},q)$ and soft-component parameters $(T,n)$. Numbers without uncertainties are adopted from a comparable hadron species with greater accuracy. 
	}
	\label{pidparams}
	\begin{center}
		\begin{tabular}{|c|c|c|c|c|c|} \hline
			& $\bar y_t$ & $\sigma_{y_t}$ & $q$ & $T$ (MeV) &  $n$  \\ \hline
			$ \pi^\pm $     &  $2.46\pm0.01$ & $0.57\pm0.01$ & $4.1\pm1$ & $145\pm3$ & $8.5\pm0.5$ \\ \hline
			$K^\pm$    & $2.65$  & $0.57$ & $4.1$ & $200$ & $14$ \\ \hline
			$K_s^0$          &  $2.65\pm0.01$ & $0.57\pm0.01$ & $4.1\pm0.1$ & $200\pm5$ & $14\pm2$ \\ \hline
			$p$        & $2.99\pm0.01$  & $0.47$ & $5.0$ & $210\pm10$ & $14\pm4$ \\ \hline
			$\Lambda$       & $2.99\pm0.01$  & $0.47\pm0.01$ & $5.0\pm0.05$  & $210$ & $14$ \\ \hline	
		\end{tabular}
	\end{center}
\end{table}

Soft-component model parameter $T \approx 145$ MeV for pions is consistent with that for unidentified hadrons found to be universal over all A-B collision systems and collision energies~\cite{alicetomspec}. The values for higher-mass hadrons are substantially greater. L\'evy exponent $n \approx 8.5$ for pions is also consistent with that for unidentified hadrons at 5 TeV and has a  $\log(\sqrt{s}/\text{10 GeV})$ energy dependence~\cite{alicetomspec}. Exponent $n$ values for more-massive hadrons are not well-defined because the hard-component fraction is much greater than for pions. Varying $n$ then has little impact on the overall spectra.

Table~\ref{otherparamsx} shows PID parameters $z_{0i}$ and $\tilde z_i = z_{hi}/ z_{si}$ for five hadron species determined from PID spectrum data in Ref.~\cite{pidpart1}. While $z_0$ was found to be independent of \ppb\ centrality within uncertainties the \ppb\ $\tilde z_i(n_s)$ exhibit significant centrality dependence as shown in Fig.~8 of Ref.~\cite{pidpart1}. It is notable that the $\tilde z_i(n_s)$ depend only on hadron mass, not on strangeness or baryon identity. Measurements of individual centrality trends for $z_{si}(n_s)$ and $z_{hi}(n_s)$ are presented in Sec.~IV of Ref.~\cite{pidpart1}.  Individual fractions $z_s(n_s)$ and $z_h(n_s)$ may also be derived from model parameters $\tilde z_i(n_s)$ and $z_{0i}$ via the relation
\bea \label{zsix}
z_{si}(n_s) &=& \frac{1 + x(n_s) \nu(n_s)}{1 + \tilde z_i(n_s)x(n_s) \nu(n_s)} z_{0i},
\eea
with $z_{hi}(n_s) = \tilde z_i(n_s)z_{si}(n_s)$. 5 TeV \ppb\ geometry (centrality) parameters $x(n_s)$ and $\nu(n_s)$ are determined in Refs.~\cite{alicetommpt,tommpt,tomglauber} based on ensemble-mean \mmpt\ data. For \pp\ collisions $\nu \equiv 2 N_{bin} / N_{part} \rightarrow 1$. The $\tilde z_i = z_{hi}/ z_{si}$ values included in Table~\ref{otherparamsx} represent averages over \ppb\ centrality. Given the TCM expression in Eq.~(\ref{pidspectcm}) the correct rescaling via $\bar \rho_{si} = z_{si}(n_s) \bar \rho_{s}$ results in data spectra coinciding with $\hat S_0(y_t)$ as $y_t \rightarrow 0$ for all centralities. 

\begin{table}[h]
	\caption{\label{otherparamsx}
		TCM model parameters for identified hadrons from 5 TeV \ppb\ collisions in Ref.~\cite{pidpart1}. Numbers without uncertainties are adopted from a comparable hadron species with greater accuracy. Values for $\tilde z_i = z_{hi} / z_{si}$ are averages over \ppb\ centrality. Parameters $ \bar p_{ts}$ and $\bar p_{th}$ are determined by model functions $\hat S_0(y_t)$ and $\hat H_0(y_t)$ defined by Table~\ref{pidparams}.  
	}
	\begin{center}
		\begin{tabular}{|c|c|c|c|c|} \hline
			&   $z_0$    &  $\tilde z_i$ &   $ \bar p_{ts}$ (GeV/c)  & $ \bar p_{th}$ (GeV/c)  \\ \hline
			$ \pi^\pm$        &   $0.82\pm0.01$  & $0.88\pm0.05$  & $0.40\pm0.02$ &    $1.15\pm0.03$  \\ \hline
			$K^\pm $   &  $ 0.128\pm0.002$   &  $2.7\pm0.2$ &  $0.60$&  $1.34$   \\ \hline
			$K_s^0$        &  $0.064\pm0.002$ &  $2.7\pm0.2$ &  $0.60\pm0.02$ &   $1.34\pm0.03$  \\ \hline
			$p $        & $ 0.065\pm0.002$    &  $5.6\pm0.2$ &  $0.73\pm0.02$&   $1.57\pm0.03$   \\ \hline
			$\Lambda $        &  $0.034\pm0.002$    & $6.5\pm0.5$ &   $0.76\pm0.02$ &    $1.65\pm0.03$ \\ \hline	
		\end{tabular}
	\end{center}
\end{table}

The results above for 5 TeV \ppb\ collisions can be compared with final results for 13 TeV \pp\ collisions presented in Table~\ref{otherparamsxx}. With the exception of pion $\tilde z_i$ the $z_{xi}$ values determined for the two collision systems are consistent within data uncertainties. The significant change for pion $\tilde z_i$ is explained in Sec.~\ref{fracsumm}.

\subsection{TCM variation with collision system and energy} \label{collenergy}

Based on determination of PID TCM model functions for 5 TeV \ppb\ collisions summarized above, prediction of PID model parameters for 13 TeV \pp\ collisions proceeds as follows. As noted, previous analysis suggests that 5 TeV \ppb\ collisions are linear superpositions of \pn\ collisions. Relevant parameters for 5 TeV \ppb\ collisions then approximate 5 TeV \pp\ collisions. In Sec.~VI of Ref.~\cite{alicetomspec} the energy dependence of nonPID spectra from non-single-diffractive (NSD) \pp\ collisions is summarized based on TCM spectrum analysis from 17 GeV to 13 TeV. Those trends are used to extrapolate 5 TeV TCM parameters to 13 TeV.

Table~\ref{engparam} (upper six rows) shows \pp\ TCM nonPID spectrum parameters over a range of energies~\cite{alicetomspec}. The lowest two rows show updated numbers from Ref.~\cite{tomnewppspec}. Those studies demonstrate that parameter variations follow simple functional forms. For instance, L\'evy exponent $n$ varies as $1/n \approx 0.0475 \sqrt{\ln(\sqrt{s} / 10 ~\text{GeV})}$, and $1/q \propto \ln(\sqrt{s} / 6~ \text{GeV})$, the latter being a measure of jet energy spectrum width on rapidity~\cite{alicetomspec}. Parameters $\bar y_t$ and $\sigma_{y_t}$ increase slowly and linearly with $\ln(\sqrt{s})$. The absolute values for $\sigma_{y_t}$ and $q$ may depend on event selection method and resulting bias~\cite{tomnewppspec}. This table and Ref.~\cite{alicetomspec} demonstrate the predictivity of the TCM.

\begin{table}[h]
	\caption{Unidentified-hadron spectrum TCM parameters for NSD \pp\ collisions within $\Delta \eta \approx 2$ at several energies.
		Starred entries are estimates by interpolation or extrapolation. Unstarred entries are derived from TCM descriptions of yield, spectrum or spectrum-ratio data. The upper entries are from Ref.~\cite{alicetomspec}. The updated lower entries are from Table I of Ref.~\cite{tomnewppspec}. The values for $\bar \rho_s$ are for NSD \pp\ collisions. 
	}
	\label{engparam}
	\begin{center}
		\begin{tabular}{|c|c|c|c|c|c|c|c|} \hline
			Eng.\. (TeV) & T\. (MeV) & $n$ & $\bar y_t$ & $\sigma_{y_t}$ & $q$ & $100\alpha$ & $\bar \rho_s$ \\ \hline
			0.017  & 145  & 27 & 2.55$^*$ & 0.40$^*$  & 6.75$^*$  & 0.07$^*$ & 0.45 \\ \hline
			0.2  & 145  & 12.5 & 2.59 & 0.435  & 5.15  & 0.6 & 2.5 \\ \hline
			0.9  & 145  & 9.82$^*$ & 2.62$^*$ & 0.53$^*$  &  4.45$^*$  & 1.0$^*$ & 3.61 \\ \hline
			2.76  & 145  & 8.83$^*$ & 2.63$^*$ & 0.56$^*$  &  4.05$^*$  & 1.2$^*$ & 4.55 \\ \hline
			7.0  & 145  & 8.16$^*$ & 2.64 & 0.595  & 3.80$^*$  & 1.4$^*$ & 5.35  \\ \hline
			13.0  & 145  & 7.8 & 2.66 & 0.615  & 3.65  & 1.5 & 5.87 \\ \hline \hline
			5.0  & 145  & 8.5 & 2.63 & 0.58  & 4.0$$  & 1.45$$ & 5.0  \\ \hline
			13.0  & 145  & 7.8 & 2.66 & 0.60  & 3.8  & 1.70 & 5.8 \\ \hline
		\end{tabular}
	\end{center}
\end{table}

Table~\ref{engparamsxx} shows PID TCM model-function parameters extrapolated from 5 TeV \ppb\ collisions and based on measured energy dependence of nonPID spectra derived from NSD \pp\ collisions. Those predictions are the basis for the current analysis of PID spectra for 13 TeV \pp\ collisions. One may compare with final results in Sec.~\ref{paramfinal}.

\begin{table}[h]
	\caption{TCM model parameters for identified hadrons from 13 TeV \pp\ collisions: hard-component parameters $(\bar y_t,\sigma_{y_t},q)$ and soft-component parameters $(T,n)$. These are {\em predicted} values based on PID values from 5 TeV \ppb\ collisions and the energy dependence of nonPID NSD \pp\ values. 
	}
	\label{engparamsxx}
	\begin{center}
		\begin{tabular}{|c|c|c|c|c|c|} \hline
			& $\bar y_t$ & $\sigma_{y_t}$ & $q$ & $T$ (MeV) &  $n$  \\ \hline
			$ \pi^\pm $     &  $2.48\pm0.02$ & $0.59\pm0.01$ & $3.9\pm0.1$ & $145\pm3$ & $7.8\pm0.5$ \\ \hline
			$K^\pm$    & $2.67\pm0.02$  & $0.59\pm0.01$ & $3.9\pm0.1$ & $200\pm5$ & $14\pm3$ \\ \hline
			$p$        & $3.00\pm0.02$  & $0.49\pm0.01$ & $4.8\pm0.4$ & $210\pm5$ & $14\pm3$ \\ \hline
		\end{tabular}
	\end{center}
\end{table}

\section{$\bf p$-$\bf p$ PID TCM Parameter estimation} \label{pppidtcm}

PID TCM spectrum parameter estimation requires two steps: (a) refine soft- and hard-component model parameters based on predictions developed in the previous section and (b) estimate parameters $z_{si}(n_s)$ and $z_{hi}(n_s)$ based on direct analysis of PID spectra. Task (b) requires estimation and possible correction of PID spectrum systematic biases, especially possible data \mbox{cross-talk} between pions and protons arising from $dE/dx$ analysis~\cite{pidpart1}.

\subsection{Proton spectrum inefficiency correction} \label{correct}

A method for diagnosing and correcting proton detection inefficiencies is described in Sec.~III B of Ref.~\cite{pidpart1}. The method is based on the TCM relation between spectrum trends at low $p_t < 0.5$ GeV/c and spectrum trends near the mode of the hard component. The method relies on accurate determination of the nonPID TCM for \pp\ collisions in Ref.~\cite{alicetomspec}, especially coefficient $\alpha$ which determines $\bar \rho_s$ and $\bar \rho_h$ given measured $\bar \rho_0$. Based on correspondence of Eq.~(\ref{pidspectcm}) with spectrum data as $y_t \rightarrow 0$ accurate values for $z_{si}(n_s)$ may be inferred there. Given  those trends, values for $\tilde z_i$ and $z_{0i}$ can be inferred from Eq.~(\ref{zsix}), where $\tilde z_i$ is in this case a centrality average. The combination $z_{hi}(n_s) \approx \tilde z_i z_{si}(n_s)$ then {\em predicts} the amplitude of the hard component in Eq.~(\ref{pidspectcm}) (second line) and thus the complete PID spectrum for protons for example.

\begin{figure}[h]
	\includegraphics[width=1.62in]{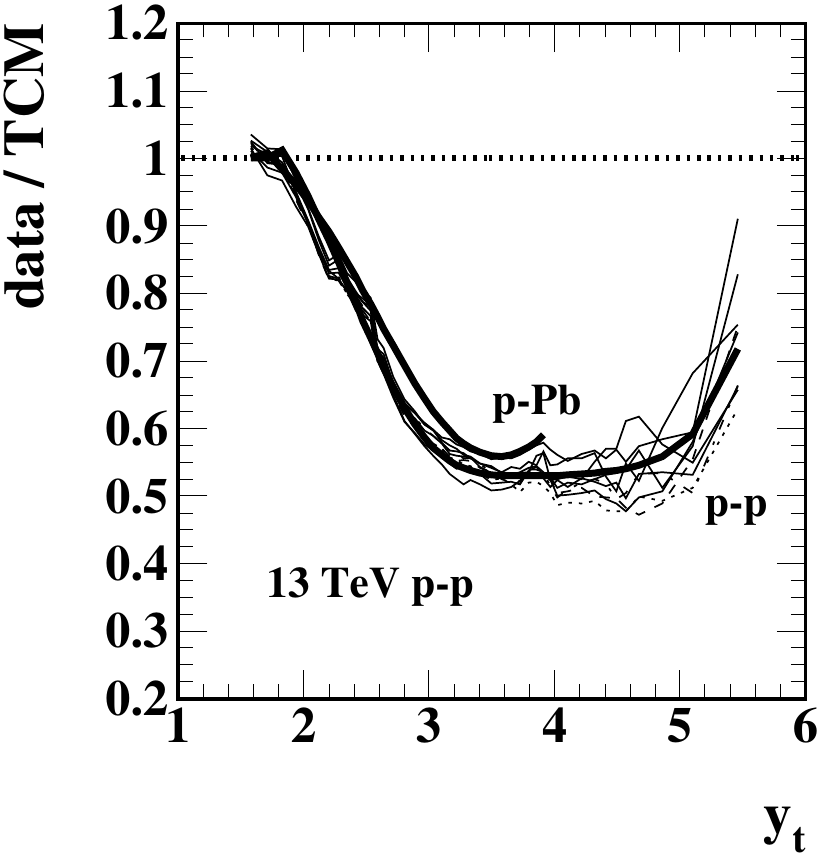}
	\includegraphics[width=1.68in]{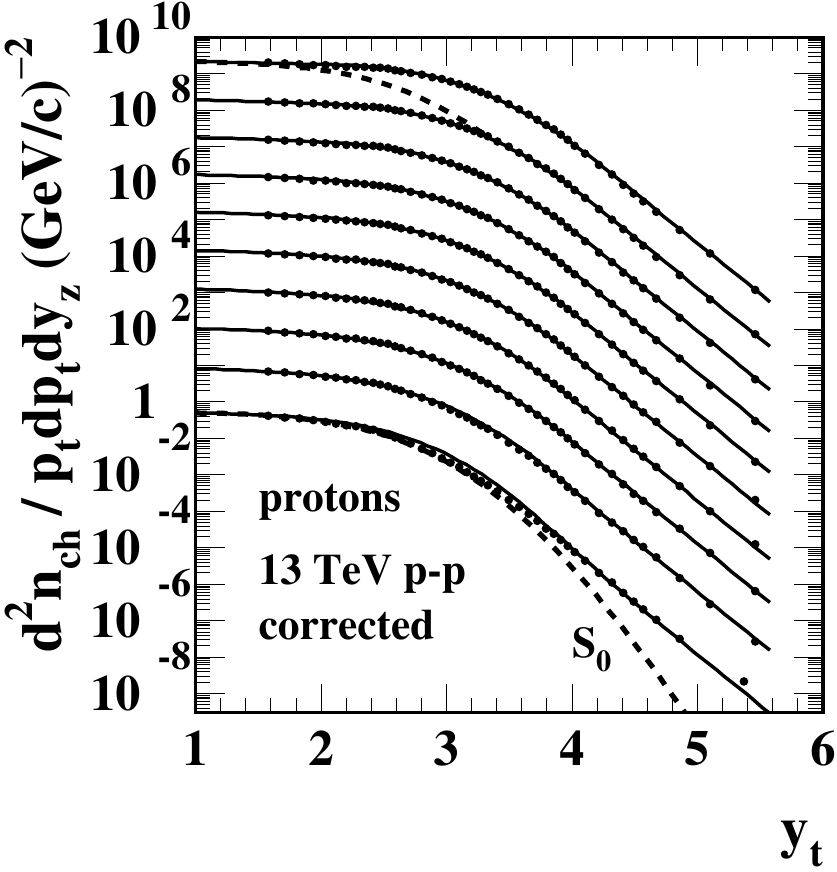}
	\caption{\label{eppsprotons}
		Left: Ratios of uncorrected identified-proton \pt\ spectra as reported in Ref.~\cite{alicepppid} to TCM spectra defined in the present study (dotted curves) for ten event classes of 13 TeV \pp\ collisions. The bold solid curves are proton efficiency models for \pp\ and \ppb~\cite{pidpart1} and collisions. The model for 13 TeV \pp\ collisions is defined by Eq.~(\ref{effpp}).
		Right: Proton spectra as in Fig.~\ref{piddata} (c) but corrected based on Eq.~(\ref{effpp}) (points). The TCM (curves) remains unchanged from Fig.~\ref{piddata} (c).
		} 
\end{figure}

Figure~\ref{eppsprotons} (left) shows ratios of uncorrected proton spectrum data for ten centrality classes from Ref.~\cite{alicepppid} to TCM proton spectra (predictions) based on Eqs.~(\ref{pidspectcm}) (second line) and (\ref{zsix}) with fixed $\tilde z_{i}(n_s) \rightarrow \tilde z_{i} = 5.6 \pm 0.2$~\cite{pidpart1} derived from spectrum trends for $y_t < 0.5$ GeV/c where proton inefficiency is not an issue.  See Fig.~\ref{pchx} (left). 

The \pp\ data-model comparison in Fig.~\ref{eppsprotons} (left) implies systematic suppression of protons. The efficiency correction derived for 5 TeV \ppb\ protons in Ref.~\cite{pidpart1} is
\bea \label{effppb}
\epsilon_p(\ppb) \hspace{-.05in} &=& \hspace{-.05in} \left\{0.58+(1\hspace{-.02in} - \hspace{-.02in}\tanh[(p_t \hspace{-.02in}- \hspace{-.02in}0.45)/0.95])/2\right\}
\\ \nonumber &\times& (1+0.003 p_t^3).
\eea
The longer bold solid curve for 13 TeV \pp\ data,
\bea \label{effpp}
\epsilon_p(\pp) \hspace{-.05in} &=& \hspace{-.05in} 0.88 \left\{0.60+(1\hspace{-.02in} - \hspace{-.02in}\tanh[(p_t \hspace{-.02in}- \hspace{-.02in}0.49)/0.65])/2\right\}~~~~
\\ \nonumber &\times& (1+0.00008 p_t^3),
\eea
is assumed to describe, {\em independent of \pp\ event class}, a proton instrumental inefficiency arising from $dE/dx$ PID measurements reported in Ref.~\cite{alicepppid}. The inefficiency appears significant only above 0.5 GeV/c ($y_t \approx 2$). The \pp\ correction is consistent with that applied to \ppb\ proton spectra (shorter bold solid) in Ref.~\cite{pidpart1}.  The same correction is applied consistently to ten \pp\ event classes.

The inefficiency may result from systematic bias within the $dE/dx$ analysis. ``In the regions where signals from several [hadron] species overlap [e.g.\ $> 0.5$ GeV/c for protons] [the] $dE/dx$ [distribution] is fit with two Gaussian distributions.... The [Gaussian] fit of the overlapping species is then integrated in the signal region [encompassing the signal species, i.e.\ protons] and subtracted from the [total integrated] signal~\cite{alicepppid}.'' The similar proton inefficiencies inferred for \pp\ (present study), \ppb~\cite{pidpart1} and \pbpb~\cite{pbpbpid} spectrum data suggest that a common proton inefficiency results from the same biased $dE/dx$ procedure applied to data in Refs.~\cite{aliceppbpid,alicepbpbpidspec,alicepppid}. 

Figure~\ref{eppsprotons} (right) repeats Fig.~\ref{piddata} (c) but the proton spectra have been corrected with the \pp\ function appearing in  Fig.~\ref{eppsprotons} (left) and defined by Eq.~(\ref{effpp}). The corrected spectra are utilized as such in the remainder of this study. 

\subsection{Estimating $\bf z_{si}(n_s)$ and $\bf z_{hi}(n_s)$ from PID spectra} \label{zxidirect}


Based on the structure of Eq.~(\ref{pidspectcm}) nearly model-independent estimates for soft-component coefficients $z_{si}(n_s)$ may be obtained from PID spectra $\bar \rho_{0i}(y_t,n_s)$ via
\bea \label{zsi}
\lim_{y_t \to 0} \bar \rho_{0i}(y_t,n_s) / \bar \rho_s \hat S_{0i}(y_t) &\approx &  
\bar \rho_{si} / \bar \rho_s = z_{si}(n_s),
\eea
where $\hat S_{0i}(y_t)$ is a soft-component model defined by parameters $T_i$ and $n_i$, and $\bar \rho_s$ is obtained for 13 TeV \pp\ collisions from $\bar \rho_0$ and $\alpha \approx 0.017$ appearing in Table~\ref{engparam}.

Given such measured values of $z_{si}(n_s)$, corresponding hard-component coefficients $z_{hi}(n_s)$ may be inferred via
\bea \label{zhi}
\lim_{y_t \to \bar y_t}  \left[\bar \rho_{0i}(y_t,n_s) \hspace{-.03in} - \hspace{-.03in} z_{si}\bar \rho_s \hat S_{0i}(y_t)\right]\hspace{-.03in} / \bar \rho_h \hat H_{0i}(\bar y_t) \hspace{-.07in} &\approx &  \hspace{-.07in} z_{hi}(n_s),~~~~
\eea
where $\bar y_t$ denotes the mode of the {\em data} spectrum hard component, $\bar \rho_h(n_s) = \alpha \bar \rho_{s}^2(n_s)$ is derived from $\alpha$ and $\bar \rho_0$ as noted above, and unit-normal hard-component models $\hat H_{0i}(y_t)$ are determined by parameters in Table~\ref{engparamsy} (via iteration).  That procedure assumes data and model hard components of {\em fixed shape}, with model factorized as
\bea \label{factorize}
H_i(y_t,n_s) &\rightarrow & z_{hi}(n_s) \bar \rho_h(n_s) \hat H_{0i}(y_t)
\eea
as in Eq.~(\ref{pidspectcm}), so data hard component $H_i(y_t,n_s)$ gives
\bea
H_i(\bar y_t,n_s)/ \bar \rho_h(n_s) \hat H_{0i}^*(\bar y_t) &\approx& z_{hi}(n_s) .
\eea
Accurate estimates for $z_{hi}(n_s)$ require that $\hat H_{0i}^*(\bar y_t)$ values  are inferred from {\em properly normalized} model functions. 

Two issues emerge for estimation of $z_{si}(n_s)$ and $z_{hi}(n_s)$ from spectrum data reported in Ref.~\cite{alicepppid}: (a) Pion spectra are strongly distorted and may contain contributions from misidentified protons, and (b) proton hard-component shapes vary systematically with \nch, in which case $\hat H_{0i}(y_t) \rightarrow \hat H_{0i}(y_t,n_s)$ with varying $\hat H_{0i}(\bar y_t,n_s)$ that requires specialized treatment as described in Sec.~\ref{protonalt}.

\begin{figure}[h]
	\includegraphics[width=1.67in]{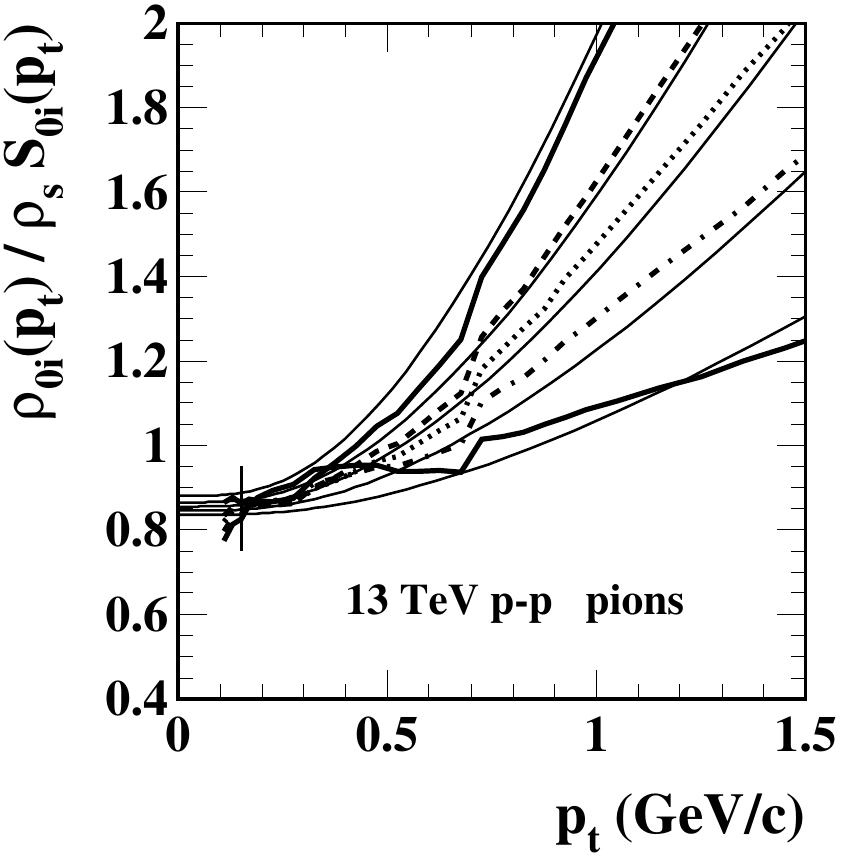}
		\includegraphics[width=1.63in]{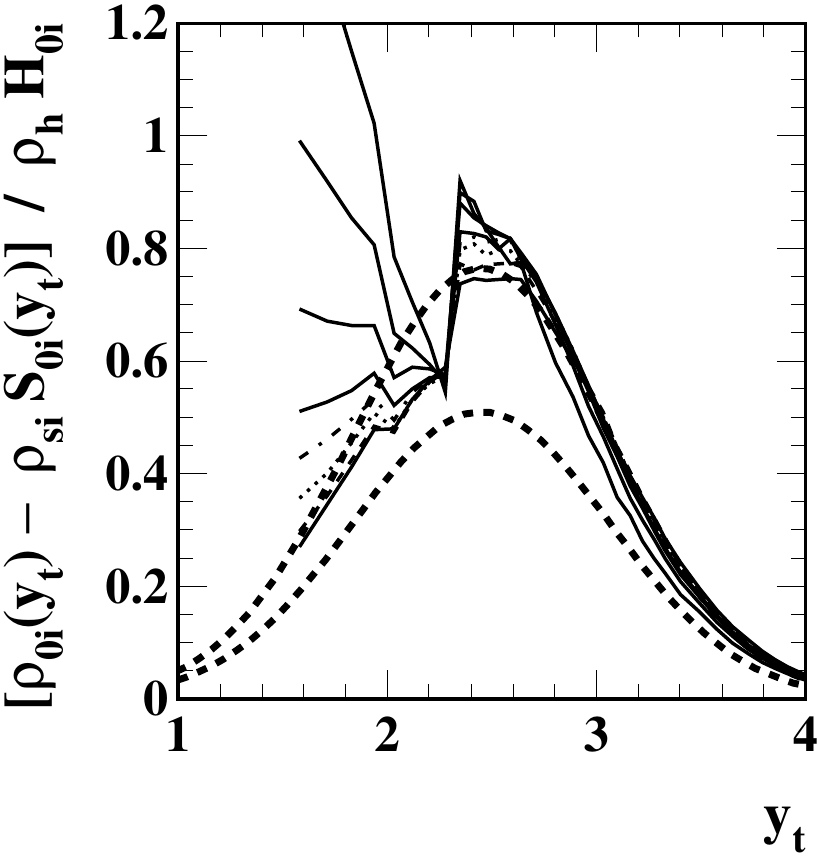}
	\caption{\label{pichx}
		Left: Pion spectrum data in the form of Eq.~(\ref{zsi}) (left) for event classes $n$ = 1, 3, 5, 7, 9 (varying line styles) and the TCM treated equivalently (thin sold). Values of $z_{si}(n_s)$ should be estimated at the vertical line but the pion data do not permit accurate estimates.
Right: Pion spectrum data in the form of Eq.~(\ref{zhi}) (left) for event classes $n$ = 1-8 (varying line styles). The lower bold dashed curve is an estimate based on charge conservation [inverted triangles in Fig.~\ref{zxi} (right)]. The upper bold dashed curve corresponds to $z_{hi}(n_s) \approx 0.75$ for pion spectra from 5 TeV \ppb\ collisions. Anomalous structure in the pion hard component is 
clearly evident.
	} 
\end{figure}

Figure~\ref{pichx} (left) shows the expression on the left of Eq.~(\ref{zsi}) for pion spectrum data (curves of varying line style) from 13 TeV \pp\ collisions for event classes 1, 3, 5, 7 and 9. For pion spectra with a resonance contribution special treatment is required corresponding to the procedure described in Sec.~III A of Ref.~\cite{pidpart1}. In  Eq.~(\ref{zsi}) model function $\hat S_{0i}(y_t)$ is replaced by $S_{0i}'(y_t)$ including a resonance contribution model. The vertical bar indicates the point at 0.15 GeV/c ($y_t \approx 1$) where hard-component contributions are small and values for $z_{si}(n_s)$ ought to be inferred. However, the pion data exhibit severe distortions that preclude reliable estimates of $z_{si}(n_s)$ from data in the left panel requiring an alternative procedure. Values of $z_{si}(n_s)$ are instead estimated based on charge conservation as described in Sec.~\ref{fracsumm}.

Figure~\ref{pichx} (right) shows spectrum data (curves of varying line style) for event classes $n\in [1,8]$ corresponding to the expression on the left of Eq.~(\ref{zhi}). As noted, values of  $z_{si}(n_s)$ are obtained based on charge conservation as described in Sec.~\ref{fracsumm}.  Fixed value $\hat H_{0i}(\bar y_t)^* = 0.262$ is determined from properly normalized $\hat H_{0i}(y_t)$. As is clear already from the left panel the pion spectra are strongly distorted. Based on results from 5 TeV \ppb\ spectra reported in Refs.~\cite{pidpart1,pidpart2} one expects pion hard components in the format of Fig.~\ref{pichx} (right) to nearly coincide with a fixed $\hat H_0(y_t)$. It is therefore not possible to extract reliable $z_{hi}(n_s)$ values from these data. The lower bold dashed curve is $z_{hi}(n_s)\hat H_{0i}(y_t) / \hat H_{0i}^*(\bar y_t)$ with $z_{hi}(n_s) \approx 0.50$ corresponding to the inverted triangles in Fig.~\ref{zxi} (right) representing charge conservation. The excess above that curve, especially its \nch\ dependence, is similar to the proton trend in Fig.~\ref{pchx} (right). The upper bold dashed curve corresponds to $z_{hi}(n_s) \approx 0.75$ for pion spectra from 5 TeV \ppb\ collisions as reported in Ref.~\cite{pidpart1} (e.g.\ its Fig.~5, right). 

\begin{figure}[h]
	\includegraphics[width=1.67in]{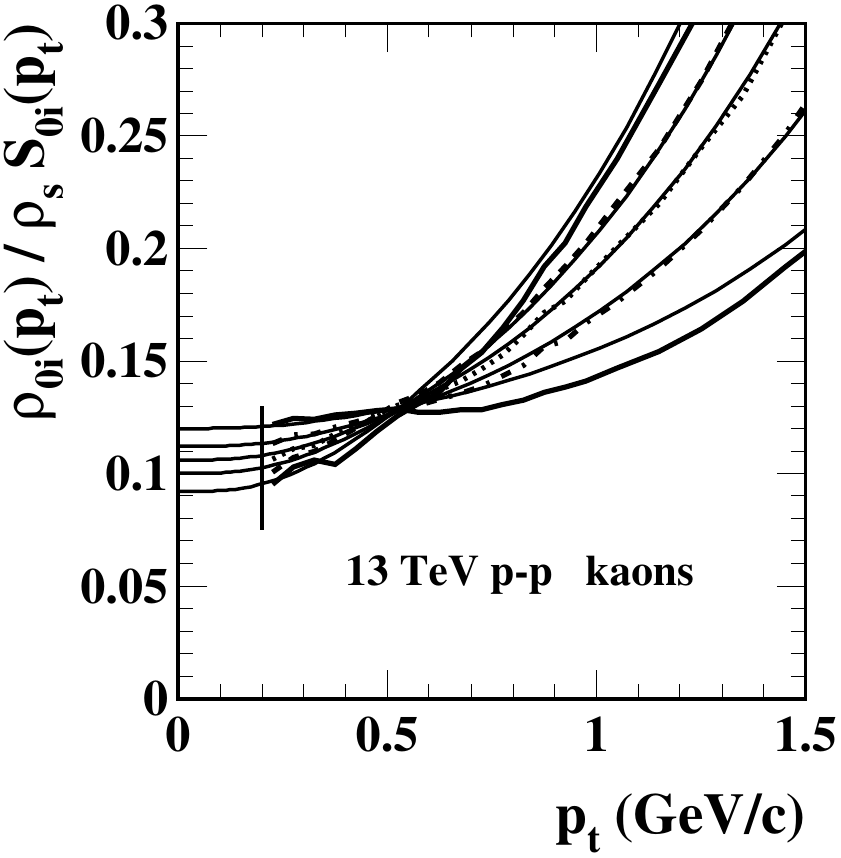}
		\includegraphics[width=1.63in]{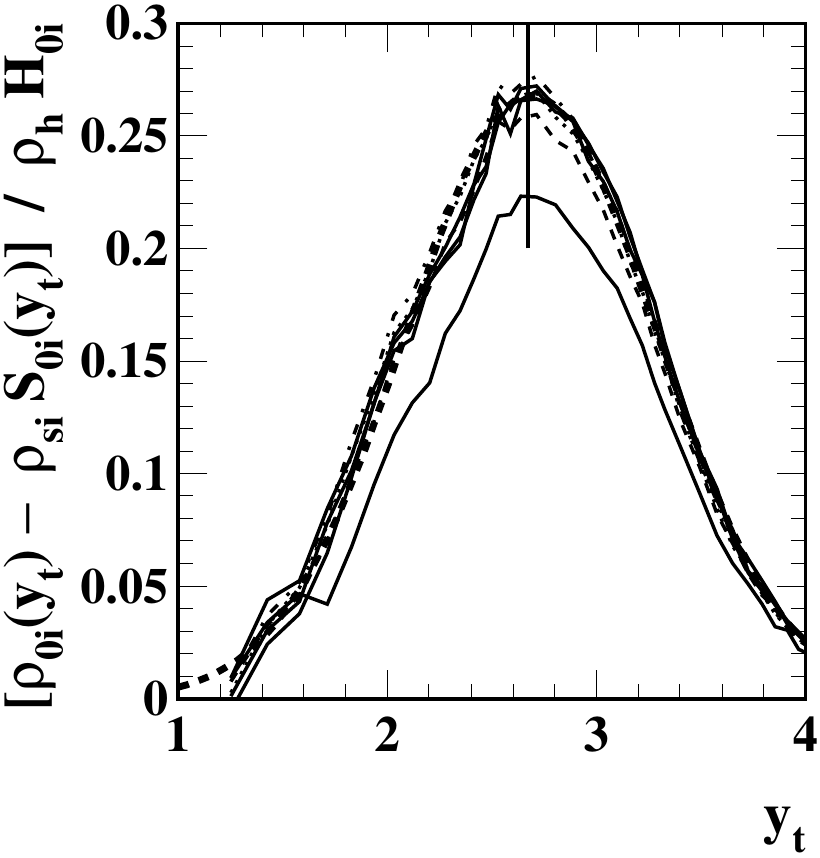}
	\caption{\label{kchx}
	Left: Kaon spectra and TCM treated as for pions in Fig.~\ref{pichx} (left). Estimates of  $z_{si}(n_s)$ are successfully obtained at the vertical line.
	Right:  Kaon spectra and TCM treated as for pions in Fig.~\ref{pichx} (right). Estimates of  $z_{hi}(n_s)$ are successfully obtained at the vertical line.
	} 
\end{figure}

\begin{figure}[h]
	\includegraphics[width=1.67in]{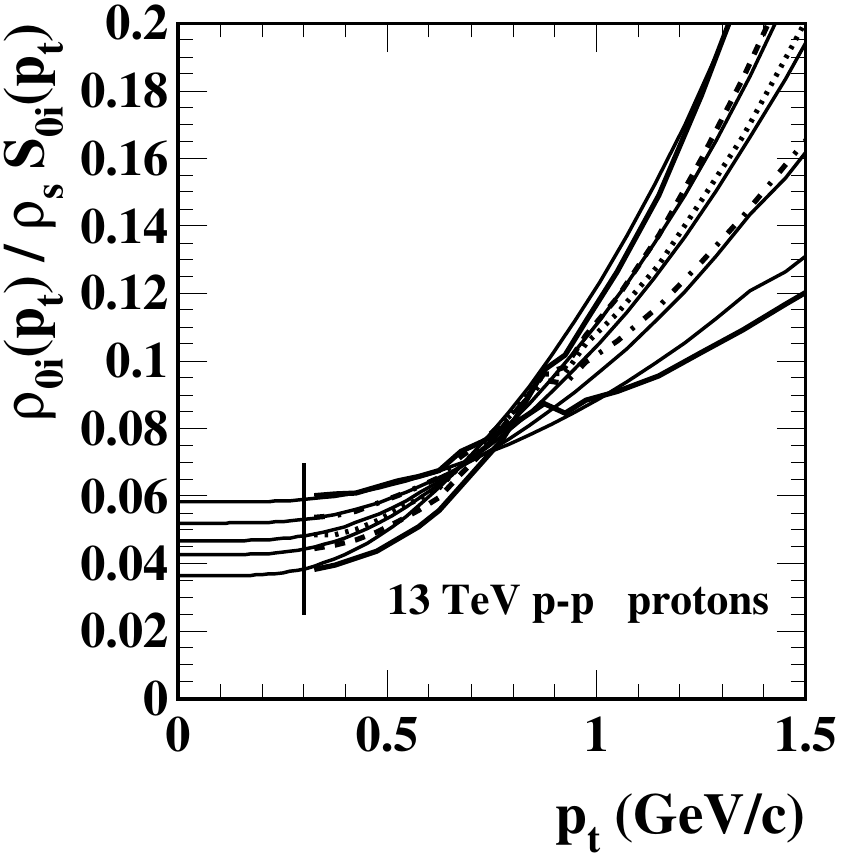}
		\includegraphics[width=1.63in]{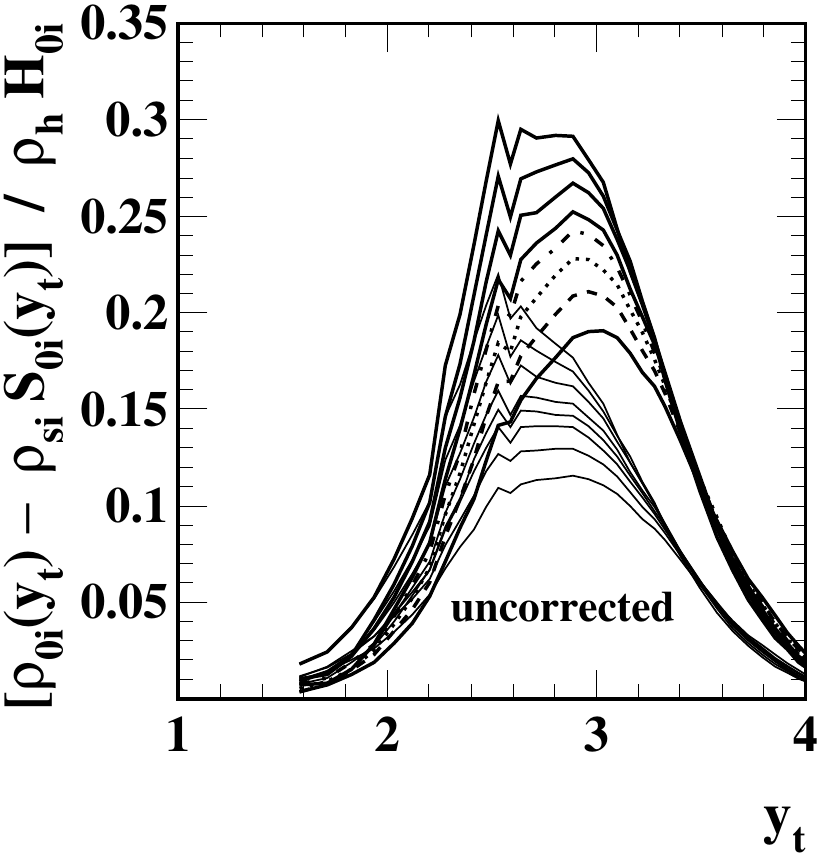}
	\caption{\label{pchx}
	Left: Corrected proton spectra and TCM treated as for pions in Fig.~\ref{pichx} (left). Estimates of  $z_{si}(n_s)$ are successfully obtained at the vertical line.
Right:  Corrected proton spectra (varying line styles) treated as for pions in Fig.~\ref{pichx} (right). Estimates of  $z_{hi}(n_s)$ and other peak parameters are described in Sec.~\ref{protonalt}. Results for uncorrected spectra appear as the thin solid curves. 
	} 
\end{figure}

Figure~\ref{kchx} (left) shows results for charged kaons also based on Eq.~(\ref{zsi}).  In contrast to \ppb\ charged-kaon data in Refs.~\cite{pidpart1,pidpart2} these \pp\ spectra extend down to 0.2 GeV/c (vertical line) and provide accurate estimates for $z_{si}(n_s)$. Spectrum data are denoted by bold curves of several line styles. Corresponding results for the TCM (based on inferred $z_{si}(n_s)$ values) are denoted by thin solid curves included here to demonstrate the accuracy of the inferred $z_{si}(n_s)$ parameter values.

Figure~\ref{kchx} (right) shows  results for charged kaons based on Eq.~(\ref{zhi}).  Fixed value $\hat H_{0i}(\bar y_t)^* = 0.248$ is determined. In  this case, although nonstatistical distortions also appear the amplitudes are small enough to allow usable estimates for hard-component fraction $z_{hi}(n_s)$ at  data modes near $y_t \approx 2.65$ (vertical line).  The bold dashed curve is $z_{hi}(n_s)\hat H_{0i}(y_t) / \hat H_{0i}^*(\bar y_t)$ with $z_{hi}(n_s)$ for event class 5.

Figure~\ref{pchx} (left) shows proton spectra from 13 TeV \pp\ collisions in relation to the expression on the left of Eq.~(\ref{zsi}). The vertical bar indicates the point at 0.3 GeV/c where the hard-component contribution is negligible and values for $z_{si}(n_s)$ are inferred. The proton data have been corrected as described in Sec.~\ref{correct}, but the correction does not affect $z_{si}(n_s)$ inferred below 0.5 GeV/c. As for kaons the corresponding TCM results are denoted by thin solid curves included to demonstrate accuracy of $z_{si}(n_s)$ values. Note that for kaons and protons the $z_{si}(n_s)$ inferred at low \pt\ are descending with increasing \nch\ whereas the hard (jet) component is increasing with increasing \nch\ leading to a crossover in spectrum trends near 0.6 GeV/c. In principle there should be no such crossover for pions given their $z_{si}(n_s)$ trend.

Figure~\ref{pchx} (right) shows corrected proton spectrum hard components (several line styles) corresponding to the expression on the left of Eq.~(\ref{zhi}). Fixed value $\hat H_{0i}(\bar y_t)^* = 0.29$ is determined. There is a significant artifact near $y_t = 2.5$ ($p_t \approx 0.85$ GeV/c); however, usable estimates for $z_{hi}(n_s)$ are still possible. The peak modes shift substantially to higher \yt\ with increasing \nch\ just as observed for \ppb\ baryon data in Refs.~\cite{pidpart1,pidpart2}. Also included are results for uncorrected proton spectra (thin solid curves). Corrected and uncorrected data coincide below \yt\ = 2.0 consistent with the correction defined by Eq.~(\ref{effpp}).

It is notable that baryon hard components with modes near \yt\ = 3 fall more rapidly below their modes and are thus negligible below \yt\ = 1.5 ($p_t \approx 0.30$ GeV/c) whereas meson hard components with modes near \yt\ = 2.6 fall less rapidly below the mode and are therefore non-negligible at \yt = 1  ($p_t \approx 0.15$ GeV/c). The greater accuracy of proton $z_{si}(n_s)$ estimates facilitates correction of proton detection inefficiency as described in Sec.~\ref{correct}.

\subsection{Summary of  $\bf z_{si}(n_s)$ and $\bf z_{hi}(n_s)$ estimates} \label{fracsumm}

\begin{figure}[h]
	\includegraphics[width=3.3in]{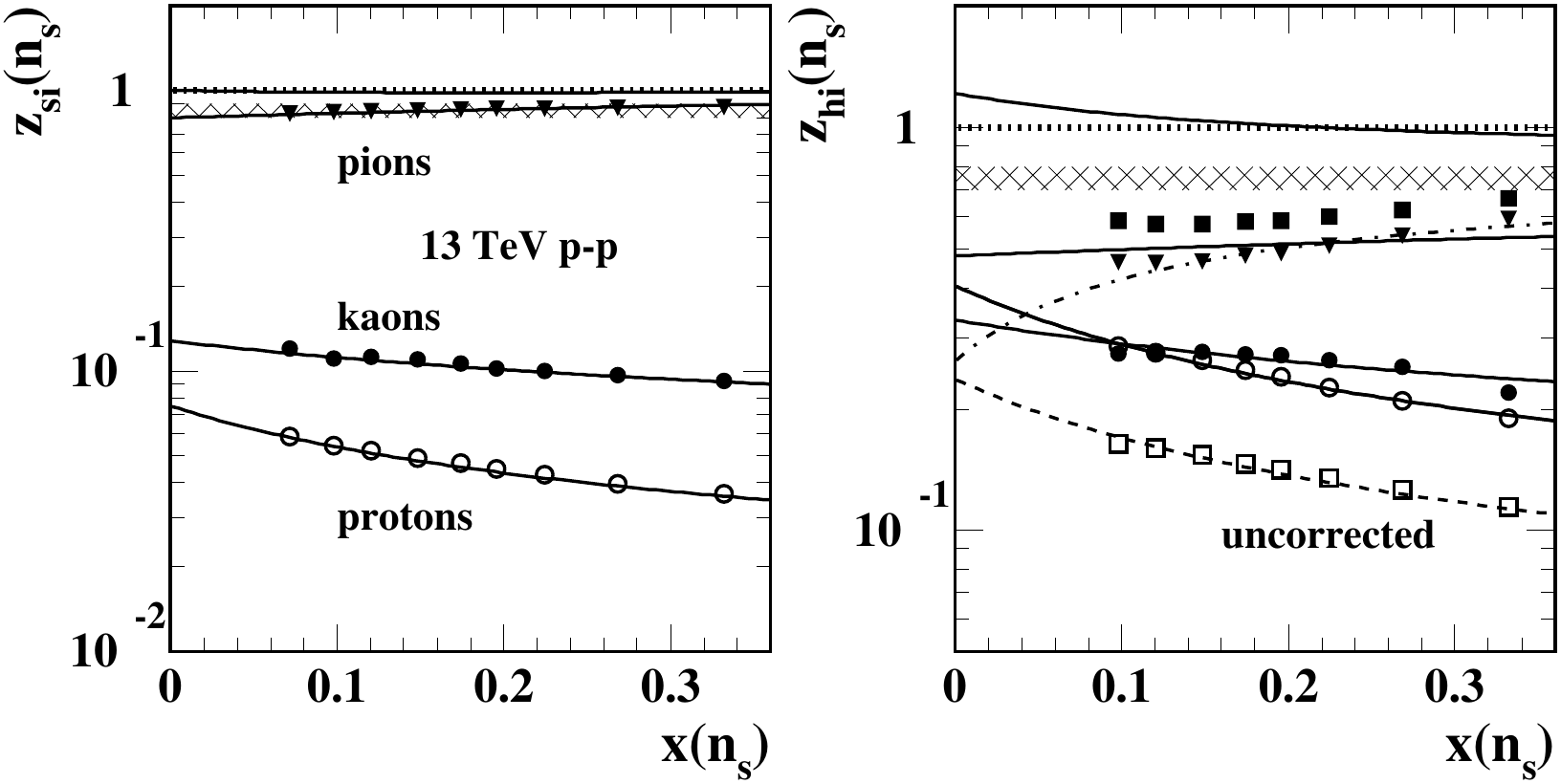}
	\caption{\label{zxi}
		Left: $z_{si}(n_s)$ estimates obtained from Eq.~(\ref{zsi}) for kaons (solid dots) and corrected protons (open circles). Values for pions (inverted triangles) are obtained from kaon and proton data assuming strict charge conservation. Three lower curves are Eq.~(\ref{zsix}) with parameters appearing in Table~\ref{otherparamsxx}. The uppermost solid curve is the sum of three lower curves.
		Right: $z_{hi}(n_s)$ estimates obtained from Eq.~(\ref{zhi}) for kaons (solid dots) and corrected protons (open circles). Values for pions (inverted triangles) are obtained from measured kaon and proton data assuming strict charge conservation.  The three lower solid curves are $z_{hi}(n_s) = \tilde z_i z_{si}(n_s)$ with parameters appearing in Table~\ref{otherparamsxx}. The uppermost solid curve is their sum approximating charge conservation. Open squares are uncorrected proton values. Solid squares are pion estimates based on {\em uncorrected} proton values and charge conservation. The hatched band represents 5 TeV \ppb\ pion values~\cite{pidpart1}.
	} 
\end{figure}

Figure~\ref{zxi} (left) shows measured values of $z_{si}(n_s)$ for kaons (solid dots) and protons (open circles) plotted vs hard/soft ratio $x(n_s)$. No correction is required for protons for $y_t < 2$ ($p_t < 0.5$ GeV/c). Because of the distortions in pion spectra pion $z_{si}(n_s)$ estimates based on the spectra themselves are denoted only by the hatched band. The solid curves for kaons and protons are Eq.~(\ref{zsix}) with $\nu \rightarrow 1$ and with $\bar z_i$ and $z_{0i}$ values as reported in Table~\ref{otherparamsxx} that best accommodate the corresponding $z_{si}(n_s)$ data. The inverted triangles are estimates for pion $z_{si}(n_s)$ based on charge-conservation sum rule $\sum_i z_{si}(n_s) = 1$. The pion solid curve is based on $z_{0i} \approx 0.80$ and $\tilde z_i \approx 0.60$ and is consistent, within data uncertainties, with charge conservation. The uppermost solid curve is the sum of lower three curves.

Figure~\ref{zxi} (right) shows measured $z_{hi}(n_s)$ values for  kaons (solid dots) and corrected protons (open circles) plotted vs hard/soft ratio $x(n_s)$. The hatched band represents 5 TeV \ppb\ pion values corresponding to $\tilde z_i \approx 0.88$ from Ref.~\cite{pidpart1}. Compare with pion data hard components in Fig.~\ref{pichx} (right). The inverted triangles are estimates for pion $z_{hi}(n_s)$ based on charge conservation with kaons and {\em corrected} protons. The dash-dotted curve is the charge complement to kaon and corrected-proton solid curves. The pion, kaon and proton solid curves are $z_{hi}(n_s) \approx \tilde z_i z_{si}(n_s)$ based on parameter values in Table~\ref{otherparamsxx}, where $\tilde z_i \approx 0.6$ is determined by pion inverted triangles (i.e.\ charge conservation).  The open boxes are $z_{hi}(n_s)$ estimates for {\em uncorrected} protons as in Fig.~\ref{pchx} (right), the amplitudes evaluated at the mode locations for corrected data. The solid squares are pion values based on charge conservation using the {uncorrected} proton data. Those results suggest that some protons may be misidentified as pions in $dE/dx$ analysis of Ref.~\cite{alicepppid}.

This overall approach to $z_{xi}$ estimation in the face of substantial data bias is based on estimates of $z_{si}(n_s)$ below $p_t \approx 0.5$ GeV/c for kaons and protons (corrected or not) being sufficiently accurate. Charge conservation is then used to estimate $z_{xi}$ values for pions (inverted triangles). The resulting $z_{hi}(n_s)$ values for pions then provide an estimate for pion $\tilde z_i$ that is not possible using only pion $z_{si}(n_s)$. Based on these \pp\ data there is not sufficient information to determine systematic variation of $\tilde z_i(n_s)$ with \nch\ as was done for \ppb\ data in Ref.~\cite{pidpart1}. In the present study those parameters are assumed constant. Solid curves in left and right panels then represent final TCM parametrizations for the $z_{xi}(n_s)$.

The fractions $z_{si}(n_s)$ and $z_{hi}(n_s)$ apply to densities $\bar \rho_s$ and $\bar \rho_h$ that by definition sum to measured total charge density $\bar \rho_0$. Results in the left panel correspond to PID spectrum soft components $\bar \rho_{si} = z_{si} \bar \rho_s$ that satisfy charge conservation. Thus, {\em observed} hard-component charge densities  $\bar \rho_{hi} = z_{hi} \bar \rho_h$ must do so as well. However, in the event of a substantial proton inefficiency the misidentified charge density should appear as one or more other charged hadron species. Correcting proton spectra without performing a complementary correction to other hadron species must then result in excess charge as indicated by the solid squares in the right panel. That and the hatched band suggest that the excess charge appears as pions, but given the distortions in Fig.~\ref{pichx} (right) a direct pion correction procedure is not readily accessible.

\section{PID TCM from $\bf z_{si}(n_s)$ and $\bf z_{hi}(n_s)$ data} \label{tcmfinal}

In the figures below, PID \pt\ spectrum data from Fig.~\ref{piddata} are replotted in left panels in the rescaled form of Eq.~(\ref{xi}) (as densities on pion \pt\ with additional factor $1/p_t$) vs pion \yt\ and compared to TCM soft component model $\hat S_{0i}(m_t)$ (bold dashed curves). In right panels spectra are plotted (as densities on \yt) in the  form of Eq.~(\ref{yi}) and compared to TCM hard component models $\hat H_{0i}(y_t)$. For the latter case spectra as densities on \pt\ or \mt\ are transformed to densities on $y_{t\pi}$ via Jacobian factor $m_{t\pi} p_t / y_{t\pi}$ where $m_{t\pi}^2 = p_t^2 + m_\pi^2$ and $y_{t\pi} = \ln[(m_{t\pi} + p_t)/m_\pi]$. 

In what follows values for kaon and (corrected) proton $z_{si}(n_s)$ and $z_{hi}(n_s)$ may be obtained from Eq.~(\ref{zsix}) via Table~\ref{otherparamsxx} or from direct measurements as described in Sec.~\ref{zxidirect} and as they appear in Fig.~\ref{zxi}. Pion $z_{xi}(n_s)$ values are represented by inverted triangles (charge conservation) in Fig.~\ref{zxi}. TCM soft- and hard-component model parameters are iterated from the predictions in Table~\ref{engparamsxx} based on results below and then summarized in Sec.~\ref{paramfinal}.

\subsection{TCM for 13 TeV $\bf p$-$\bf p$ meson spectra}

In Ref.~\cite{pidpart1} 5 TeV \ppb\ spectrum hard components for mesons are observed to shift {\em down} on \yt\ with increasing \ppb\ centrality whereas hard components for baryons shift {\em up} on \yt. For 13 TeV \pp\ collisions meson hard components do not shift significantly with increasing \nch\ but baryon hard components do shift substantially to higher \yt\ requiring specialized analysis described in Sec.~\ref{protonalt}.

\begin{figure}[h]
	\includegraphics[width=3.3in]{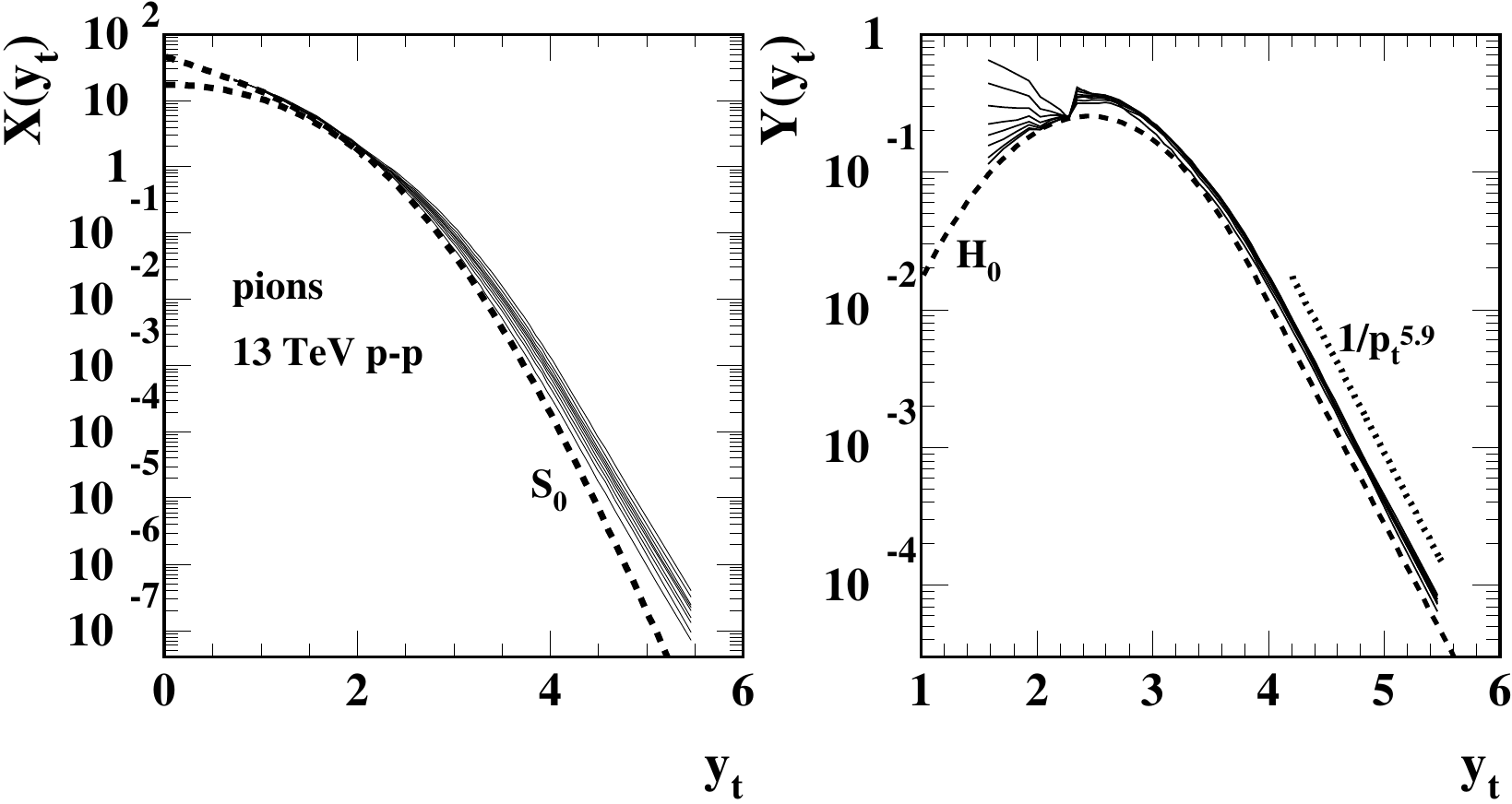}
	\caption{\label{pionsx}
		Left: Pion spectra for $n \in [1,9]$ rescaled to the form of Eq.~(\ref{xi}) (solid) compared to TCM model function $S_0'(y_t)$ (upper dashed) that is modified to accommodate resonances. The lower dashed curve is unmodified unit-normal model function $\hat S_0(y_t)$ (L\'evy distribution).
		Right: Pion spectrum hard components for $n \in [1,8]$ in the form of Eq.~(\ref{yi}) (solid) compared to TCM model function $\hat H_0(y_t)$ (dashed).
	} 
\end{figure}

Figure~\ref{pionsx} (left) shows charged pion $\pi^+ + \pi^-$ spectra from Fig.~\ref{piddata} (a). Published data spectra  have been divided by \pt\ to be consistent with the definition $\bar \rho_0(p_t) = d^2 \bar n_{ch} / p_t dp_t dy_z$ used in the present study. The spectra are then rescaled by soft-component density $\bar \rho_{si} = z_{si}(n_s) \bar \rho_s$  with $ z_{si}(n_s)$ as solid triangles in Fig.~\ref{zxi} (left). 

Rescaled spectra $X_i(y_t)$ can then be compared with soft-component model $S_{0i}'(y_{t}) = f(y_t)\hat S_{0i}(y_{t})$ (upper dashed curve) that incorporates resonance model $f(y_t)$ from Ref.~\cite{pidpart1}. The lower dashed curve (at low \yt) is $\hat S_{0i}(y_t)$ (density on $m_{t}$) with $T = 145$ MeV and $n = 8.0$. 

Figure~\ref{pionsx} (right) shows quantity $Y_i(y_t)$ from Eq.~(\ref{yi}), transformed to a density on \yt\ for direct comparison with $\hat H_0(y_t)$ as defined by Eq.~(\ref{h00}), reconfigured to the more precise form 
\bea \label{zhix}
  \left[\bar \rho_{0i}(y_t,n_s) \hspace{-.03in} - \hspace{-.03in} z_{si}(n_s)\bar \rho_s S_{0i}'(y_t)\right]\hspace{-.03in} / z_{hi}(n_s) \bar \rho_h  &\approx &  \hat H_{0i}(y_t)~~~~~
\eea
using inferred values $z_{si}(n_s)$ and $z_{hi}(n_s)$ as solid triangles in Fig.~\ref{zxi}. The bold dashed curve is $\hat H_0(y_t)$ with model parameters  $(\bar y_t,\sigma_{y_t},q)$ for pions as in Table~\ref{engparamsy}. Anomalous structure in the pion hard component is clearly evident. A possible source of  the large pion data-TCM discrepancy, misidentified protons, is discussed in Sec.\ref{pionproton}.

\begin{figure}[h]
	\includegraphics[width=3.3in]{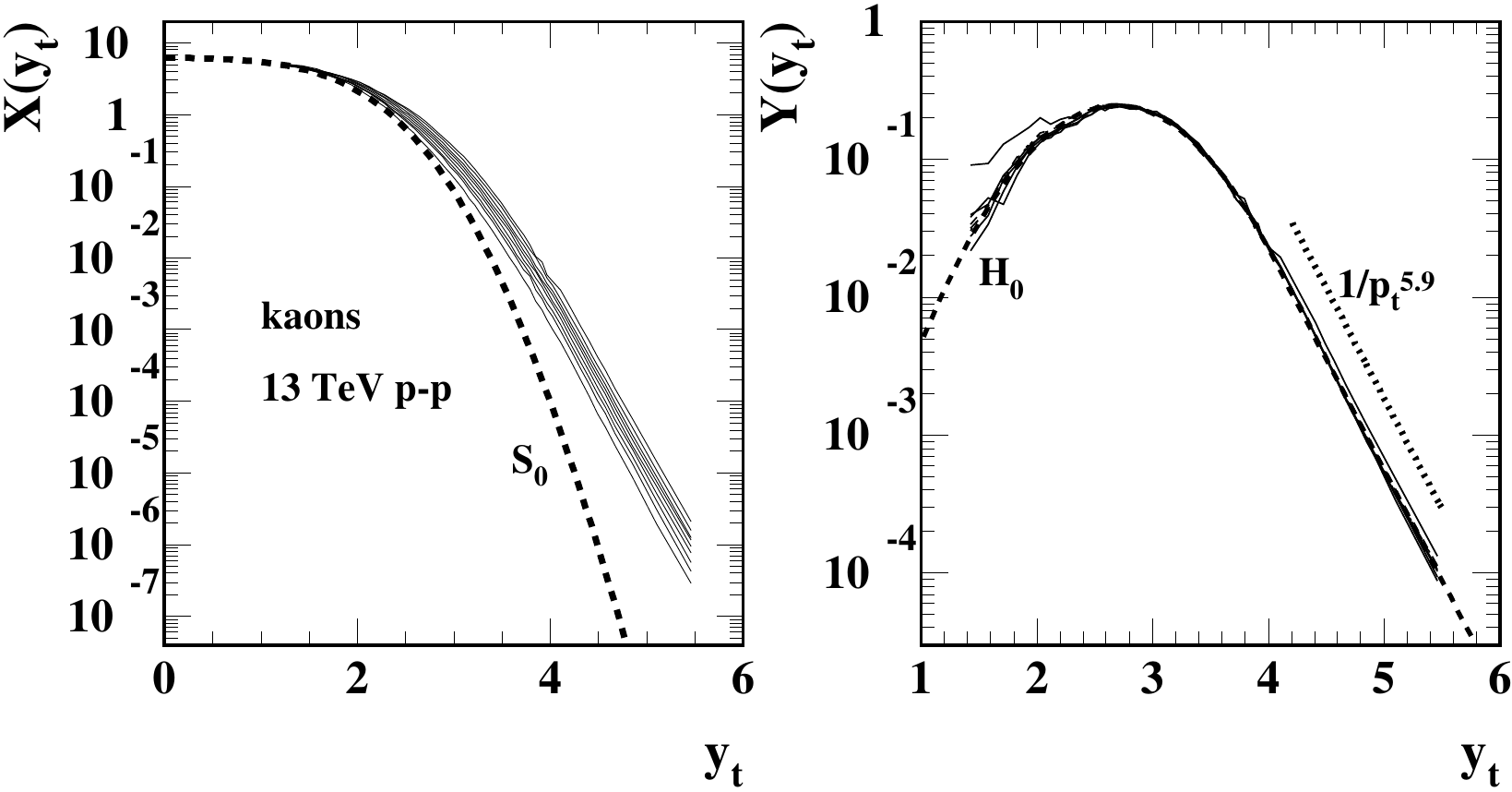}
	\caption{\label{kchxx}
		Left: Kaon spectra for $n \in [1,9]$ rescaled to the form of Eq.~(\ref{xi}) (solid curves) compared to TCM model function $\hat S_0(y_t)$ (dashed curve).
Right: Kaon spectrum hard components for $n \in [1,8]$ in the form of Eq.~(\ref{yi}) (solid curves) compared to TCM model function $\hat H_0(y_t)$ (dashed curve).
	} 
\end{figure}

Figure~\ref{kchxx} shows charged-kaon $K^+ + K^-$ spectra from Fig.~\ref{piddata} (b) treated in the same manner as for charged pions. For both meson species the hard-component exponential parameter $q \approx 3.7$ corresponds to $1/p_t^n$ power law exponent $n = 5.9$ as illustrated by the dotted lines. In this plotting format the linear trend above \yt\ = 4 and its lack of variation with event class is quantitatively apparent, in contrast to the format of Fig.~1 in Ref.~\cite{alicepppid}.

\subsection{TCM for 13 TeV $\bf p$-$\bf p$ proton spectra} \label{protonalt}

\begin{figure}[h]
	\includegraphics[width=3.3in]{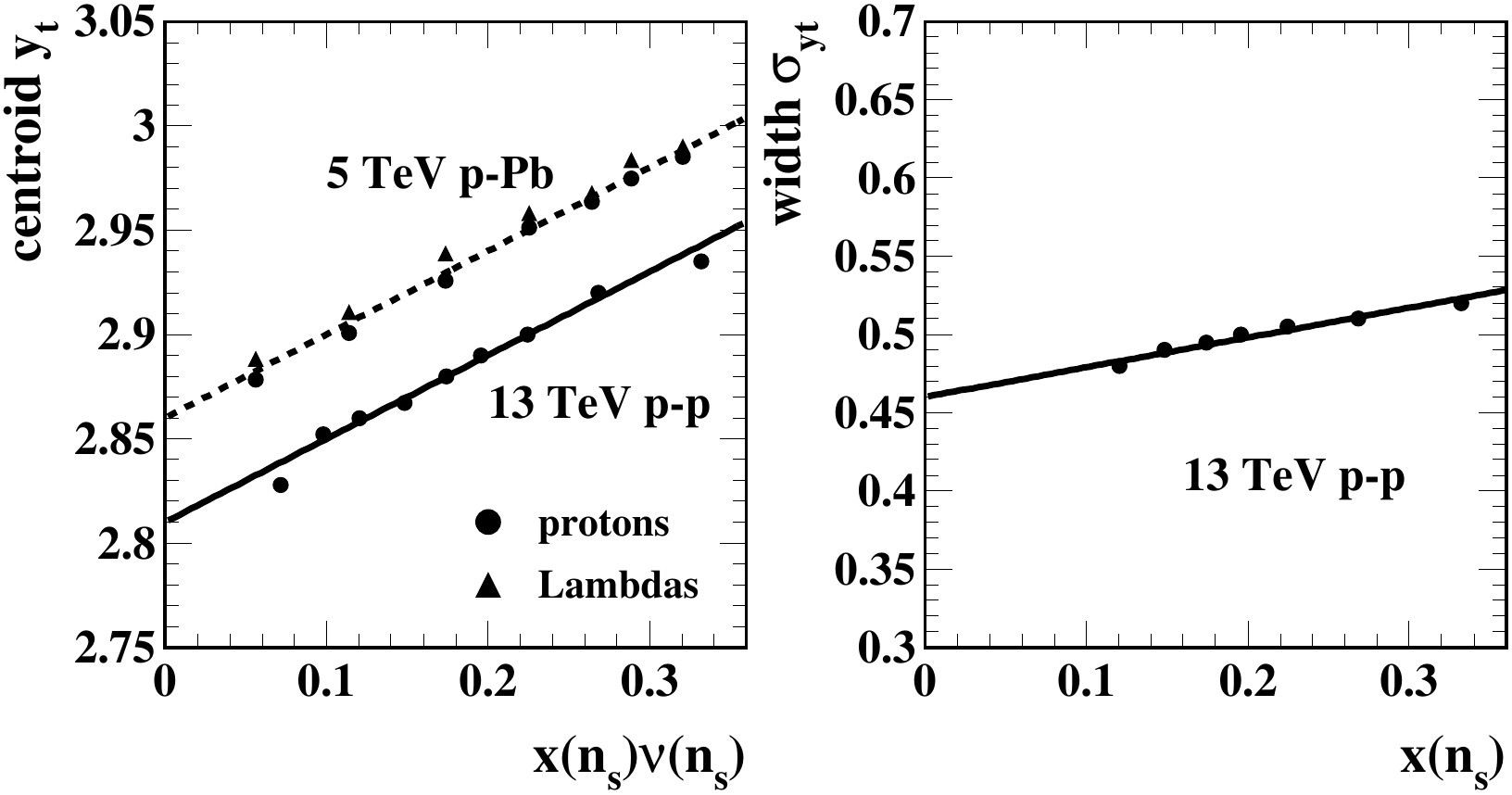}
	\caption{\label{centroids}
		Left: TCM hard-component centroids $\bar y_t$ for proton and Lambda spectra from 5 TeV \ppb\ collisions as reported in Ref.~\cite{pidpart2} (upper points) and for proton spectra from 13 TeV \pp\ collisions (lower points) as determined in the present study. The lines are defined in the text.
		Right: Hard-component widths $\sigma_{y_t}$ above the mode for proton spectra from 13 TeV \pp\ collisions. The line is defined in the text.
	} 
\end{figure}

As observed in Fig.~\ref{pchx} (right) the peak modes for corrected proton hard components are observed to shift substantially on \yt, and peak amplitudes [described by $z_{hi}(n_s)$] also vary substantially with \pp\ \nch.  The peak width {\em above the mode} increases significantly as well. Those variations are not caused by the correction itself which has the same form for all event classes. The mode shifts and amplitude variations are correlated such that data hard components approximately coincide with a fixed power-law (straight-line) trend at higher \yt. TCM model function $\hat H_{0i}(y_t;\bar y_t,\sigma_{y_t})$ can accommodate those characteristics if peak mode $\bar y_t(n_s)$ shifts with coefficient $z_{hi}(n_s)$ according to
\bea  \label{barytns}
\bar y_t(n_s) &=& \bar y_{t0} + (1/q) \ln[ z_{hi0} / z_{hi}(n_s)]
\eea
following Eq.~(\ref{h00}) (second line), where $\bar y_{t0}$ and $z_{hi0}$ correspond to a reference \nch\ class (e.g.\ $n = 3$).
Hard-component data should then be described by
\bea \label{pidhcnewx}
H_i(y_t,n_s) &\approx& z_{hi}(n_s)\bar \rho_h\hat H_{0i}[y_t;\bar y_t(n_s),\sigma_{y_t}(n_s)],~~
\eea
where $\bar y_t(n_s)$ and $\sigma_{y_t}(n_s)$ vary as shown in Fig.~\ref{centroids}. The $\hat H_{0i}(y_t,n_s)$ model normalization then becomes an issue.

The values of $z_{hi}(n_s)$ inferred from Fig.~\ref{pchx} (right) invoke Eq.~(\ref{zhi}) with the fixed value $\hat H_{0i}^*(\bar y_t) = 0.29$ for all centrality classes. The {\em measured} peak amplitudes are then represented by
$z_{hi}^*(n_s) \approx z_{hi}(n_s)\hat H_{0i}(\bar y_t,n_s) / \hat H_{0i}^*(\bar y_t)$ as an approximation to $z_{hi}(n_s)$, where $\hat H_{0i}(\bar y_t,n_s)$ represents the peak amplitude of a properly-normalized variable $\hat H_{0i}(y_t,n_s)$ model for event class $n_s$. The required form for the TCM is $z_{hi}(n_s)\hat H_{0i}(\bar y_t,n_s) \approx z_{hi}^*(n_s)\hat H_{0i}^*(\bar y_t)$ as the coefficient for {\em unit}-amplitude model $H_{0i}(y_t,n_s)$.

Figure~\ref{centroids} (left) shows centroid variation $\bar y_t(n_s)$ (upper points) for protons (corrected) and Lambdas from 5 TeV \ppb\ collisions vs hard/soft ratio $x\nu$ as presented in Ref.~\cite{pidpart2}. The inferred baryon values for $\bar y_t(n_s)$ follow a linear trend (dashed) on centrality measure $x\nu$
\bea
\bar y_t(n_s) &=& 2.86 + 0.4 x(n_s)\nu(n_s).
\eea
The lower points show the centrality trend for centroid variation corresponding to measured $z_{hi}(n_s)$ values for corrected 13 TeV \pp\ proton spectra in Fig.~\ref{zxi} (right) and Eq.~(\ref{barytns}).  The inferred 13 TeV values for $\bar y_t(n_s)$ also follow a linear trend (solid) on hard/soft ratio $x$
\bea
\bar y_t(n_s) &=& 2.81 + 0.4 x(n_s)~~~\text{($\nu = 1$)}.
\eea

Figure~\ref{centroids} (right) shows model widths $\sigma_{y_t}(n_s)$ that accommodate the proton data. The straight line is $\sigma_{y_t}(n_s) = 0.46 + 0.19 x(n_s)$. No comparable proton peak width variation was observed for 5 TeV \ppb\ collisions in Ref.~\cite{pidpart2}. It is notable that the several parameters controlling variable-TCM PID hard-component models consistently vary linearly with hard/soft (jet/nonjet) ratio $x\nu$ (\ppb) or $x$ (\pp) within data uncertainties.

\begin{figure}[h]
	\includegraphics[width=3.3in]{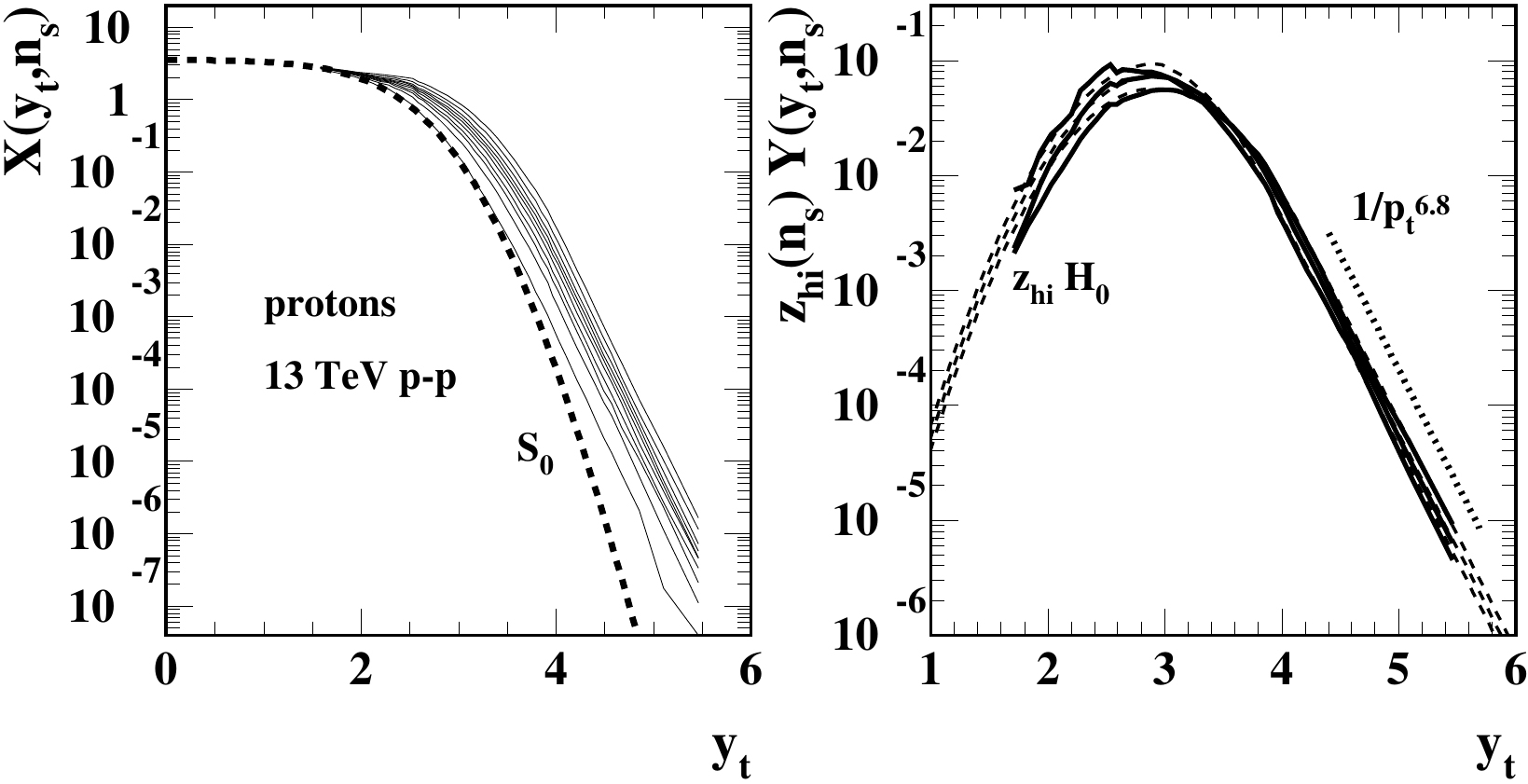}
	\caption{\label{pchxx}
		Left: Proton spectra for $n \in [1,9]$ rescaled to the form of $X_i(y_t,n_s)$ in Eq.~(\ref{xi}) (solid) compared to TCM model function $\hat S_0(y_t)$ (dashed).
Right: Proton spectrum hard components for $n \in [1,8]$ as $z_{hi}(n_s) Y_i(y_t,n_s)$ (solid) for $n = 1, 5, 9$ compared to the hard-component model in the form $z_{hi}(n_s)\hat H_0(y_t,n_s)$ (dashed).
	} 
\end{figure}

Figure~\ref{pchxx} shows corrected proton $p + \bar p$ spectra from Fig.~\ref{eppsprotons} (right) processed just as for charged pions. The TCM model functions in the right panel (dashed) are constructed as described above: {\em Unit}-amplitude (not normalized) model functions $H_0(y_t,n_s)$ are generated with centroid determined by Eq.~(\ref{barytns}) and width above the peak mode varying as in Fig.~\ref{centroids} (right). The width below the mode is fixed at $\sigma_{y_t} = 0.50$. Final amplitudes are then determined by applying factors $z_{hi}(n_s)^*\hat H_0(\bar y_t)^*$ (with $\hat H_0(\bar y_t)^* = 0.29$), shown as the dashed curves in Fig.~\ref{pchxx} (right), to approximate $z_{hi}(n_s)\hat H_0( y_t,n_s)$.

\subsection{Possible proton-pion cross-talk} \label{pionproton}

Section~\ref{correct} describes a method to correct inefficiency for proton spectra. One can test the possibility that missing protons are misidentified as pions. While such a test would be difficult to perform directly on data due to differing acceptances and data point positions the test can be performed on the TCM defined on a common ``continuum'' (100 points equally spaced on \yt) in the form
\bea \label{crosst}
\bar \rho_{0\pi}'(y_t,n_s) &=& \bar \rho_{0\pi}(y_t,n_s) + [1 - \epsilon_p(y_t)]\bar \rho_{0p}(y_t,n_s),~~
\eea
where unprimed functions are TCM predictions for ideal data, primes indicate TCM representations of uncorrected data and $\epsilon_p(y_t)$ is defined by Eq.~(\ref{effpp}). Uncorrected proton spectra are well represented by $\bar \rho_{0p}'=\epsilon_p\bar \rho_{0p}$ which inverts the correction applied to published proton data.

\begin{figure}[h]
	\includegraphics[width=1.65in]{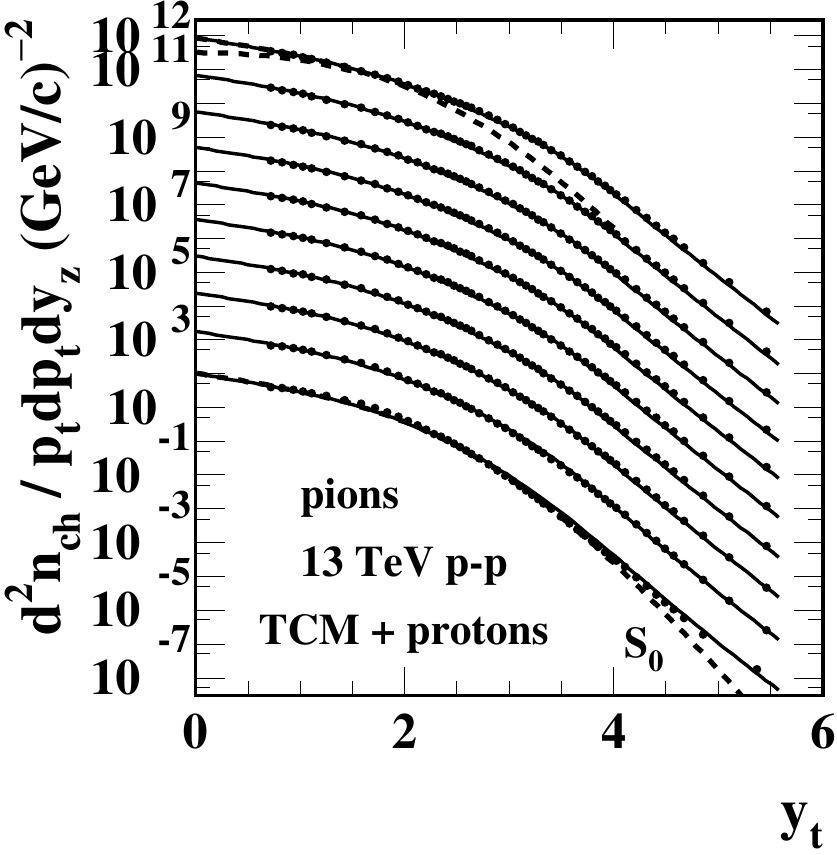}	\includegraphics[width=1.65in]{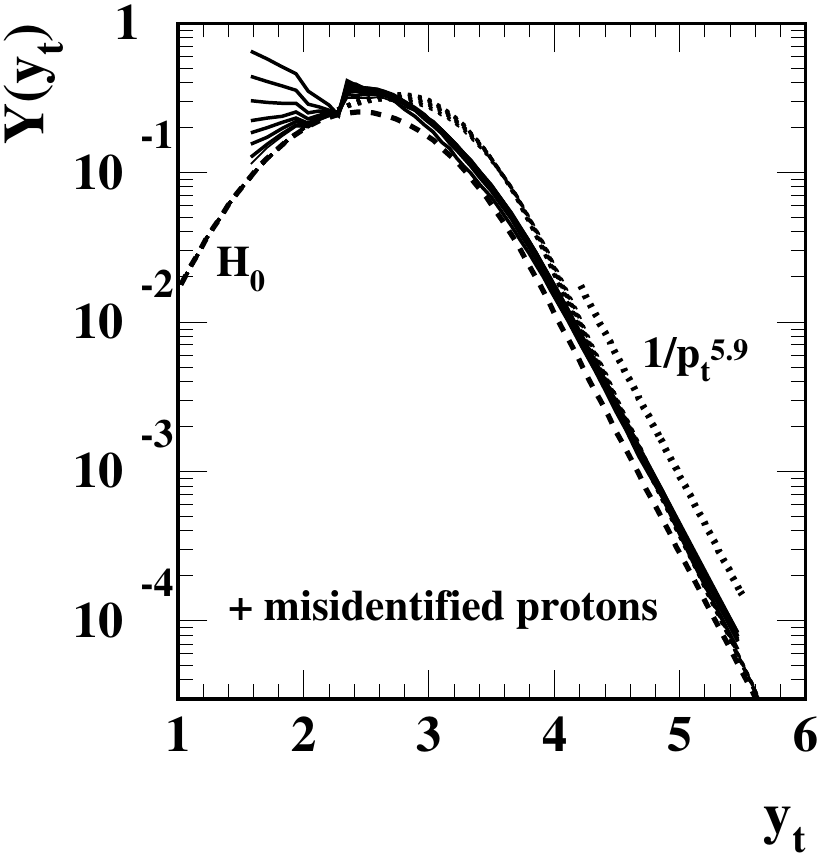}
	\caption{\label{crosstalk}
		Left: Published pion spectra as in Fig.~\ref{piddata} (a) (points) compared to modified TCM spectra $\bar \rho_{0\pi}'(y_t,n_s)$ (curves) defined by Eq.~(\ref{crosst}) showing improved agreement.
		Right: Pion data hard components (solid) as in Fig.~\ref{pionsx} (right) compared to modified TCM hard components (dotted) defined by Eq.~(\ref{crosst}) with $\bar \rho_{0}(y_t) \rightarrow \hat H_0(y_t)$. Much of the data-TCM (dashed) disagreement appears to be explained by a proton contribution.
	} 
\end{figure}

Figure~\ref{crosstalk} (left) shows uncorrected pion data (points) compared to a modified TCM prediction (curves) based on Eq.~(\ref{crosst}). Whereas uncorrected pion data in Fig.~\ref{piddata} (a) fall well above the TCM prediction for ideal data above $y_t \approx 2$ this modified TCM simulating added (missing?) protons provides a much-improved data description.

Figure~\ref{crosstalk} (right) repeats Fig.~\ref{pionsx} (right) but adds TCM hard components (dotted) modified according to Eq.~(\ref{crosst}), with $\bar \rho_{0x}(y_t,n_s)  \rightarrow \hat H_{0x}(y_t,n_s)$ and proton hard components $\hat H_{0p}(y_t,n_s)$ multiplied by factors $z_{hi}(n_s) / 0.25$ to represent the hard-component amplitude variation in Fig.~\ref{pchxx} (right). Although the pion data are not perfectly described two aspects are clear: (a) the general magnitude of the pion data-TCM discrepancy is well-approximated, and (b) the pion-data power-law slope above \yt\ = 4 seems closer to the proton exponent 6.8 than the pion exponent 5.9 describing the pion dashed curve. This exercise suggests that the major source of disagreement between pion TCM and data is the missing protons requiring the correction described in Sec.~\ref{correct}.

\subsection{Final 13 TeV $\bf p$-$\bf p$ TCM parameters} \label{paramfinal}

Table~\ref{engparamsy} presents final TCM model-function parameters for results presented here in Sec.~\ref{tcmfinal}. The uncertainties for proton parameters $\bar y_t$ and $\sigma_{y_t}$ indicate the range of variation of those parameters for the variable-TCM hard component as described in Sec.~\ref{protonalt}. Corresponding ensemble-mean \pt\ values $\bar p_{ts}$ and $\bar p_{th}$ for soft and hard spectrum components corresponding to these model parameters are presented in Table~\ref{otherparamsxx}.

\begin{table}[h]
	\caption{TCM model parameters for identified hadrons from 13 TeV \pp\ collisions: hard-component parameters $(\bar y_t,\sigma_{y_t},q)$ and soft-component parameters $(T,n)$. These are {\em final} values based on 13 TeV \pp\ results reported above. The first two proton values are averages over event classes. Detailed proton parameter variations are described in Sec.~\ref{protonalt}.
	}
	\label{engparamsy}
	\begin{center}
		\begin{tabular}{|c|c|c|c|c|c|} \hline
			& $\bar y_t$ & $\sigma_{y_t}$ & $q$ & $T$ (MeV) &  $n$  \\ \hline
			$ \pi^\pm $     &  $2.46 \pm 0.02$ & $0.60 \pm 0.02$ & $3.7\pm 0.1$ & $145\pm 2$ & $8.0\pm 0.3$ \\ \hline
			$K^\pm$    & $2.68 \pm 0.02$  & $0.60\pm 0.02$ & $3.7\pm 0.1$ & $200\pm 5$ & $14\pm 2$ \\ \hline
			$p$        & $2.90\pm 0.05$  & $0.50\pm 0.03$ & $4.6\pm 0.2$ & $210\pm 5$ & $14\pm 2$ \\ \hline
		\end{tabular}
	\end{center}
\end{table}

Table~\ref{otherparamsxx} presents final TCM $z_{0i}$ and $\tilde z_i$ parameters for results presented here in Sec.~\ref{tcmfinal}. Those parameters then determine the solid curves in Fig.~\ref{zxi} that can in principle generate the required $z_{si}(n_s)$ and $z_{hi}(n_s)$ parameter values. Note that the $\tilde z_i$ values are approximately proportionality to hadron mass as noted in Sec.~V A of Ref.~\cite{pidpart1}, and the $z_{0i}$ values are consistent with statistical-model predictions~\cite{statmodel}. Thus, the TCM for 13 TeV \pp\ collisions is predicted quantitatively {\em within data uncertainties} by previous results from other collision systems.

\begin{table}[h]
	\caption{TCM species fraction coefficients $z_0$ and $\tilde z_i$ for identified hadrons from 13 TeV \pp\ collisions. These are final values from the present study. The soft- and hard-component ensemble-means $\bar p_{ts}$ and $\bar p_{ts}$ correspond to model parameters in Table~\ref{engparamsy} and will be employed in a follow-up study. The large uncertainty for proton $\bar p_{th}$ corresponds to variation of the proton hard-component centroid in Sec.~\ref{protonalt}.
	}
	\label{otherparamsxx}
	\begin{center}
		\begin{tabular}{|c|c|c|c|c|} \hline
			&   $z_0$    &  $\tilde z_i$ &   $ \bar p_{ts}$ (GeV/c)  & $ \bar p_{th}$ (GeV/c)  \\ \hline
			$ \pi^\pm$        &   $0.80\pm0.01$  & $0.60\pm0.05$  & $0.40\pm0.02$ &    $1.05\pm0.03$  \\ \hline
			$K^\pm $   &  $ 0.130\pm0.005$   &  $2.60\pm0.05$ &  $0.60\pm0.02$&  $1.46\pm0.03$   \\ \hline
			$p $        & $ 0.075\pm0.005$    &  $5.60\pm0.05$ &  $0.74\pm0.02$&   $1.55\pm0.10$   \\ \hline
		\end{tabular}
	\end{center}
\end{table}

Ensemble-mean \pt\ values in the right columns are based on model-function parameters in Table~\ref{engparamsy} and will be used in a study of hadron species transport from soft to hard component within small collision systems.

\section{TCM data description quality} \label{quality}

As in previous studies of the TCM applied to hadron spectra~\cite{pidpart2,ppbbw} the quality of data descriptions is evaluated based on Z-scores calculated with statistical uncertainties. Some modification of statistical uncertainties reported by Ref.~\cite{alicepppid} appears to be necessary.

\subsection{Evaluating data-model accuracy via Z-scores} \label{modelacc}

Spectrum data-model comparisons are often represented by data/model ratios which can be misleading, as discussed for instance in Ref.~\cite{ppprd}. A more meaningful measure of model validity is the Z-score~\cite{zscore} defined by
\bea \label{zscore}
Z_i &=& \frac{O_i - E_i}{\sigma_i} \rightarrow \frac{\text{data $-$ model}}{\text{statistical error}},
\eea
where $O_i$ is a spectrum datum, $E_i$ is the corresponding expectation (model prediction) and $\sigma_i$ is the data r.m.s. statistical uncertainty (error). Based on the Z-score definition in Eq.~(\ref{zscore}) the relation to the $\chi^2$ statistic is
\bea \label{chinu}
\chi^2 & \equiv & \sum_{i=1}^N \frac{(O_i - E_i)^2}{\sigma_i^2} =  \sum_{i=1}^N Z_i^2
\eea
for $N$ data points in a spectrum. Given model degrees of freedom $\nu = N - \text{number of fit parameters}$ one expects $\chi^2 \sim \nu$, in which case the r.m.s. value for Z scores for an acceptable fit should be $\sqrt{\nu / N}$ -- somewhat less than 1.  An important advantage of Z-scores over the integral $\chi^2$ measure is detailed {\em differential} information about the nature of any significant data-model deviations.

Data/model ratios are related to Z-scores by
\bea \label{suppress}
\frac{\text{data}}{\text{model}} - 1 &\approx& \text{Z-score} \times \frac{\text{error}}{\text{data}},
\eea
with error/model (exact) $\rightarrow $ error/data (approximate). The error/data ratio (typically $\ll 1$) can vary by orders of magnitude between different particle types and collisions systems, and even across \yt\ intervals. Interpretation of data/model ratios relative to 1 is thus problematic.

For meaningful evaluation of model description quality proper statistical uncertainties (errors) are required. The following subsection evaluates and corrects statistical uncertainties as provided by Ref.~\cite{alicepppid}.

\subsection{Statistical uncertainties} \label{statsrules}

For spectra presented on transverse momentum \pt\ the bin widths on \pt\ are typically strongly varied, with greater widths at higher \pt\ to compensate the falling spectra as in Fig.~1 of Ref.~\cite{alicepppid}. The published statistical uncertainties include sharp changes in magnitude that are not simply explained by bin width changes (e.g.\ only some appear at transitions between bin widths) and, when employed in Z-scores, tend to inject misleading structures into the result. In this study statistical uncertainties are obtained from published spectrum data as described below. Since the event number for each event class is not readily available estimated uncertainties are scaled overall for each class to best match published uncertainties.

Assuming Poisson statistics applied to the {\em total} particle number integrated within each \pt\ bin width $\delta p_t$ and $\Delta \eta$ the statistical uncertainty for charge density $\bar \rho_{0i}(p_t) \equiv d^2n_{chi} / p_{ti} dp_{ti} d\eta$ compatible with Ref.~\cite{ppprd} is
\bea \label{newstat}
\delta \bar \rho_{0i}(p_{ti},n_s) &=& \frac{\sqrt{N_{evt} \bar \rho_{0i}(p_{ti},n_s) p_{ti} \delta p_{ti} \Delta \eta}} {N_{evt} p_{ti} \delta p_{ti} \Delta \eta}
\\ \nonumber
&=&  \sqrt{ \bar \rho_{0i}(p_t,n_s) / p_{ti}\delta p_{ti}} \times \text{constant} \ll 1
\eea
where bin widths $\delta p_{ti}$ increase stepwise with \pt\ to accommodate density variations over orders of magnitude.

The left panels in each figure below present the published statistical and {\em total} systematic uncertainties (solid curves) from Ref.~\cite{alicepppid} in ratio to data. The dashed curves present the results of Eq.~(\ref{newstat}) based on published PID data spectra $\bar \rho_{0i}(p_t,n_s)$. As noted, since the event numbers are not available the constant in  Eq.~(\ref{newstat}) is adjusted for each event class to best match the published statistical uncertainties. The matching for statistical uncertainties alone is shown in right panels for event classes 1, 5, 9 of each hadron species. For protons the {\em uncorrected} data spectra (what was actually counted) are used in Eq.~(\ref{newstat}), and the resulting statistical errors then receive the same correction as the published spectra.

\begin{figure}[h]
	\includegraphics[width=3.3in]{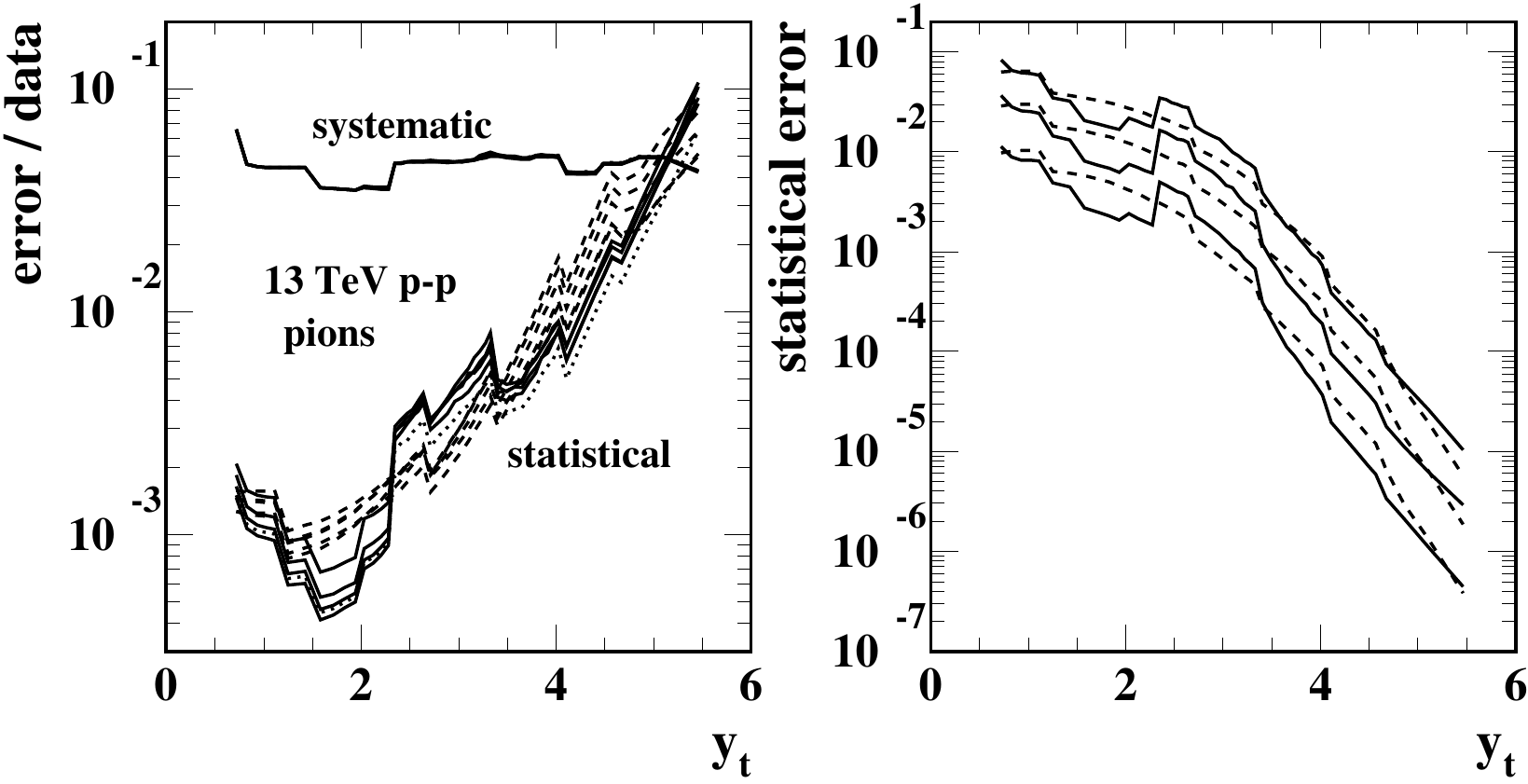}
	\caption{\label{errorcomparepions}
		Left: Statistical (lower) and total-systematic (upper) data uncertainties for pions from 13 TeV \pp\ collisions published in Ref.~\cite{alicepppid} (solid curves) compared to estimates of statistical uncertainties (dashed) inferred from published PID spectra via Eq.~(\ref{newstat}).
		Right: Comparison of published statistical uncertainties from Ref.~\cite{alicepppid} (solid) with estimates  via Eq.~(\ref{newstat}) (dashed) scaled to best match the published values.
	} 
\end{figure}

\begin{figure}[h]
	\includegraphics[width=3.3in]{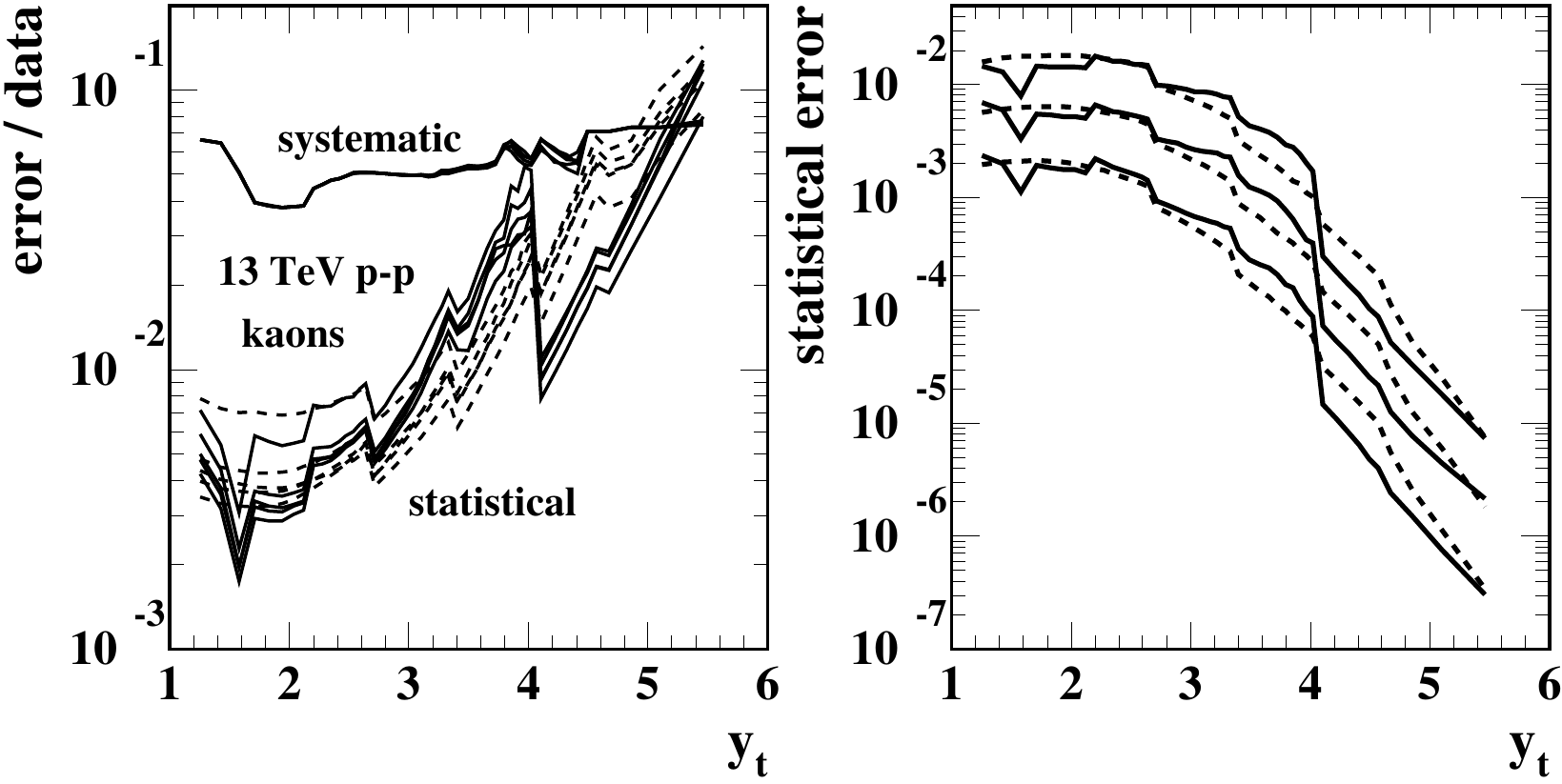}
	\caption{\label{errorcomparekaons}
Same as for Fig.~\ref{errorcomparepions} but for charged kaons.	
	} 
\end{figure}

\begin{figure}[h]
	\includegraphics[width=3.3in]{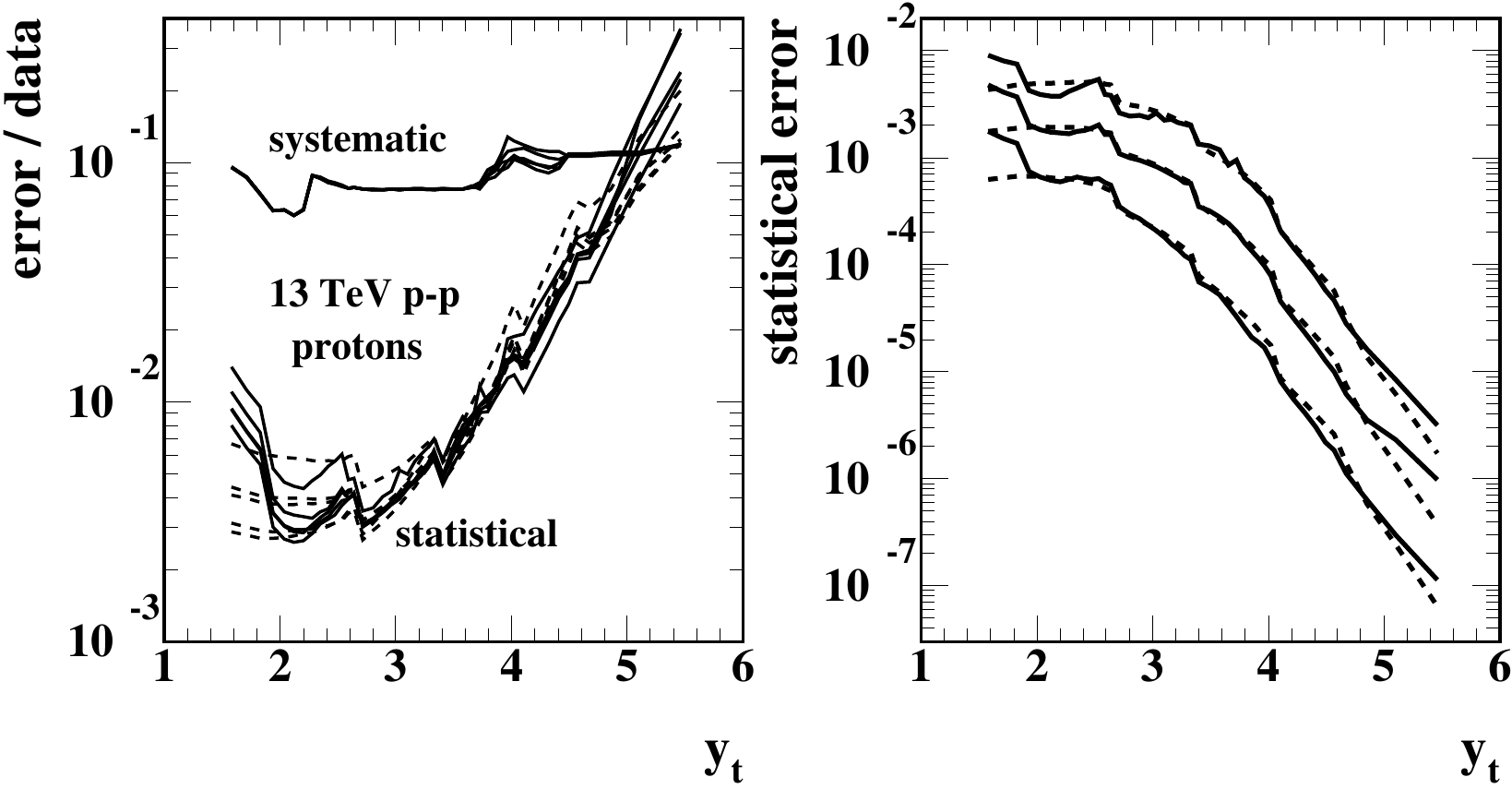}
	\caption{\label{errorcompareprotons}
Same as for Fig.~\ref{errorcomparepions} but for protons.	
	}  
\end{figure}

\begin{figure}[h]
	\includegraphics[width=1.65in]{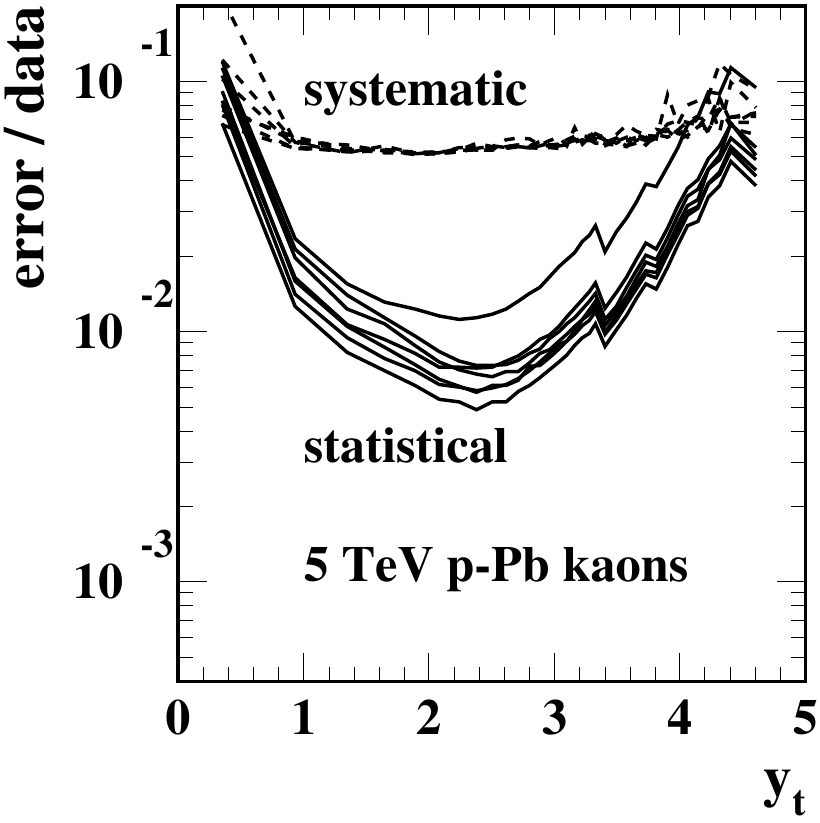}
	\includegraphics[width=1.65in]{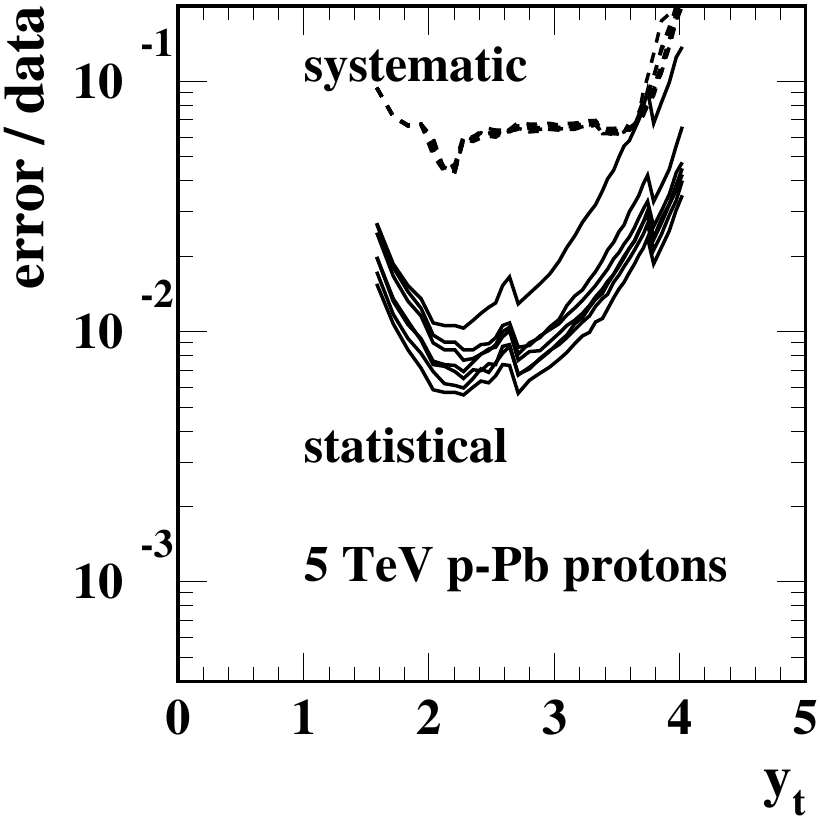}
	\caption{\label{olderrordat}
		Left: Statistical (solid) and total-systematic (dashed) uncertainties divided by spectrum data (error/data) for neutral kaons from 5 TeV \ppb\ collisions as presented in Ref.~\cite{ppbbw} based on data reported by Ref.~\cite{aliceppbpid}.
		Right: Same as left panel but for (uncorrected) protons.
		} 
\end{figure}

Figure~\ref{olderrordat} shows error/data ratios from 5 TeV \ppb\ collisions for neutral kaons (left) and (uncorrected) protons (right) as presented in Ref.~\cite{ppbbw}. It is notable that the step-wise variations in error/data ratios for 5 TeV \ppb\ data from Ref.~\cite{aliceppbpid} are compatible with what is obtained from  Eq.~(\ref{newstat}) for 13 TeV \pp\ collisions but seems inconsistent with the \pp\ results from Ref.~\cite{alicepppid}, although the two data sets are from the same collaboration.

\subsection{TCM data description}

\begin{figure}[h]
	\includegraphics[width=3.3in]{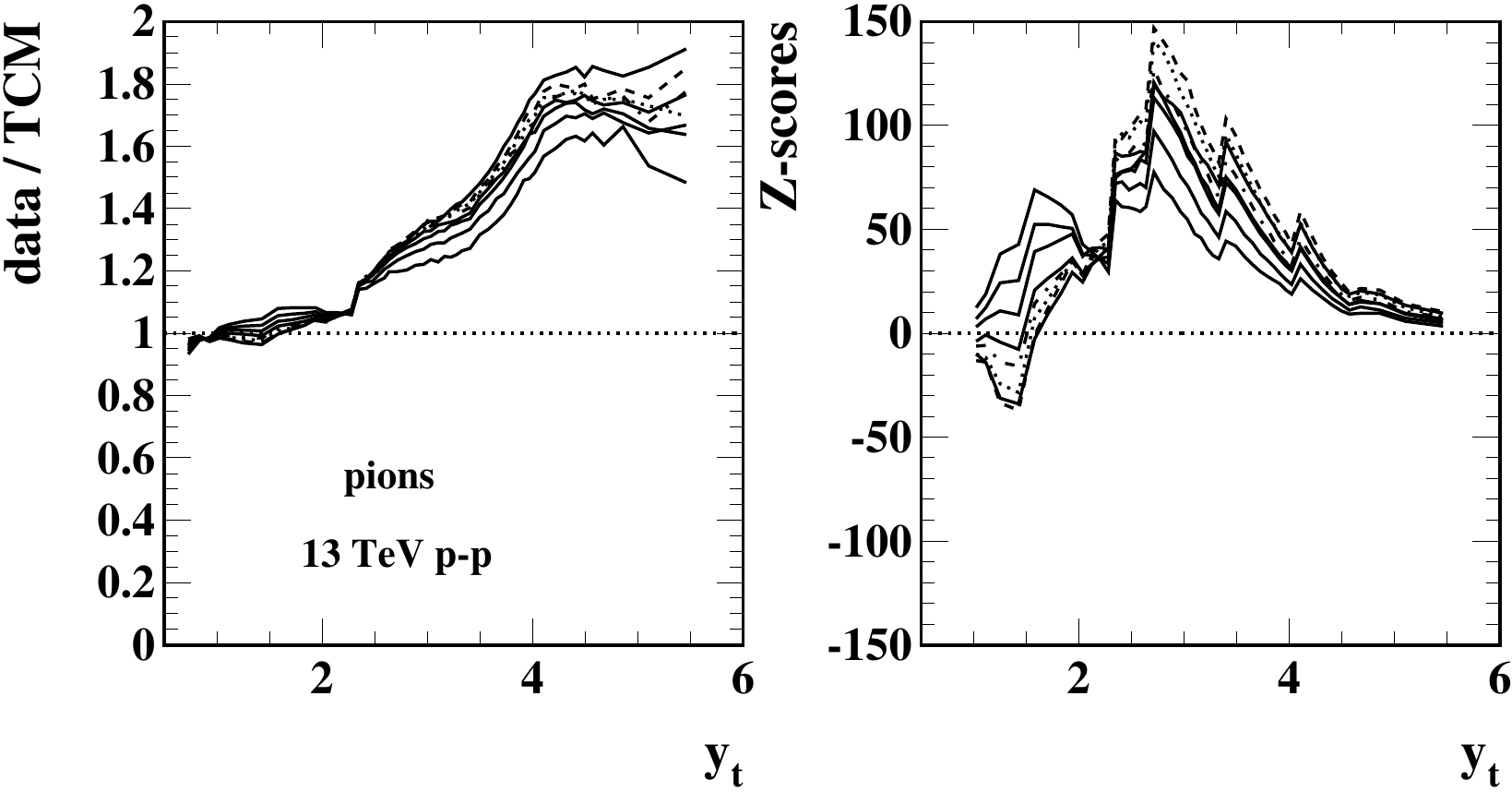}
	\caption{\label{1b1}
		Left: Data/TCM spectrum ratios for pions from ten event classes of 13 TeV \pp\ collisions.
		Right: Corresponding Z-scores exhibiting large nonstatistical excursions.
	} 
\end{figure}

\begin{figure}[h]
	\includegraphics[width=3.3in]{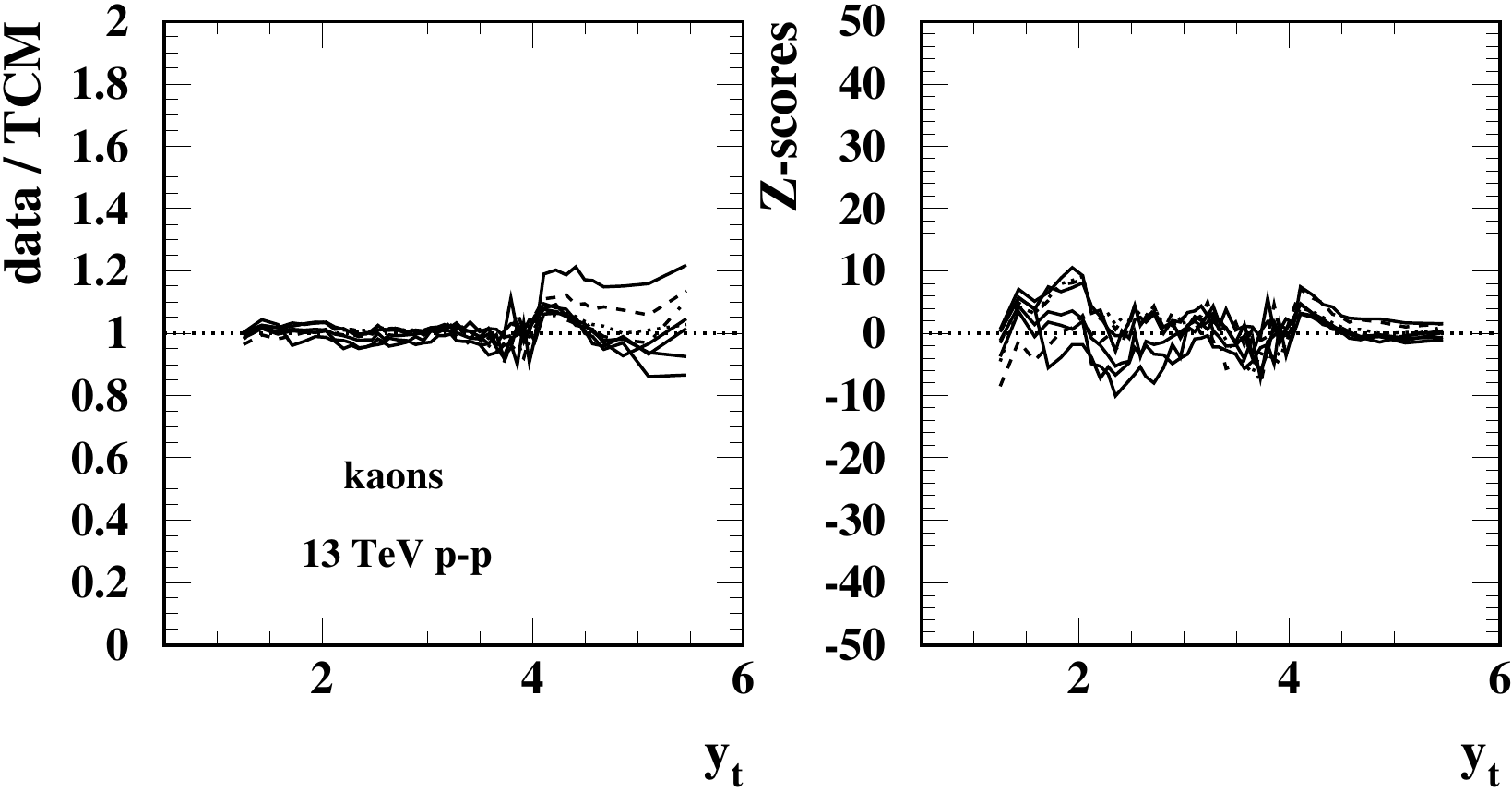}
	\caption{\label{1b2} \label{key}
		Left: Data/TCM spectrum ratios for charged kaons from ten event classes of 13 TeV \pp\ collisions.
Right: Corresponding Z-scores indicating acceptable model description.
	} 
\end{figure}

\begin{figure}[h]
	\includegraphics[width=3.3in]{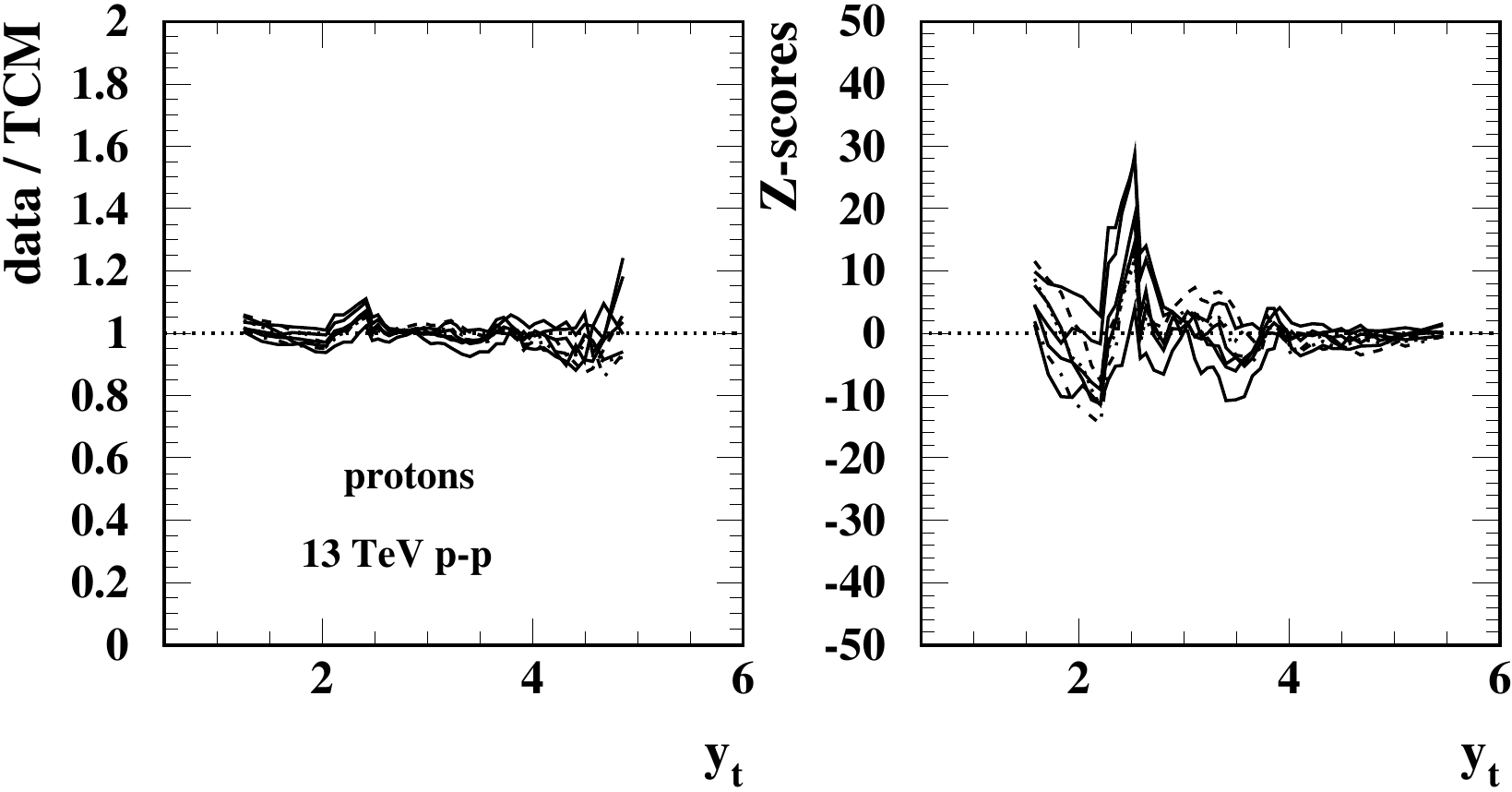}
	\caption{\label{1b3}
		Left: Data/TCM spectrum ratios for protons from ten event classes of 13 TeV \pp\ collisions.
Right: Corresponding Z-scores indicating acceptable model description. Inference of statistical errors for corrected proton spectra (observed vs corrected numbers) is described in Sec.~\ref{statsrules}.
	} 
\end{figure}

In the left panels of Figs.~\ref{1b1} -- \ref{1b3} below data/TCM ratios are plotted for three hadron species from 13 TeV \pp\ collisions. Except for pions deviations from 1 are a few percent with a few local excursions, seemingly indicating acceptable model descriptions. For pions, large discrepancies in Fig.~\ref{pichx} (right) are clearly evident.
Corresponding Z-scores are plotted in the right panels. Z-scores for pions indicate an unacceptable data description.  Large excursions over both broad and narrow \yt\ intervals suggest nonphysical pion data variations that should not be described by a physical model. That finding is consistent with results for pions in Sec.~\ref{zxidirect}. A reason for the large data-model difference is suggested in Sec.~\ref{pionproton}.

The Z-scores for kaons and (corrected) protons are consistent with an acceptable model description aside from a few sharp structures that may be attributed to local data anomalies. As can be inferred from systematic vs statistical errors in left panels of Sec.~\ref{statsrules}, use of systematic uncertainties in Z-scores would result in a factor 10 - 100 reduction below \yt\ = 4 ($p_t = 3.8$ GeV/c) which would be quite misleading.

\section{$\bf p$-$\bf p$ PID spectrum and yield ratios} \label{pidratios}

Reference~\cite{alicepppid} reports PID yield ratios (e.g.\ $p/\pi$) in its Figs.~2, 5 and 6. While such results have become conventional in high-energy nuclear physics reports one may ask why. As demonstrated in this and previous studies (e.g.\ ~\cite{ppbpid,pidpart1,pidpart2}) PID spectra include at least two major contributions (soft and hard or nonjet and jet components) with quite different behaviors. And spectrum trends for mesons are quite different from those for baryons~\cite{ppbpid,pidpart2}. Such differences can be isolated clearly and quantitatively when individual spectra are analyzed differentially. Ratios of differential PID spectra are effectively impossible to interpret unambiguously on their own. PID ratios from Ref.~\cite{alicepppid} are here interpreted in the context of the present TCM study of individual PID spectra.

\subsection{PID ratio model}

Based on Eq.~(\ref{pidspectcm}) (second line) PID {\em differential} spectrum ratios can be expressed via the TCM as
\bea \label{pidspectcmx}
\frac{\bar \rho_{0i}(y_t,n_s)}{\bar \rho_{0j}(y_t,n_s)}&\approx&  \frac{z_{si}(n_s) \bar \rho_{s} \hat S_{0i}(y_t) +   z_{hi}(n_s) \bar \rho_{h} \hat H_{0i}(y_t,n_s)}{z_{sj}(n_s) \bar \rho_{s} \hat S_{0j}(y_t) +   z_{hj}(n_s) \bar \rho_{h} \hat H_{0j}(y_t,n_s)} 
 \nonumber \\
&\rightarrow & \frac{z_{si}(n_s)\hat S_{0i}(y_t)}{z_{sj}(n_s)\hat S_{0j}(y_t)}~~~\text{at low \yt}
\nonumber \\
 &\rightarrow& \frac{z_{hi}(n_s) \hat H_{0i}(y_t,n_s)}{z_{hj}(n_s) \hat H_{0j}(y_t,n_s)}~~~\text{at high \yt}
\eea
The trend at low \yt\ depends on Eq.~(\ref{zsix}): limiting ratio $z_{0i} / z_{0j}$ for $x(n_s) \rightarrow 0$ and a {\em decreasing} trend with increasing $x(n_s)$ if species $i$ is more massive than species $j$ due to $\tilde z_i$ being proportional to hadron mass (see Fig.~8 of Ref.~\cite{pidpart1}). At high \yt\ the trend would also be decreasing since $z_{hi}(n_s) = \tilde z_i z_{si}(n_s)$ {\em if}\, model functions $\hat H_{0x}(y_t)$ are fixed, independent of event class. However, for 5 TeV \ppb\ collisions the data hard components vary substantially with centrality, mesons shifting to {\em lower} \yt\ while baryons shift to {\em higher} \yt\ with increasing \nch. The result, as demonstrated in Fig.~9 of Ref.~\cite{pidpart2}, is a change to {\em increasing} ratio values with \nch\ above some point near \yt\ = 3.2 ($p_t \approx 1.7$ GeV/c). The spectrum results for 13 TeV \pp\ spectra in Secs.~\ref{zxidirect} and \ref{protonalt} suggest an increasing trend for that system as well.
The corresponding PID TCM expression for {\em integrated} yield ratios is
\bea \label{pidyieldrat}
\frac{\bar \rho_{0i}(n_s)}{\bar \rho_{0j}(n_s)} &=& \frac{z_{si}(n_s)}{z_{sj}(n_s)} \cdot \frac{1+ \tilde z_{i} x(n_s)}{1+\tilde z_{j} x(n_s)}
\\ \nonumber
&\rightarrow& \frac{z_{0i}}{z_{0j}}~~\text{independent of $n_s$},
\eea
per Eq.~(\ref{zsix}), i.e.\ constant values independent of \nch. Approximate centrality independence of parameters $z_{0i}$ for 5 TeV \ppb\ spectra is confirmed in Sec.~V B of Ref.~\cite{pidpart1}.

\subsection{$\bf p$-$\bf p$ PID spectrum and yield ratio data}

\begin{figure}[h]
	\includegraphics[width=3.3in]{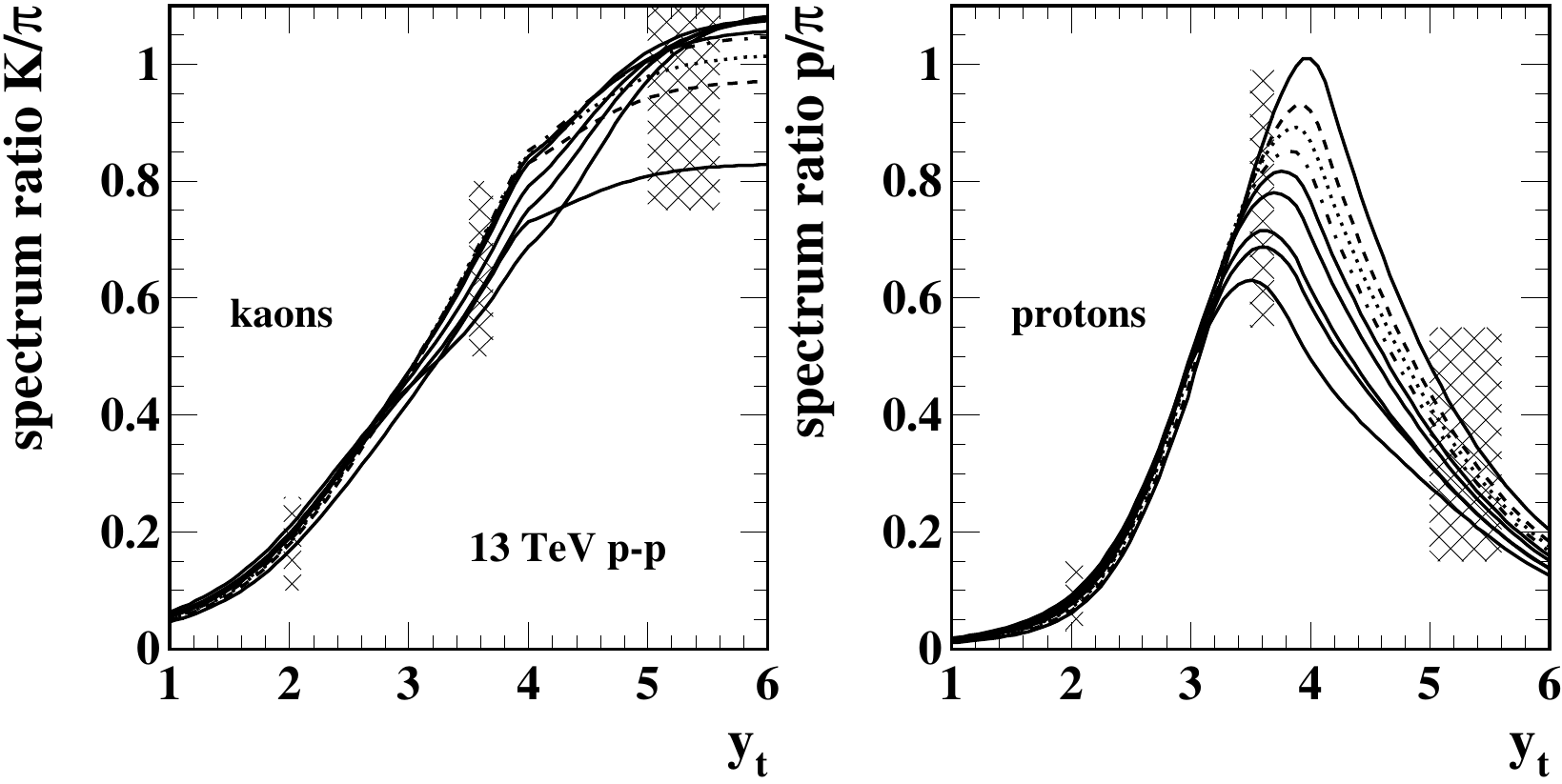}
	\caption{\label{specrat}
		Left: Charged-kaon-to-pion ratios of TCM spectra from Sec.~\ref{pppidtcm} for ten event classes of 13 TeV \pp\ collisions. The cross-hatched regions correspond to \pt\ intervals in Fig.~2 of Ref.~\cite{alicepppid}.
		Right: Similar treatment of proton-to-pion spectrum ratios.
	} 
\end{figure}

Figure~\ref{specrat} shows TCM hadron/pion spectrum ratios corresponding to Eq.~(\ref{pidspectcmx}) (first line) for charged kaons (left) and protons (right). The line styles vary with event class from most-central as solid, dashed, dotted and dash-dotted, with solid thereafter. The three hatched areas relate to \pt\ intervals associated with Fig.~2 of Ref.~\cite{alicepppid} as discussed further below. Similar spectrum ratios for 5 TeV \ppb\ collisions are considered in detail in Sec.~VIII of Ref.~\cite{ppbpid} and Sec.~IV of Ref.~\cite{pidpart2}. In either case the prominent peak near \pt\ = 3 GeV/c ($y_t \approx 3.75$) for either $p/\pi$ or $\Lambda / K$ ratios is attributed to a {\em relative shift} between hard-component modes for mesons vs baryons as is evident in Sec.~\ref{tcmfinal}. In contrast, no comparable peak structure appears for $K/\pi$ spectrum ratios.
Baryon/meson peak structures near 3 GeV/c are thus dominated by hard components  associated with jet production~\cite{hardspec,fragevo,ppquad,mbdijets}.  

The structure of PID spectrum ratios can be simply explained quantitatively within a TCM context in terms of Eq.~(\ref{pidspectcmx}) taking $K/\pi$ as an example. In the low-\yt\ and low-\nch\ limit $z_{si}(n_s) / z_{sj}(n_s) \rightarrow z_{0i} / z_{0j} \approx 0.16$ for $K/\pi$. The ratio of soft-component models at low \yt\ is somewhat less than 1 because of the pion resonance contribution and the different slope parameters $T$. The combination is consistent with a ratio value $\approx 0.1$ as in Fig.~\ref{specrat} (left).

At high \yt\ $z_{hi}(n_s) / z_{hj}(n_s) \approx 0.25 / 0.50 \approx 0.5$ (averaged over event classes). Pion and kaon hard-component models  have the same widths but are displaced relatively per $\bar y_t = 2.46$ vs 2.68 for pions vs kaons. The ratio of model functions at high \yt\ (approximate power laws) then goes as $p_t^{5.9} / (p_t - \delta p_t)^{5.9} \approx 1+ 5.9 \delta p_t / p_t$. But $\delta p_t = m_t \delta y_t \approx 0.2 m_t$ so the ratio  goes as $1 + 5.9 \times 0.2 m_t / p_t \approx 2$. Combined with $z_{hi}(n_s) / z_{hj}(n_s) \approx 0.5$ that yields an expected ratio value $\approx 1$ at high \yt\ as in Fig.~\ref{specrat} (left). 

The $p/\pi$ ratio trends on \yt\ can also be understood as follows: At low \yt\ the ratio limiting case per Eq.~(\ref{pidyieldrat}) is $z_{0i} / z_{0j} = 0.075 / 0.80 \approx 0.09$, but as for the $K/\pi$ ratio the combination of resonance contribution to pions and a much harder proton soft-component model leads to a model ratio at low \yt\ much less than one, the product being consistent with the $\approx 0.025$ value in Fig.~\ref{specrat} (right). 

The high-\yt\ limiting case per Eq.~(\ref{pidspectcmx}) includes the ratio $z_{hi} / z_{hj} \approx 0.25/0.5 = 0.5$. However the structure of the model-function ratio $\hat H_{0i}(y_t)/\hat H_{0j}(y_t)$ is determining and results from a combination of higher centroid for protons (2.46 vs 2.9 for $\pi$ vs $p$) and a broader peak and harder tail for pions (0.60 vs 0.50) and (3.7 vs 4.6). Thus, with increasing \yt\ above 2 the proton peak dominates near and just above its centroid (i.e.\ the mode of the ratio peak), but beyond \yt = 4 the width and harder tail of the pion peak prevails leading to rapid reduction of the $p/\pi$ ratio.

While the detailed structure of PID spectra summarized in Sec.~\ref{tcmfinal} can be used to interpret quantitatively the structure of corresponding spectrum ratios the latter cannot be used to interpret the former. Differences (differential analysis) make information more accessible whereas ratios suppress it. From the present study it is safe to conclude that the structure of PID spectrum ratios as in Fig.~\ref{specrat} is dominated by minimum-bias jets.

\begin{figure}[h]
	\includegraphics[width=3.3in]{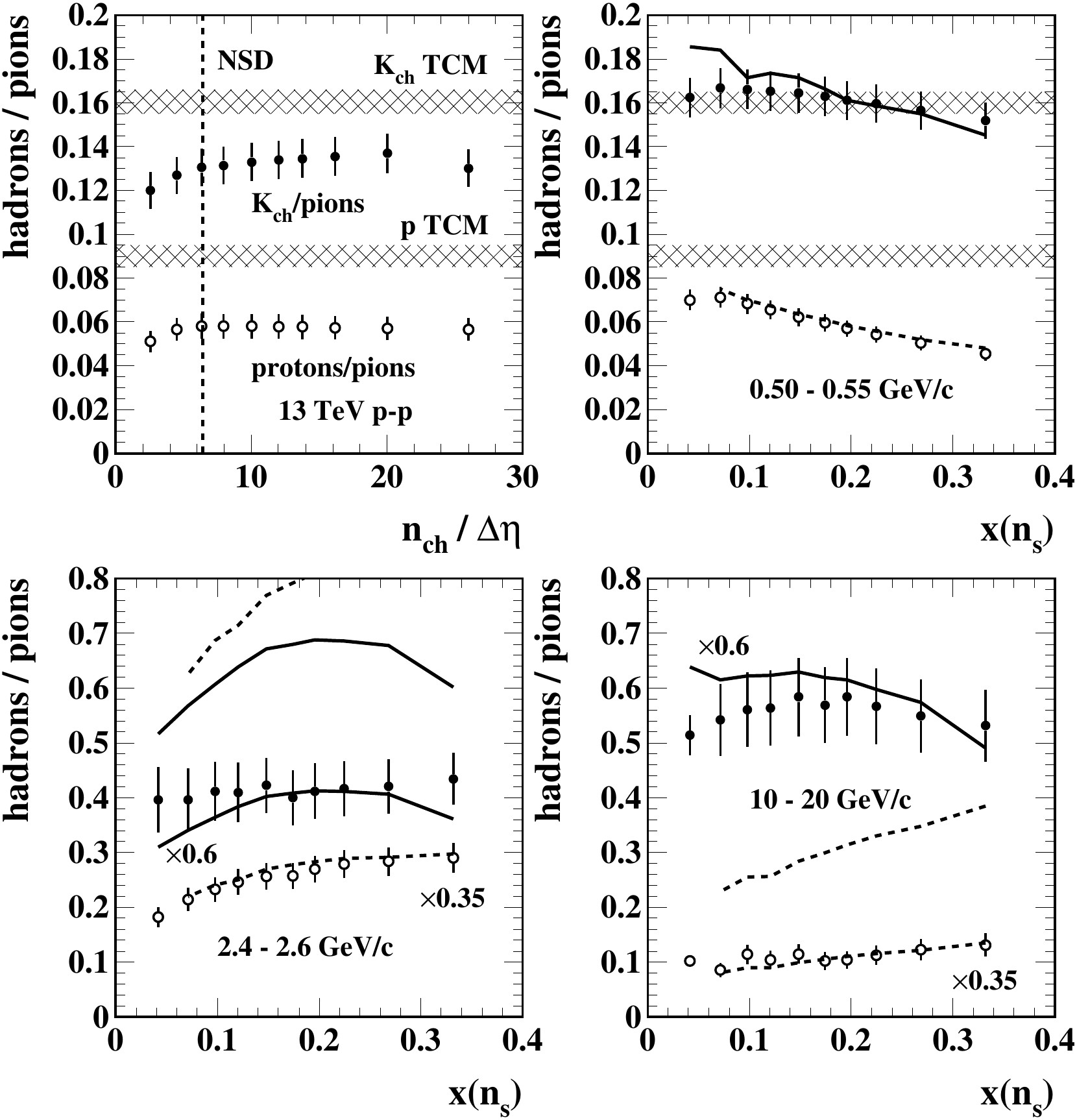}
\put(-140,155) {\bf (a)}
\put(-23,230) {\bf (b)}
\put(-140,105) {\bf (c)}
\put(-23,105) {\bf (d)}
	\caption{\label{intrat}
	(a) Ratios of integrated yields (points) for charged-kaons-to-pions (upper) and protons-to-pions (lower) compared to TCM estimates for those ratios based on the present study (hatched). Panel (a) corresponds to Fig.~5 of Ref.~\cite{alicepppid}. The vertical dotted line indicates $\bar \rho_0$ for 13 TeV NSD \pp\ collisions.
	(b-d) hadron/pion yield ratios evaluated within three \pt\ intervals as in Fig.~2 of Ref.~\cite{alicepppid} for charged kaons (solid dots) and protons (open circles). The curves are TCM predictions for kaons/pions (solid) and protons/pions (dashed). In (c,d) the lower curves for each line style are the upper curves (predictions) rescaled by the indicated factors. The bars represent published total systematic uncertainties. In panel (d) the unrescaled kaon curve falls above the plot area.
	} 
\end{figure}
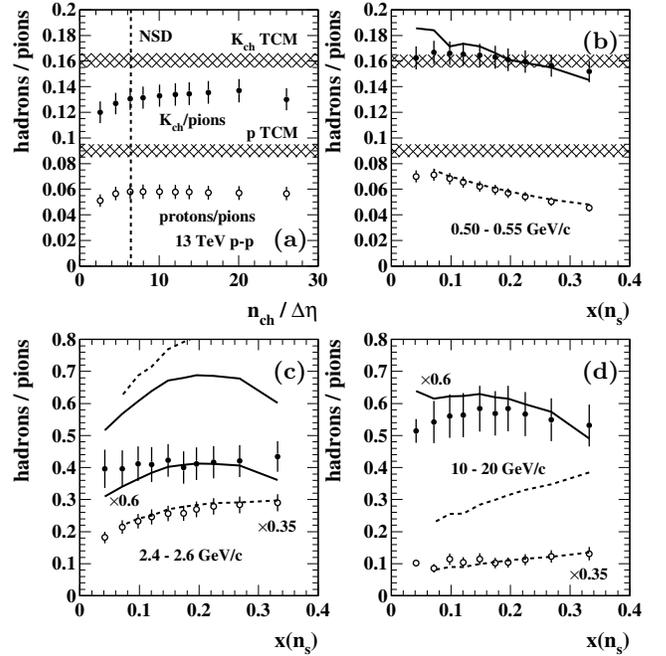

Figure~\ref{intrat} (a) shows integrated yield ratios (points) from 13 TeV \pp\ collisions as presented in Fig.~5 of Ref.~\cite{alicepppid}. According to TCM Eq.~(\ref{pidyieldrat}) those trends should be constant, with values $z_{0i} / z_{0j}$ corresponding to entries in Table~\ref{otherparamsxx}. In principle there is no jet dependence to integrated yields.
The constant values predicted by Eq.~(\ref{pidyieldrat}) (see Table~\ref{otherparamsxx}) are $K/\pi \approx 0.130 / 0.80 \approx 0.16$ and $p/\pi \approx 0.075 / 0.80 \approx 0.09$ as represented by the hatched bands in panel (a).  The NSD value $\bar \rho_0 \approx 6.4$ for 13 TeV \pp\ collisions is indicated by the vertical dashed line. Some suppression may occur for lower \nch\ values.

Figure~\ref{intrat} (b-d) shows PID yield ratios for 13 TeV \pp\ collisions evaluated within specific \pt\ intervals [0.50,0.55] GeV/c, [2.4,2.6] GeV/c and [10,20] GeV/c (hatched bands in Fig.~\ref{specrat}) for charged kaons (solid dots) and protons (open circles) as shown in Fig.~2 of Ref.~\cite{alicepppid}. Also shown are results when the same procedure is applied to TCM PID spectra from Sec.~\ref{pppidtcm} for charged kaons (solid) and protons (dashed). In panel (b) no proton correction is required. In panels (c) and (d) the upper curve for solid or dashed is the TCM prediction and the lower curve corresponds to additional factor 0.6 or 0.35. Those factors may be interpreted in the context of the proton efficiency issue addressed in Sec.~\ref{correct} where the apparent proton efficiency is about 0.55 above \yt\ = 3 ($p_t \approx 1.4$ GeV/c). It is possible that the missing protons are not ``lost'' but instead misidentified as another hadron species during $dE/dx$ analysis. The present results suggest that protons are misidentified as pions (see Sec.~\ref{pionproton}) resulting in lowered kaon/pion ratios as well as proton/pion ratios.

In panel (b) the expected behavior at low \pt\ is exhibited: almost linear decrease with jet/nonjet ratio $x(n_s)$ corresponding to Eq.~(\ref{zsix}). The decrease is stronger for protons since parameter $\tilde z_i$ in that equation is simply proportional to hadron mass~\cite{pidpart1}. In panel (c) near the transition point at \yt\ $\approx$ 3.3 the trends are ambiguous. In panel (d) corresponding to  exponential (on \yt) tails of hard components the kaon/pion trend is consistent with decrease especially for larger $x(n_s)$ (see Fig.~\ref{specrat}, left) whereas the proton/pion trend is increasing as expected because of the shifting proton hard component.

\begin{figure}[h]
	\includegraphics[width=3.3in]{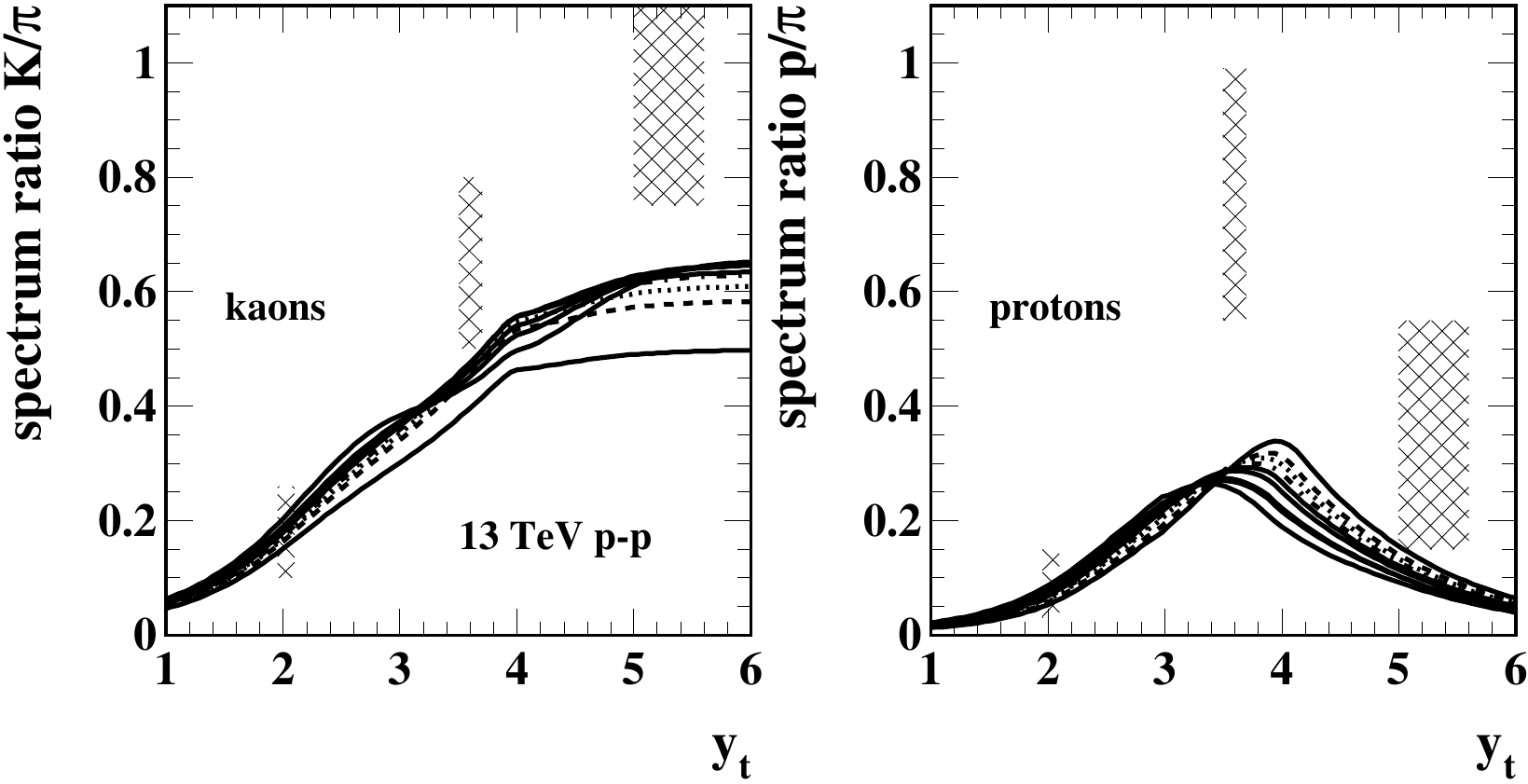}
	\caption{\label{3ebad}
	Same as Fig.~\ref{specrat} but TCM proton spectra correspond to {\em uncorrected} proton data, and TCM pion spectra are produced with $\tilde z_i = 0.9$ rather than 0.6 to emulate data hard components in Fig.~\ref{pichx} (right), both reflecting published data.
	} 
\end{figure}

Figure~\ref{3ebad} illustrates an alternative procedure wherein the TCM description is modified to describe spectra as published rather than predicting unbiased spectra based on low-\pt\ trends. The {\em inverse} proton correction is applied to TCM proton spectra, and $\tilde z_i$ for pions is increased from 0.6 to 0.9 to be consistent (on average) with the pion hard-component data in Fig.~\ref{pichx} (right). The combination results in substantial reduction of kaon/pion spectrum ratios and large reduction of proton/pion ratios consistent with the factors 0.6 and 0.35 in Fig.~\ref{intrat}.

Figure~\ref{3ebad} can be compared with Fig.~2 of Ref.~\cite{aliceppbpid}. For that 5 TeV \ppb\ study the upper limit for charged-kaon data is $y_t \approx 3.6$ corresponding to 2.5 GeV/c and for proton data is $y_t \approx 3.75$ corresponding to 3 GeV/c. There is quantitative agreement between 5 TeV \ppb\ ratios and 13 TeV \pp\ ratios within uncertainties that seem to confirm common $dE/dx$ biases for the two data volumes.

\begin{figure}[h]
	\includegraphics[width=3.3in]{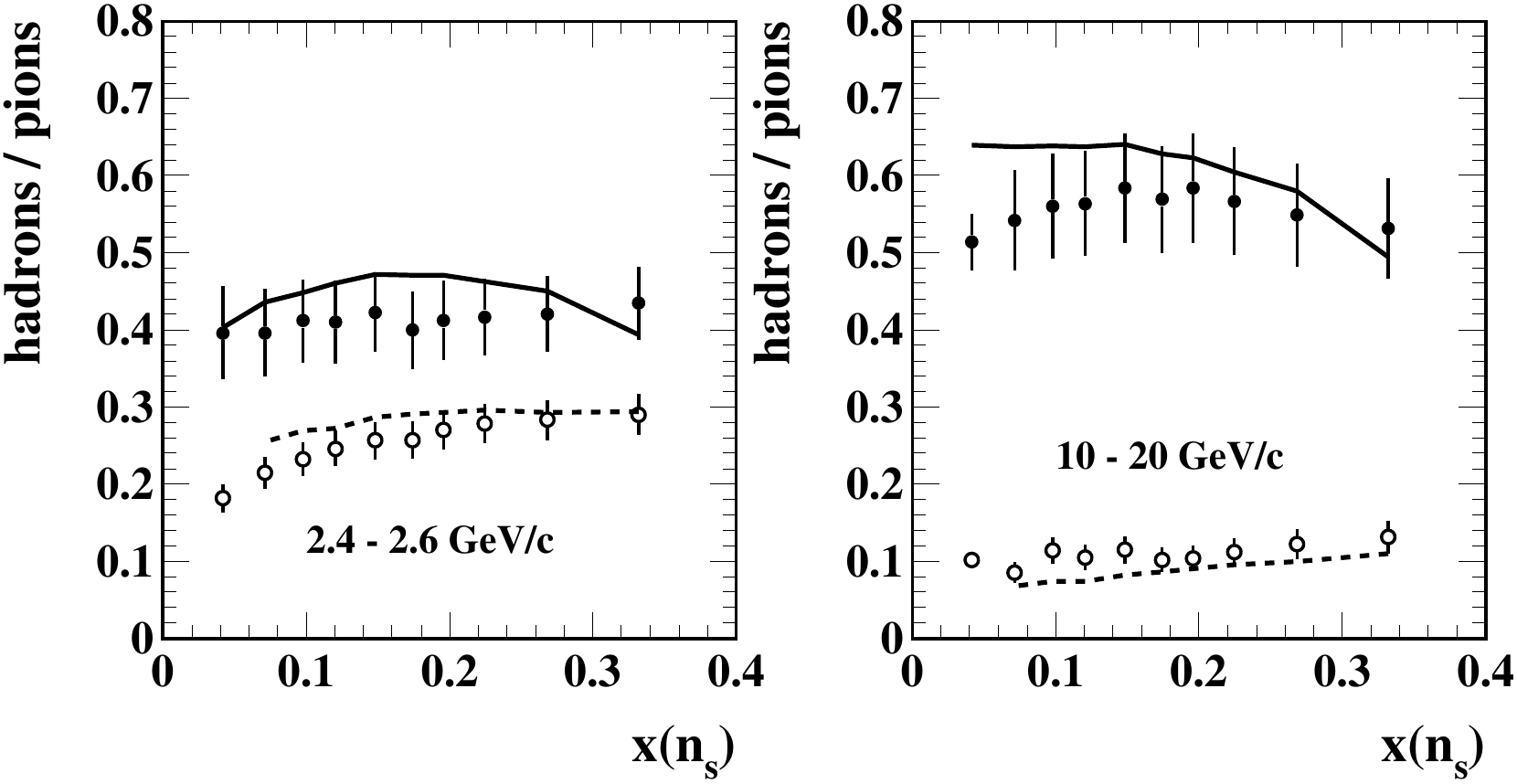}
	\caption{\label{3cbad}
	Same as Fig.~\ref{intrat} (c) and (d) except TCM pion and proton spectra are modified as in Fig.~\ref{3ebad} caption to approximate published PID spectrum data from Ref.~\cite{alicepppid}. The TCM curves are not rescaled.
	} 
\end{figure}

Figure~\ref{3cbad} shows results corresponding to panels (c) and (d) of Fig.~\ref{intrat}, but, because the TCM is in this case altered to describe uncorrected pion and proton spectra as described above, no reduction factors are required for the TCM curves. The ratio data from Ref.~\cite{alicepppid} are then described within their systematic uncertainties.

\begin{figure}[h]
	\includegraphics[width=3.3in]{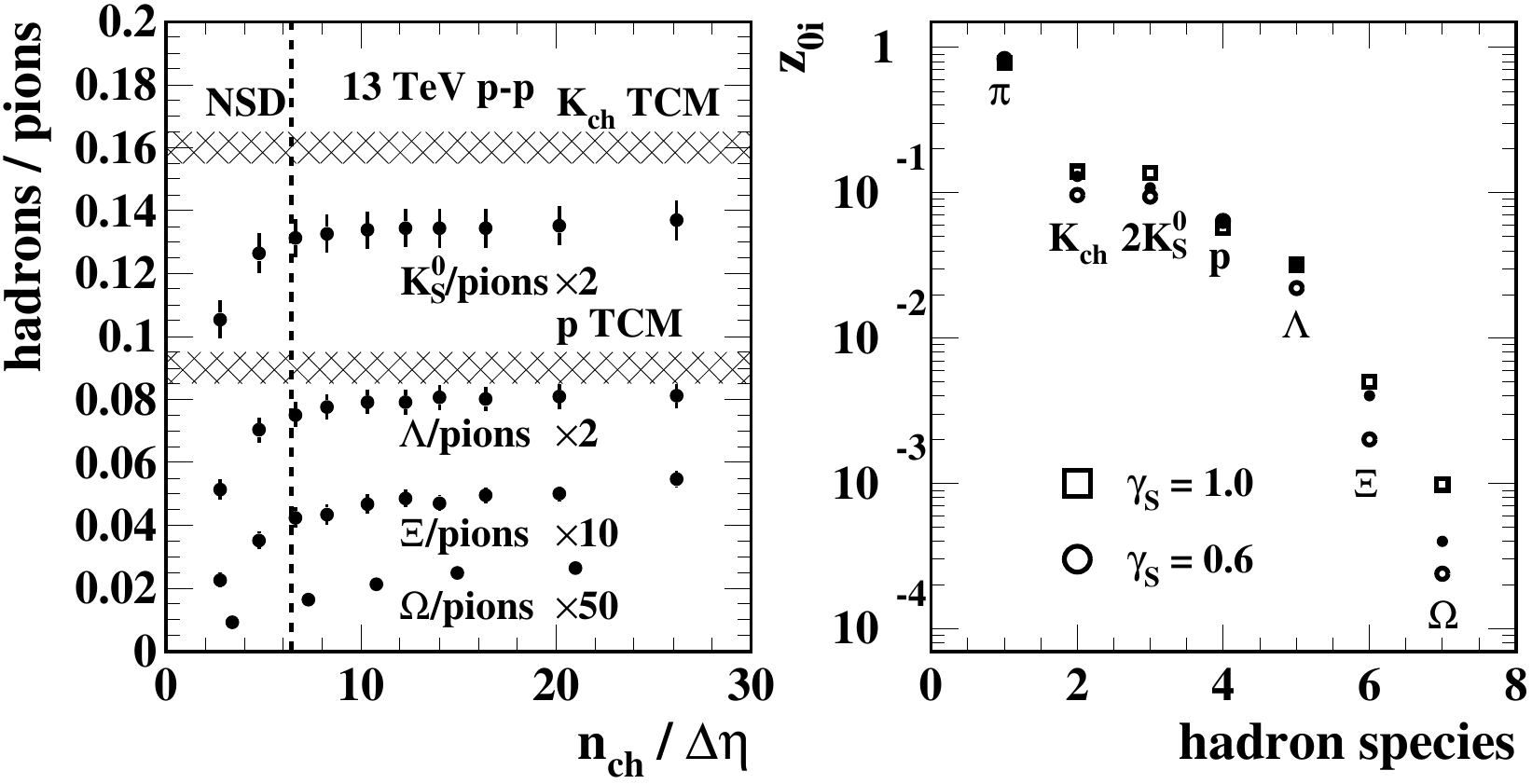}
	\caption{\label{strange}
		Left: Ratios of integrated yields (points) for strange-hadrons-to-pions from 13 TeV \pp\ collisions  compared to TCM estimates for kaons/pions and protons/pions (hatched) based on the present study. That panel corresponds to Fig.~6 of Ref.~\cite{alicepppid}.  The vertical dotted line indicates $\bar \rho_0$ for 13 TeV NSD \pp\ collisions.
		Right: Values of TCM parameter $z_{0i}$ (solid dots) inferred directly from 13 TeV \pp\ PID spectrum data in Sec.~\ref{pppidtcm} Table~\ref{otherparamsxx} or from PID ratio data (e.g.\ left panel) assuming $z_{0i} = 0.80$ for pions. The open symbols are from Ref.~\cite{statmodel}, a statistical model corresponding to the grand-canonical limit with $T = 170$ MeV and strangeness-suppression parameter $\gamma_\text{S} = 0.6$ (circles) and 1.0 (squares).
	} 
\end{figure}

Figure~\ref{strange} (left) shows integrated-yield data ratios (points) from 13 TeV \pp\ collisions for several species of strange hadrons in ratio to pions as presented in Fig.~6 of Ref.~\cite{alicepppid}. Although the increasing trends on charge density $\bar \rho_0 = n_{ch} / \Delta \eta$ have been described as reflecting strangeness ``enhancement'' the saturation values for {\em larger} event multiplicities are consistent with statistical-model predictions~\cite{statmodel} as demonstrated in the right panel.

Figure~\ref{strange} (right) shows a comparison between ratio data in Figs.~\ref{intrat} (a) and \ref{strange} (left) and statistical-model estimates from Ref.~\cite{statmodel}, the latter reporting {\em predictions} of hadron abundances from \pp\ collisions for LHC energies (10 TeV) prior to the start of LHC operations. Its Table II (employed here) corresponds to the grand-canonical limiting case. Estimates relating to strangeness suppression parameter $\gamma_\text{S} = 0.6$ and 1.0 are provided. The quantity reported is equivalent to $\bar \rho_{0i} / \bar \rho_0 = z_{0i}$ in the present notation. That parameter could be evaluated independently for each hadron species according to the procedure in Sec.~\ref{zxidirect} by extrapolating measured quantity $z_{si}(n_s)$ to zero \nch\ (no jet contribution) to determine each $z_{0i}$. But since integrated-spectrum ratios in Figs.~\ref{intrat} (a) and \ref{strange} (left) are equivalent to $z_{0i}/z_{0j}$ as in Eq.~(\ref{pidyieldrat}), and the denominators are pion $z_{0j} \approx 0.80$ reported in Table~\ref{otherparamsxx}, the products produce the solid dots in Fig.~\ref{strange} (right). The statistical-model predictions are denoted by open boxes for $\gamma_\text{S} = 1.0$ and open circles for $\gamma_\text{S} = 0.6$. The PID data all fall within those limits.

As perspective for such a comparison, quantity $z_{0i}$ with $n_{ch} \rightarrow 0$ corresponds to a TCM soft component alone with no jet contribution. Each hadron species in that limit {\em requires a unique soft-component description} with slope parameters $T = 145$ MeV for pions, 200 MeV for kaons and 210 MeV for protons as in the present study. The statistical-model predictions of Ref.~\cite{statmodel} are based on a common temperature  $T = 170$ MeV for all hadrons. Attempting to enforce a single temperature value on the TCM would result in rejection of the model per standard statistical measures (e.g.\ Z-scores). Permitting separate $T$ assignments within the TCM leads to data described within statistical uncertainties (Sec.~\ref{quality}). The statistical model responds to a small subset (spectrum integrals) of  information carried by particle data. The full complement of spectrum and two-particle correlation data imposes strong constraints on models, falsifying some and forcing others to converge on what may be a proper physical description of high-energy nuclear collisions. 

\subsection{Competing interpretations for PID ratios}

Reference~\cite{alicepppid} presents several conjectures concerning PID yields and their ratios. The general themes relate to (a) hydrodynamic models and collective flow, (b) universal scaling of hadrochemistry with charge density $\bar \rho_0$ and (c) collision-energy dependence of hadron production. 

(a) Particle ratio trends appear similar to those previously described by hydrodynamical models: ``The $p_T$-differential particle [spectrum] ratios [i.e.\ its Fig.~2] exhibit an evolution with multiplicity, similar to that observed in pp collisions at $\sqrt{s} = 7$ TeV, which is {\em qualitatively} described by some of the hydrodynamical and pQCD-inspired models... [emphasis added].'' In a study of PID spectra from 5 TeV \ppb\ collisions Ref.~\cite{aliceppbpid} interpreted {\em differential} spectrum ratios (e.g.\ Fig.~\ref{specrat} of the present study) in the context of \pbpb\ spectrum ratios (right panels in Fig.~2 of Ref.~\cite{aliceppbpid}) as follows: Arguing by analogy, there is ``significant enhancement [of baryon/meson ratios] at intermediate $p_T \sim 3$ GeV/c, {\em qualitatively reminiscent} of that measured in \pbpb\ collisions. The latter are {\em generally discussed in terms of collective flow or quark recombination} [emphasis added].'' 

(b) It is asserted that PID ratio trends such as those in Figs.~5 and 6 of Ref.~\cite{alicepppid} suggest that hadrochemistry exhibits a universal scaling with charge density. The abstract states ``This [PID ratio trends]...extends to strange and multistrange hadrons, suggesting that, at LHC energies, {\em particle hadrochemistry scales with particle multiplicity} [i.e.\ $\bar \rho_0$] {\em the same way under different collision energies and colliding systems} [emphasis added].''

(c) Concerning collision-energy dependence the abstract promises to ``...isolate the center-of-mass energy dependence of light-flavour particle production.'' Presumably that relates particularly to PID ratio data from a range of collision energies in Figs.~5 and 6 of Ref.~\cite{alicepppid}. And referring to Fig.~2 ``...the $p_T$-integrated [over small \pt\ intervals] hadron-to-pion yield ratios measured in pp collisions at two different center-of-mass energies are consistent when compared at similar multiplicities.''  The PID TCM context reported here illuminates those points. 

(a) TCM analysis of differential PID \pt\  spectra as reported in Refs.~\cite{ppbpid} and \cite{pidpart2} clearly demonstrates that the prominent peak near \mbox{3} GeV/c ($y_t \approx 3.75$) in {\em baryon}/meson ratios is associated with {\em jet-related} spectrum hard components and results from relative shifts of peak modes for baryons vs mesons, relating then to quantitative differences in jet  formation. No such behavior is observed for {\em meson}/meson ratios because of the hard-component trends demonstrated in Sec.~\ref{tcmfinal} of the present study and in Sec.~VI of  Ref.~\cite{pidpart1}. Prominent features of PID spectra and their various ratios are thus {\em quantitatively} explained in terms of measured jet properties.

(b) The ``universal scaling'' observation relates to Eq.~(\ref{pidyieldrat}) of the present study: Ratios of {\em fully-integrated} PID spectra have the simple limiting value $z_{0i} / z_{0j}$. Such ratios are predicted by statistical models as in Ref.~\cite{statmodel}, and see Fig.~\ref{strange} (right). The close correspondence between statistical models and PID ratio data would not be apparent from Fig.~5 of Ref.~\cite{alicepppid} because of data biases therein of tens of percent (e.g.\ see Sec.~\ref{correct} and Figs.~\ref{3ebad} and \ref{3cbad}). What is missed from such integrals and their simple ratios is the complex interplay between jet and nonjet spectrum components of PID spectra as illustrated in Sec.~\ref{zxidirect} and Ref.~\cite{pidpart1}. 

(c) As to energy dependence of ratios, Ref.~\cite{statmodel} states (in 2009) that ``...at LHC the [yield] ratios between different [hadron] species become essentially energy-independent...'' which is {\em qualitatively} consistent with Figs.~5 and 6 of Ref.~\cite{alicepppid} within substantial systematic uncertainties. But the proper {\em quantitative} connection requires correction of substantial biases as in Secs.~\ref{correct} and \ref{fracsumm} combined with Eq.~(\ref{pidyieldrat}) of the present study.

\section{BW model fits to $\bf p$-$\bf p$ PID spectra} \label{bwave}

The BW model for high-energy nuclear collisions as described for instance in Ref.~\cite{blastwave}  (for S-S collisions at $\sqrt{s_{NN}} \approx 19$ GeV), and as applied in Ref.~\cite{alicepppid} to data from \pp\ collisions at $\sqrt{s} \approx 13$ TeV, assumes that hadron emission from a particle source as observed in a comoving frame should be isotropic and follow a Boltzmann exponential on hadron energy. Any deviation from those trends of a particle distribution on energy (e.g.\ \mt) and angle as measured in an A-B collision center-of-momentum (CM) frame is then interpreted to reveal particle source motion (i.e.\ a velocity field) attributed to a flowing collision system~\cite{cooper}. Application of the BW model as fits to \mt\ spectra from several hadron species is seen as both validating the assumption of ``collectivity'' (i.e.\ a source velocity field) and as determining the properties (state of motion) of the flowing particle source. A critique of such assumptions is reported in Ref.~\cite{ppbbw}.

\subsection{Argument for BW relevance to small systems}

An argument in Ref.~\cite{alicepppid} in support of the BW model applied to \pp\ spectrum data proceeds as follows: ``In large collision systems such as Pb-Pb, multiplicity-dependent modifications of hadron $p_T$ spectra [any changes?] {\em can be interpreted as the hydrodynamical radial expansion of the system} [emphasis added] and studied in the context of the Boltzmann-Gibbs Blast-Wave model. ... As the trends observed in the evolution of particle spectra measured in pp collisions are highly reminiscent to those in \ppb\ and \pbpb, it is interesting to check whether the Blast-Wave model can be extended to describe pp collisions. ... Now for the first time, we can study the evolution of $\langle \beta_T \rangle$, $T_{kin}$ and $n$ in pp collisions as a function of the collision energy.'' The phrase ``highly reminiscent'' is similar to usage above connected with baryon/meson ratios ``...{\em qualitatively reminiscent} of that measured in \pbpb\ collisions.'' There is no reference to specifics and no quantification. Emphasis is therefore inserted above to make clear a fundamental assumption relating to the BW model: {\em Any} variation of spectrum shapes with \nch\ or A-B centrality may be interpreted to indicate hydrodynamic flows in {\em any} collision system.

\subsection{Interpretation of BW model fit results}

Table~\ref{flowparams} presents BW model fit parameters from Ref.~\cite{alicepppid} for \pt\ spectra from 13 TeV \pp\ collisions. Several physical inferences are presented on the basis of those results, referring in part to Figs.~3 and 4 of that paper. Points for later comment are denoted by letters [X].

\begin{table}[h]
	\caption{Blast-wave parameters for simultaneous fits of pion, charged-kaon and proton spectra from 13 TeV \pp\ collisions~\cite{alicepppid}.
	}	\label{flowparams}
	\begin{center}
		\begin{tabular}{|c|c|c|c|c|} \hline
			$n$ &   $n_{ch} / \Delta \eta$ &   $\langle \beta_T \rangle$     & $T_{kin}$ (GeV)  & $n$   \\ \hline
			1	   &   26   &  0.49  & 0.163   &  1.5  \\ \hline
			2	 &  20  &  0.445   &  0.174  & 1.7   \\ \hline
			3	 & 16.2  & 0.41 & 0.179  &   2.0   \\ \hline
			4	 & 13.8 &  0.38 &  0.182  &   2.3    \\ \hline
			5	 &  12  &  0.36  &  0.182  &   2.5  \\ \hline
			6	 &  10  & 0.325  &   0.184 &  2.9   \\ \hline
			7	 & 8    & 0.29  & 0.184  &  3.5   \\ \hline
			8	 & 6.3  & 0.25  & 0.184  &   4.2   \\ \hline
			9	 & 4.5  & 0.20  &  0.181 &   5.7   \\ \hline
			10	 & 2.55  & 0.12  & 0.174  &  11.6   \\ \hline
		\end{tabular}
	\end{center}
\end{table}

``At larger multiplicities...$T_{kin}$ decreases [with nch] and becomes similar to that measured in \ppb\ collisions [at 5 TeV], [A] suggesting that the system decouples [meaning?] at lower temperature and thus is longer-lived. ...
The average expansion velocity $\langle \beta_t \rangle$ increases with $\langle dN_{ch}/d\eta \rangle$ and its values are consistent for pp collisions at different $\sqrt{s}$ as well as with the corresponding values for p-Pb collisions, [B] indicating that small systems become more explosive [meaning?] at larger multiplicities. In contrast to this, $\langle \beta_t \rangle$ measured in \pbpb\ collisions is lower than that in smaller systems for the common [same] $\langle dN_{ch}/d\eta \rangle$ range.... [C] This indicates that the size of the colliding system might have significant effects [meaning?] on the final state particle dynamics.'' Referring to BW parameter $n$ [D] ``...in pp and p-Pb collisions, large $n$ suggests high pressure gradients which lead to larger $\langle \beta_t \rangle$, while in \pbpb\ collisions, $n \sim 1$ could be interpreted as lower pressure gradient and thus smaller expansion velocity.'' Labeled points [X] are discussed in Sec.~\ref{bwconclude}.

\subsection{BW data description quality: Z-scores}

Just as for TCM fit quality in Sec.~\ref{quality} this subsection examines BW model fit quality for the same \pp\ PID spectrum data. Whereas the TCM is required to describe all available spectrum data and {\em is not  fitted to individual spectra}, it is common practice to limit BW model fits to restricted \pt\ intervals based on several arguments. For instance, as a result of resonance decays contributing to pion spectra ``...one can choose to omit the low-$p_T$ pions.'' It is acknowledged that since ``...there is a strong dependence of Bast-Wave parameters on the fitting range, it is important to consider the same $p_T$ range...in order to obtain a consistent comparison between different colliding systems.'' But one may well ask how can parameter values be interpreted physically for arbitrarily defined fit intervals? In Ref.~\cite{aliceppbpid} the fitting intervals are explicitly defined by ``the available data at low $p_T$ and {\em based on the agreement with the data at high $p_T$} [emphasis added].'' The same intervals are used in Ref.~\cite{alicepppid}: 0.5 - 1 GeV/c, 0.2 - 1.5 GeV/c and 0.3 - 3 GeV/c respectively for pions, charged kaons and protons, with corresponding \yt\ intervals [2,2.7], [1.2,3.1] and [1.5,3.75] denoted by arrows.

\begin{figure}[h]
	\includegraphics[width=3.3in]{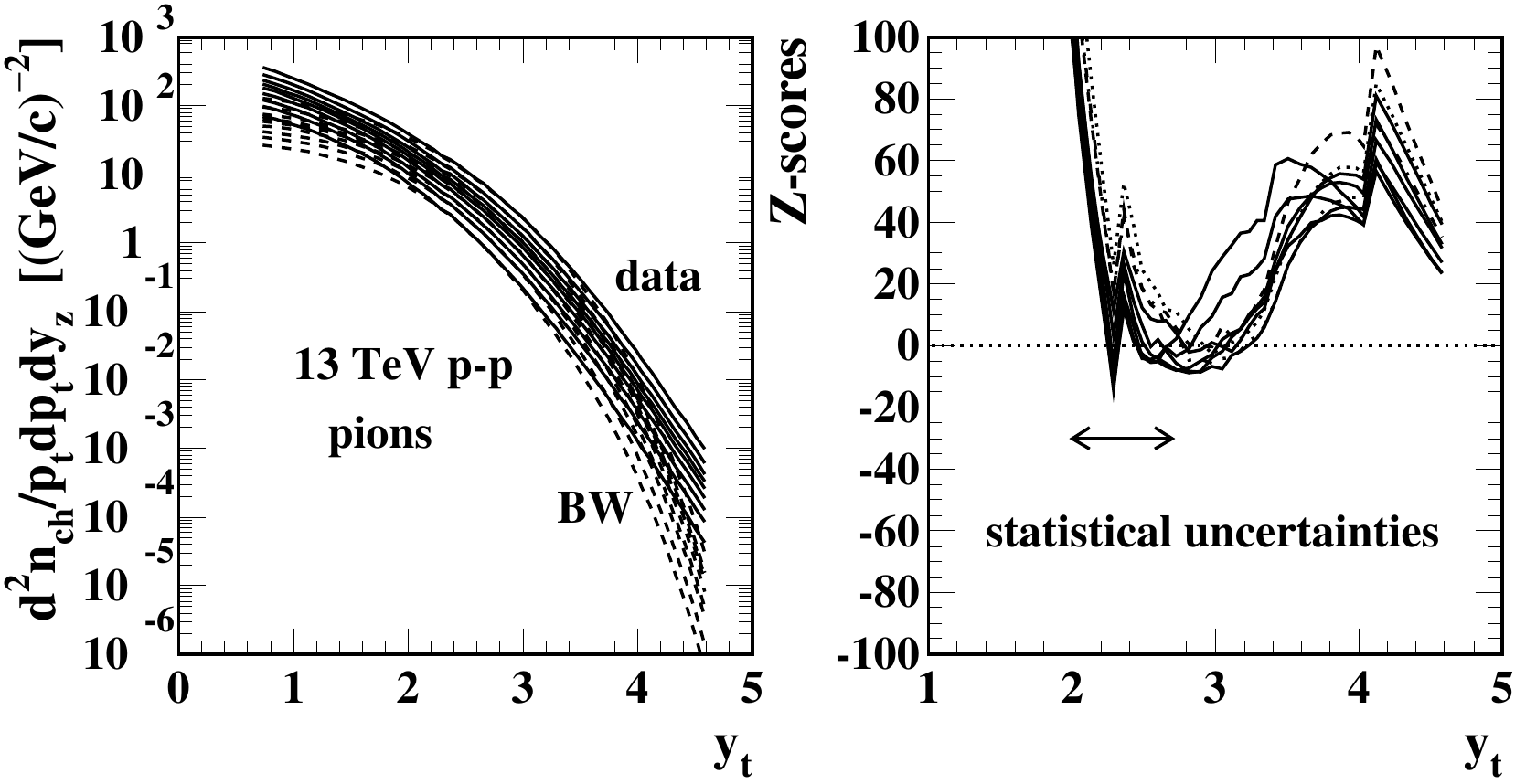}
	\caption{\label{9a}
		Left: Identified pion spectra for 13 TeV \pp\ collisions (solid) and corresponding BW model fits (dashed) from Ref.~\cite{alicepppid}.
		Right: Z-scores (based on statistical uncertainties) for data and model in the left panel following the procedure in Sec.~\ref{modelacc}. The arrow indicates the imposed fit interval.
	} 
\end{figure}

\begin{figure}[h]
	\includegraphics[width=3.3in]{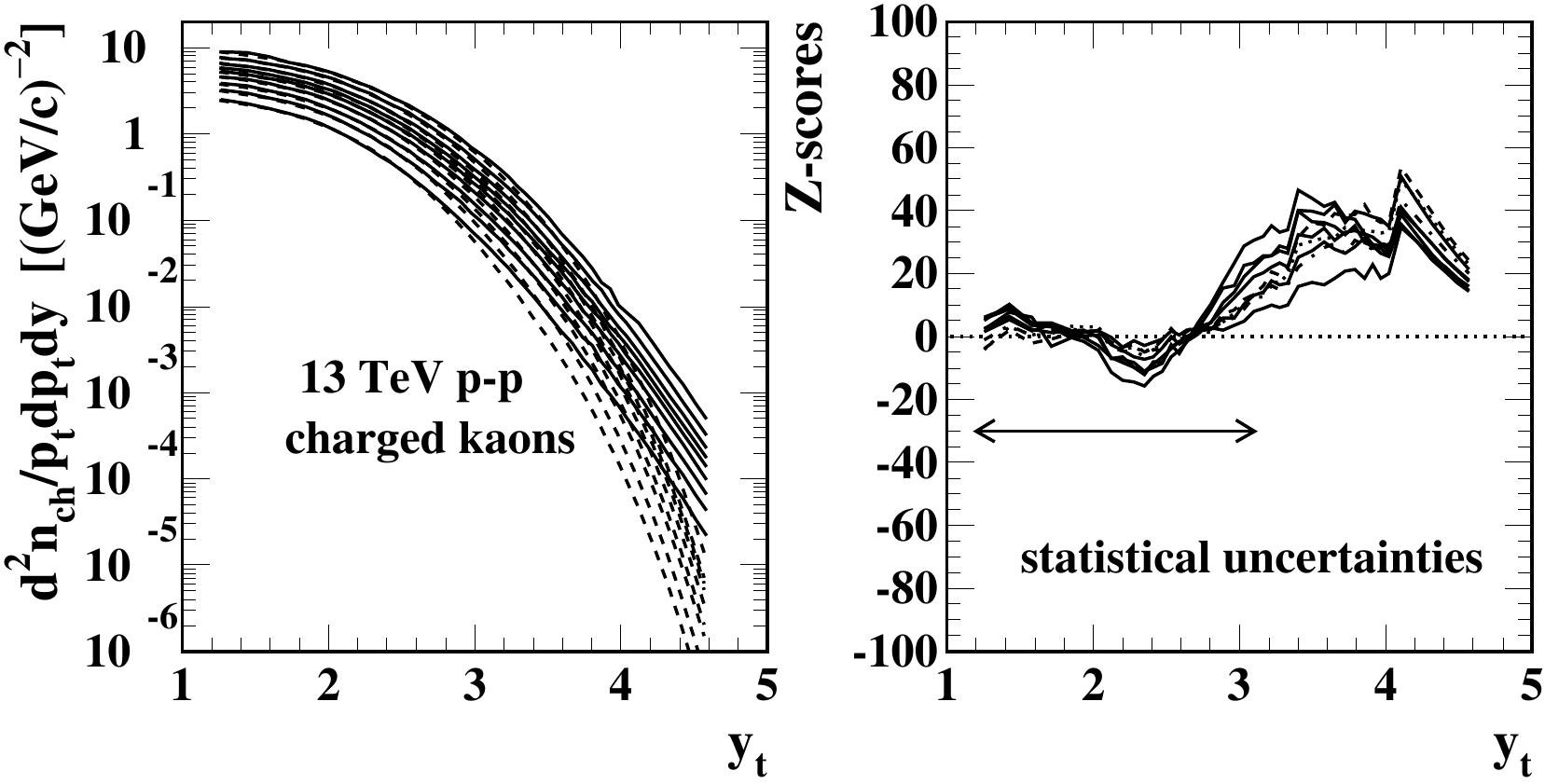}
	\caption{\label{9b}
Same as Fig.~\ref{9a} except for charged kaons.
	} 
\end{figure}

\begin{figure}[h]
	\includegraphics[width=3.3in]{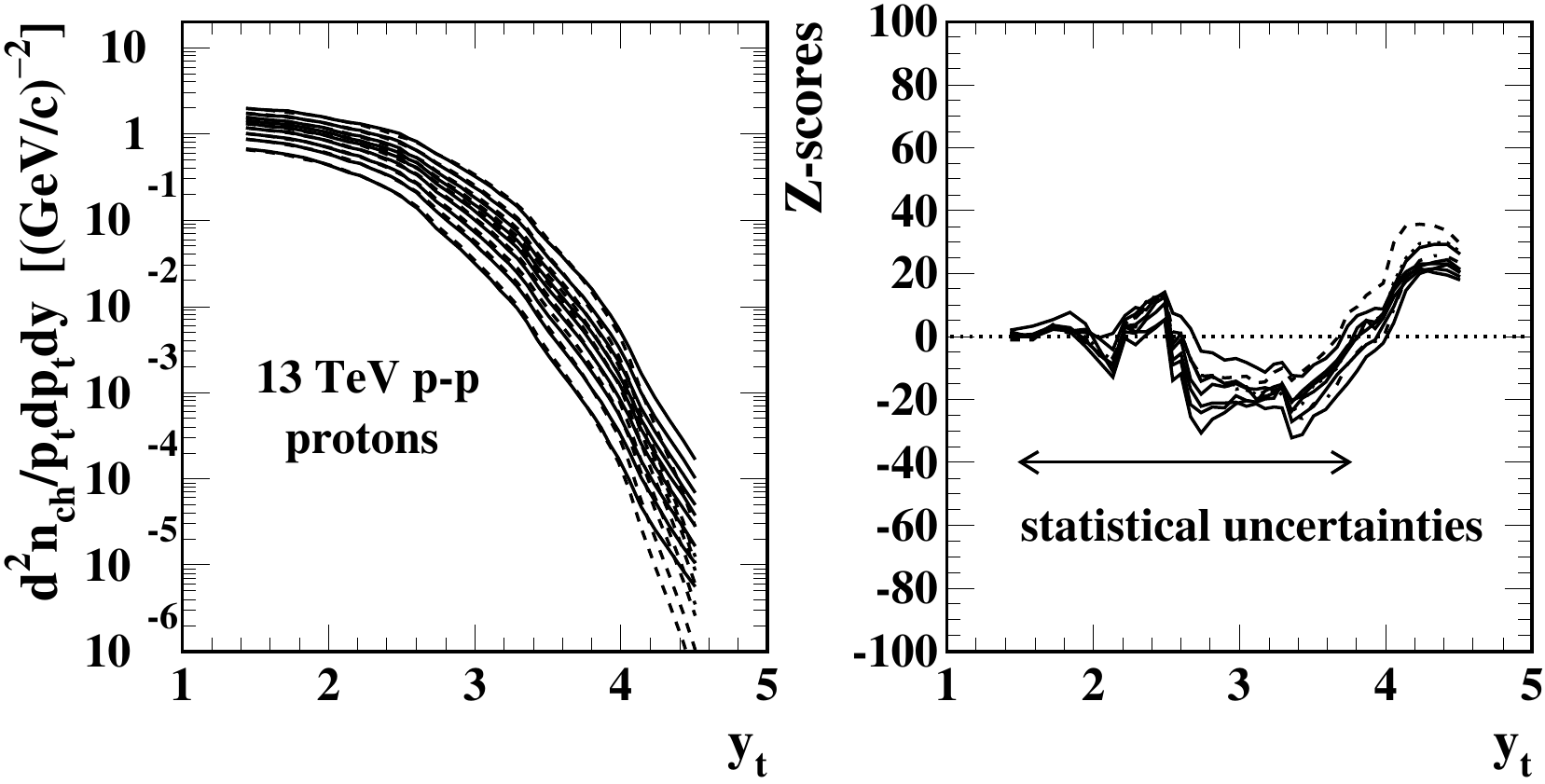}
	\caption{\label{9c}
Same as Fig.~\ref{9a} except for {\em uncorrected} protons.
	} 
\end{figure}

In Figs.~\ref{9a} through \ref{9c} below the left panels include data spectra (solid) extending from the lower bounds of data acceptances to a \pt\ value ($\approx 6.5$ GeV/c or $y_t \approx 4.5$) where calculation of the BW model becomes unstable. The BW model itself (dashed), using parameter values from Table~\ref{flowparams}, is then defined on data \pt\ values over the same interval for direct comparison.

The right panels present Z-scores corresponding to BW model fits, the line styles varying with descending event multiplicity as solid, dashed, dotted and dash-dotted, with solid thereafter as for other figures in this study. Z-scores are as defined in Sec.~\ref{modelacc}. Recall that Z-score magnitudes $\approx 1$ indicate an acceptable model fit. In each case the Z-scores falsify the BW model for these data.

In Fig.~\ref{9c} the BW model is compared to {\em uncorrected} (i.e.\ published) \pp\ proton spectra which presumably were the subject of the actual model fits in Ref.~\cite{alicepppid}. It is then interesting to compare those fit results with the corrected proton spectra as reported in Sec.~\ref{correct}.

\begin{figure}[h]
	\includegraphics[width=3.3in]{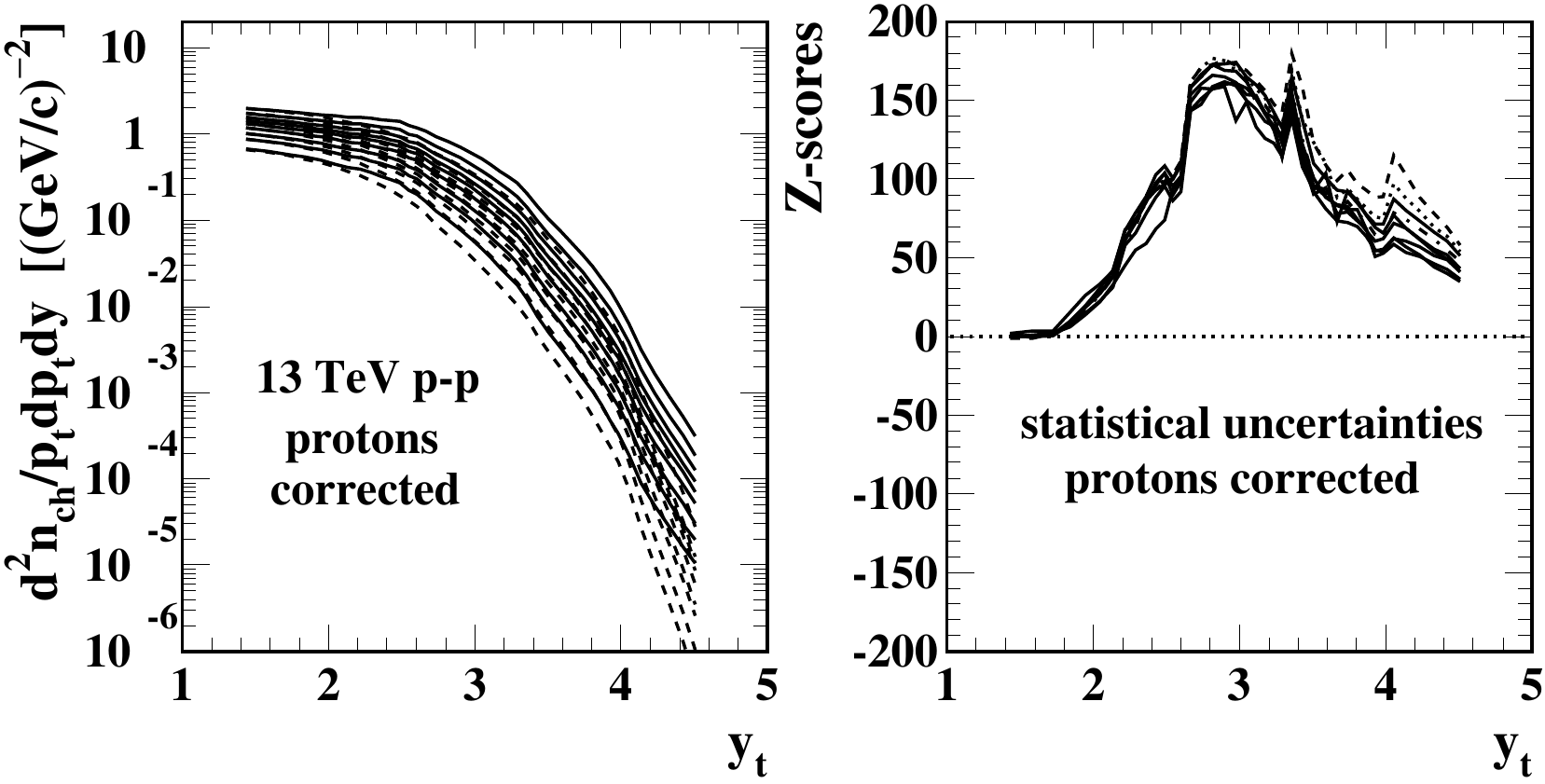}
	\caption{\label{9ccorr}
Same as Fig.~\ref{9c} except for {\em corrected} protons.
	} 
\end{figure}

Figure~\ref{9ccorr} shows the same proton BW fits now compared to corrected proton spectra. The Z-scores as expected demonstrate much greater disagreement. The basis for the proton correction is reported both in Sec.~\ref{correct} (for \pp\ collisions) and in Sec.~III A of Ref.~\cite{pidpart1} (for \ppb\ collisions). In the latter study the agreement of corrected proton spectra with Lambda spectra (with similar hadron mass) is notable. If the corrected proton spectra were included in the BW model fits one should expect major changes in fitted parameter values, introducing even more uncertainty as to their interpretation.

\subsection{Questioning the significance of BW model fits} \label{bwconclude}

Usually omitted from the conventionally expressed BW model context is the contribution of minimum-bias jets to hadron production and therefore to single-particle spectra and two-particle correlations. The TCM as inferred {\em empirically} from \pp\ spectrum data~\cite{ppprd} includes a hard (jet) component that is quantitatively compatible with measured jet properties~\cite{fragevo,jetspec2,mbdijets}. Jet contributions to hadron production violate basic assumptions for the BW model in that (a) jet fragment spectra deviate strongly from a Boltzmann exponential on \mt, and (b)  hadrons related to an energetic parton (i.e.\ a correlated jet) deviate strongly from an isotropic angular distribution. 

Responses to items [X] follow: Fitted parameter $T_{kin}$ decreases with increasing \nch\ [A] ``...suggesting that the [collision] system decouples at lower temperature and thus is longer-lived.'' Or, the one-component (monolithic) BW model is attempting to describe a competition between two hadron production mechanisms (jet and nonjet) each of which is quite simple. The TCM description of spectrum data reflects the {\em observed} property that the nonjet (soft) component for each hadron species is determined by a {\em fixed} slope parameter $T \approx 145,~200$ and 210 MeV for pions, kaons and baryons respectively. The decrease of $T_{kin}$ with \nch\ is the BW model response to {\em increased} jet production relative to the soft component.

The inferred expansion velocity $\langle \beta_t \rangle$ [for \pp\ collisions] increases with charge multiplicity \nch\ [B] ``...indicating that small systems become more explosive at larger multiplicities.'' On the other hand, {\em observed} dijet production in \pp\ collisions increases {\em quadratically} with the nonjet charge density~\cite{ppprd} leading to dramatic increase of \pp\ spectrum hard components and related jet manifestations in angular correlations~\cite{ppquad}. The sharply increased jet production requires BW-model $\langle \beta_t \rangle$ to increase substantially in its failed attempt to accommodate jets.

It seems ironic that larger collision systems (with nominally greater densities) exhibit smaller $\langle \beta_t \rangle$ values. In the BW context it is concluded that [C] ``...the size of the colliding system might have significant effects on the final state particle dynamics.'' On the other hand, jet production is {\em measured} to increase quadratically with mean \nn\ charge multiplicity~\cite{ppprd}. For given {\em total} \nch\ the mean \nn\ multiplicity is greatest for individual \pp\ collisions and least for \aa\ collisions where total \nch\ is supplemented by large increase in participant nucleon number $N_{part}$. \aa\ collisions then produce the lowest dijet numbers {\em per \nn\ binary collision}, explaining  the  $\langle \beta_t \rangle$ trend.

A related argument includes [D] ``...in pp and p-Pb collisions, large [BW profile parameter] $n$ suggests high pressure gradients which lead to larger $\langle \beta_t \rangle$, while in \pbpb\ collisions, $n \sim 1$ could be interpreted as lower pressure gradient and thus smaller expansion velocity [$\langle \beta_t \rangle$].'' That  is also ironic in that $\langle \beta_t \rangle$ and $n$ are {\em anti}\,correlated in Table~\ref{flowparams}. In a context where jet production is properly acknowledged, BW $\langle \beta_t \rangle$ trends are  explained in terms of an inappropriate spectrum model responding to jet contributions to spectra that are, as noted, {\em quantitatively predicted by measured jet properties}~\cite{hardspec,fragevo,jetspec2,mbdijets}. 
If analysis techniques (e.g.\ BW model fits to spectra) do not distinguish jet from nonjet contributions to data features the resulting inferences may have no scientific relevance.

\section{Systematic uncertainties} \label{sys}

PID TCM spectrum parameter estimation involves two tasks: (a) refine soft- and hard-component model parameters based on predictions developed in Sec.~\ref{spectrumtcm} and (b) estimate parameters $z_{si}(n_s)$ and $z_{hi}(n_s)$ for 13 TeV \pp\ collisions based on direct analysis of PID spectra. The second task requires evaluation and possible correction of systematic biases, especially possible cross talk between pions and protons identified via $dE/dx$ analysis~\cite{pidpart1}.

\subsection{TCM model functions $\bf \hat S_{0i}(y_t)$ and $\bf \hat H_{0i}(y_t)$} \label{modelfunc}

Assuming a {\em fixed} model, the TCM requires parameter $\alpha$ in $\bar \rho_h \approx \alpha \bar \rho_s^2$ to obtain densities $\bar \rho_s$ and $\bar \rho_h$ from measured charge density $\bar \rho_0$ and five model-function parameter values (two soft + three hard) for each hadron species. As described in the introduction to Sec.~\ref{spectrumtcm} the strategy for this analysis has been to combine PID model parameters previously determined for 5 TeV \ppb\ collisions with the energy dependence of a nonPID TCM for \pp\ collisions to predict PID parameter values for 13 TeV \pp\ collisions. Resulting predictions are presented in Table~\ref{engparamsxx} which can be compared with final values in Table~\ref{engparamsy} obtained by optimizing the TCM description quality in the present study as in Sec.~\ref{quality}. Predicted and final values are consistent within data uncertainties. However,  the pion values are subject to additional uncertainty because of the biases evident in Fig.~\ref{pionsx} (right). Generally speaking, a consistent and accurate picture of TCM model variations across small collision systems and energies has emerged from the present study and Refs.~\cite{ppprd,ppbpid,pidpart1,pidpart2}.

\subsection{PID species fractions $\bf z_{si}(n_s)$ and $\bf z_{hi}(n_s)$}

A systematic issue for fractional abundances of some hadron species is already apparent in the analysis of 5 TeV \ppb\ collisions reported in Ref.~\cite{ppbpid}. In its Sec.~6 pions exceed the TCM expectation (by about 40\%) in Fig.~4 (right) while protons are strongly suppressed (again by about 40\%) in Fig.~6 (a). Three other species agree with TCM predictions within statistical uncertainties. Reference~\cite{pidpart1} presents a much more detailed TCM analysis of \ppb\ PID spectra in which a method is devised for correcting the proton data (Sec.~III B). However,  the pion excess of Ref.~\cite{ppbpid} is not acknowledged or addressed. It is clear in Fig.~9 (right) of Ref.~\cite{pidpart1} that the sum of $z_{hi}(n_s)$ values (triangles) inferred from charged-hadron spectrum data, including {\em corrected} proton spectra, violate charge conservation (expected sum rule) by about 17\%. Analysis details explain the apparent disconnect.

In Ref.~\cite{ppbpid} quantities $z_{si}(n_s)$ and $z_{hi}(n_s)$ were generated by parameter values $z_{0i}$ and $\tilde z_i = z_h / z_s$ (assumed constant) in its Table~4. As noted in that text ``...[$\tilde z_i$] is first adjusted to achieve  coincidence of all seven normalized spectra [i.e.\ in the form $X_i(y_t)$] as $y_t \rightarrow 0$. Parameter $z_{0i}$ is then adjusted to match those rescaled spectra to unit-normal $\hat S_0(y_t)$....'' The result was a true prediction of $z_{hi}(n_s)$ values based on low-\yt\ spectrum trends. Given that prediction the biased pion hard components in its Fig.~4 (right) were revealed. In Ref.~\cite{pidpart1} a more precise method was adopted in which $z_{si}(n_s)$ and $z_{hi}(n_s)$ are separately inferred from spectra as in Sec.~\ref{zxidirect} of the present study. But that method is predicated on unbiased spectrum data (which for {\em corrected} protons and kaons is the case). Therefore, the pion bias revealed in Ref.~\cite{ppbpid} was overlooked. The pion $\tilde z_i = 0.8 \pm 0.05$ value reported in Ref.~\cite{ppbpid} was based on $\alpha \approx 0.0113$. The values inferred in Ref.~\cite{pidpart1} are based on updated estimate $\alpha \approx 0.013$ for 5 TeV. A direct comparison of the $\tilde z_i$ requires rescaling $\tilde z_i = 0.8 \times 0.0113 / 0.013 \rightarrow 0.70 \pm 0.05$ that can be compared with $\tilde z_i = 0.60 \pm 0.05$ in Table~\ref{otherparamsxx} of the present study based on charge conservation and $\tilde z_i = 0.88 \pm 0.05$ from Ref.~\cite{pidpart1} arising from the pion bias.

The present study benefits in a way from systematic distortions of pion spectra, forcing recognition that the uncertainty of inferring pion $\tilde z_i$ values, and therefore $z_{hi}$ values, from pion low-\yt\ spectrum structure was underestimated. The only reliable way to estimate pion $z_{hi}(n_s)$ values is by enforcing charge conservation among $z_{hi}$  values relative to other charged-hadron species with more-accurate $z_{hi}$ estimates. Given that revised context the $z_{xi}$ values for 13 TeV \pp\ collisions are consistent with those for 5 TeV \ppb\ collisions within data uncertainties.

\subsection{PID yield and spectrum ratios}

The biases in PID analysis via $dE/dx$ techniques presented in Refs.~\cite{aliceppbpid,alicepppid}, as revealed in Refs.~\cite{ppbpid,pidpart1} and the present study, strongly impact PID yield and spectrum ratios and inferences derived therefrom. There are two issues: (a) Even for unbiased data such ratios represent substantial discard of valuable information. Contrast the amount of information conveyed by figures in Sec.~\ref{tcmfinal} with what is available from spectrum ratios as in Fig.~\ref{specrat}. The former lead to detailed and substantive physical interpretations while the latter may form a basis for unsupported speculation. (b) Substantial biases lead to systematic errors of tens of percent, as illustrated by comparing Fig.~\ref{specrat} with Fig.~\ref{3ebad} above. How meaningful then are data-Monte Carlo comparisons such as in Fig.~2 and 5 of Ref.~\cite{alicepppid}? In its Fig.~2 PYTHIA-based Monte Carlo results (curves) near 0.23 (without factor 1.5) are close to the TCM result $\approx 0.30$ in Fig.~\ref{intrat} (d) (upper dashed curve) for {\em corrected} pion and proton data from the present study, whereas the uncorrected data ratios (points) fall near 0.10. In Fig.~5 of Ref.~\cite{alicepppid} the PYTHIA Monte Carlo returns an integrated-yield $p/\pi$ ratio $\approx 0.08$ compared to uncorrected-data ratio $\approx 0.055$ and TCM corrected-data ratio $\approx 0.09$ appearing in Fig.~\ref{intrat} (a).

\section{Discussion} \label{disc}

Several issues relating to PID spectra from 13 TeV \pp\ collisions have been addressed in the text above. In this section the predictivity of the TCM is summarized, and the centrally-important issue, claims of ``collectivity'' (flows) in small collision systems, is confronted via a summary of evidence for and against such claims.

\subsection{The TCM as a predictive data model}

In contrast to a number of approaches to spectrum description (e.g.\ BW model, Tsallis model~\cite{tommodeltests}) the TCM is a {\em predictive} model with elements that may be compared directly and quantitatively with jet measurements and fundamental QCD theory. The TCM is not fitted to individual spectra, which would require many parameter values. Consistency of fixed-TCM model parameters across collision systems and energies (e.g.\ Table~\ref{engparam}) is summarized in Sec.~\ref{modelfunc}. It is notable that the few supplementary parameters controlling {\em variable}-TCM PID hard-component models (e.g.\ Fig.~\ref{centroids} and see Ref.~\cite{pidpart2}) vary linearly with hard/soft (jet/nonjet) ratio $x\nu$ (\ppb) or $x$ (\pp) within data uncertainties. And the $z_{xi}$ coefficients are predicted by a combination of Eq.~(\ref{zsix}), $z_{0i}$ values predicted by a statistical model~\cite{statmodel} and $\tilde z_i$ values simply proportional to hadron mass~\cite{pidpart1}.

The demonstrated simplicity and accuracy of the TCM has several consequences: (a) TCM predictions and final model parameters confirm that \ppb\ collisions are linear superpositions of \pn\ collisions within data uncertainties. Any viable physical (e.g.\ Monte Carlo) model must satisfy that condition. (b) Systematic PID biases are determined quantitatively and (in the case of protons) successfully corrected. The relation between proton and pion biases is newly determined in the present study. (c) The hadrochemistry of \pp\ and \ppb\ collisions is quantitatively represented by a simple model noted in the previous paragraph. Variation of hadron abundances with \nch\ (\pp) or collision geometry (\ppb) is simply explained. (d) Variation of PID yield and spectrum ratios is described quantitatively in terms of dijet production relative to projectile-nucleon fragmentation (hard vs soft components). Finally, (e) the BW spectrum model is falsified by a conventional statistical measure (Z-scores) in direct comparison with the TCM. Full employment of information carried by particle data leads to interpretation of data systematics that excludes a flow hypothesis.

\subsection{``Collectivity'' and PID spectrum data}

It has become conventional to interpret certain data features in small systems as confirming hydrodynamic flows (collectivity). Reference~\cite{nagle} asserts  that ``...the field of relativistic heavy ion physics is in the midst of a revolution...driven by the experimental observation of flow-like features in the collisions of small hadronic systems,'' its emphasis being on two-particle correlations in small systems (see Sec.~4 of Ref.~\cite{nagle}). But each point has a countervailing response. So-called ridges in 2D angular correlations (peaks at $\phi = 0, \pi$ extending over a broad $\eta$ interval) in \pp\ and \pa~\cite{ppcms,ppbridge} are described as ``...evidence of flow-like collective behavior.'' But analysis of 2D angular correlations from high-statistics 200 GeV \pp\ collisions suggests that such data features result from few-gluon interactions, not collective flow~\cite{ppquad,gluequad}. 
Collectivity is also invoked in connection with so-called mass ordering of differential $v_2(p_t)$ data from small systems. But even in \aa\ collisions PID $v_2(p_t)$ data transformed to proper rapidity variables exhibit a common {\em fixed} boost (what produces ``mass ordering'' on linear \pt) that {\em does not vary with \aa\ centrality} down to peripheral \nn\ collisions, inconsistent with a flow scenario~\cite{quadspec}. For each  collectivity support argument presented in Ref.~\cite{nagle} there are countervailing  responses~\cite{nature,ppbbw,tommodeltests,anomalous,mbdijets,hardspec,nonjetquad,njquad,harmonics,harmonics2}.

For Ref.~\cite{alicepppid} the relevant experimental issue is PID \pt\ spectrum systematics in small vs large collision systems and inference of hydrodynamic flows in small systems based on argument by analogy: \pp\ spectra become ``harder'' with increasing \nch\ and more so for protons (``mass ordering''), but such trends are also observed for \aa\ collisions where they are seen as {\em naturally} or {\em usually}  associated with hydrodynamic flows. So, the same must be true for \pp\ collisions. But as with small-system collectivity arguments based on two-particle correlations, arguments based on PID \pt\ spectra rely on analysis techniques that discard most information carried by data. 

Figure~1 of Ref.~\cite{alicepppid} serves as an example of information discard. Within a conventional flow context {\em any} change in spectrum shape (e.g.\ deviation from a Boltzmann exponential as argued in Ref.~\cite{blastwave}) is interpreted to indicate particle source motion within a flowing medium. Plots on linear \pt\ up to 20 GeV/c permit no significant visual access to the low-\pt\ region near 1 GeV/c where the great majority of jet fragments resides. Presented spectra are dominated visually by a few high-\pt\ bins with poor statistics. Event classes scaled up by factors 2 frustrate direct comparisons. Spectrum ratios (to a minimum-bias INEL $ > 0$ reference) do show {\em qualitatively} that with increasing \nch\ spectra become ``harder'' at lower \pt\ but retain the same slopes (i.e.\ power-law exponents) at higher \pt. The first part of the statement then leads to inference of flows in \pp\ collisions as noted above. Figure~1 and related comments comprise the only attempt at analysis of differential spectrum structure in Ref.~\cite{alicepppid}. Contrast that with Secs.~\ref{pppidtcm}-\ref{pidratios} of the present study in which detailed differential spectrum structure is explored at the level of data statistical uncertainties, and the role of {\em minimum-bias jets} in spectrum evolution is quantitatively revealed.

\section{Summary}\label{summ}

The present study is a differential analysis of  identified-hadron (PID) \pt\ spectra for ten event classes derived from 143 million 13 TeV \pp\ collisions. The PID spectra are described by a two-component (soft + hard) model (TCM)  of hadron production mechanisms. The model soft component is associated with longitudinal projectile-nucleon dissociation while the hard component is quantitatively consistent with large-angle scattering of low-$x$ partons (gluons) to form a minimum-bias jet ensemble. Model parameter values are derived from those for 5 TeV \ppb\ collisions by extrapolation according to previously-determined TCM energy dependence. TCM parameters are not derived from fits to individual data spectra.

Given establishment of a TCM for 13 TeV \pp\ data as a {\em predictive} reference the following questions are addressed: (a) Does the TCM provide an accurate description of PID spectra from 13 \pp\ collisions? (b) Does \pp\ PID spectrum evolution with \nch\ (e.g.\ low-\pt\ ``hardening'' increasing with hadron mass) demonstrate the presence of collectivity (i.e.\ radial flow) in \pp\ collisions? (c) Are PID yield and spectrum ratios consistent with hydrodynamical flows? (d) What is the relation between 5 TeV \ppb\ and 13 TeV \pp\ PID spectra, especially the minimum-bias jet contribution? (e) Does the Blast-wave (BW) model applied to spectrum data confirm the presence of radial flow or even provide an adequate data description?

The principal analysis results are as follows: 
(a) The PID spectrum TCM provides an accurate (i.e.\ within statistical uncertainties) description of charged-kaon and (corrected) proton spectra. In the present study it is newly observed that missing protons are likely misidentified as pions, thus substantially biasing pion spectra.
(b) The TCM is thus statistically equivalent to \pp\ PID spectra. Qualitatively-observed spectrum evolution (e.g.\ ``hardening) is {\em quantitatively} explained by the interplay of soft (nonjet) and hard (jet-related) components according to the fundamental relation $\bar \rho_h \approx \alpha \bar \rho_s^2$. While jets are also a ``collective'' phenomenon (i.e.\ an instance of multiparticle {\em correlation}) there is no requirement to introduce a flow hypothesis.
(c) Several ratio variations with \nch\ and hadron mass that have been attributed to radial flow are actually jet manifestations quantitatively represented by the TCM spectrum hard component. Again, no flow hypothesis is required.
(d) Approximate invariance of PID spectrum hard components with \nch\ demonstrates that the jet component of hadron production in both \ppb\ and \pp\ collisions remains unmodified over a large range of hadron and jet densities. Energy evolution of hard components is as expected given measured jet energy spectra. PID data are consistent with no change in hadron species fractions ($z_{xi}$) between two collision systems and energies, consistent with statistical-model predictions.
(e) The BW model is strongly rejected by data according to the standard Z-scores statistic. The model does not provide evidence for the presence of radial flow.

Arguments for collectivity in small collision systems based on certain features appearing in 2D angular correlations and related statistics (e.g.\ Fourier coefficients, cumulants)  are outside the scope of this study. However, evidence from other responding studies, some related to the TCM as a general hadron production model, again favor an alternative description including minimum-bias jets and possibly other elementary few-gluon interactions, not hydrodynamic flows. Observation of data features in small collision systems similar to features in \aa\ collisions, attributed there to flows and QGP, may prompt challenges to the latter interpretation. The role of small systems as {\em control experiments} would then be restored.


\end{document}